\documentstyle[11pt,epsf,psfig]{article}
\newskip\humongous \humongous=0pt plus 1000pt minus 100pt
\def\caja{\mathsurround=0pt}
\def\eqalign#1{\,\vcenter{\openup1\jot \caja
       \ialign{\strut \hfil$\displaystyle{##}$&$
        \displaystyle{{}##}$\hfil\crcr#1\crcr}}\,}
\newif\ifdtup

\newcounter{eqnumber}[section]
\renewcommand{\theeqnumber}{\thesection.\arabic{eqnumber}}
\def\equn{
\refstepcounter{eqnumber}
\eqno({\rm \theeqnumber})
}

\savebox4{\rule[-2\unitlength]{0pt}{10\unitlength}%
\begin{picture}(10,10)(0,2)
\put(0,0){\line(0,1){10}}
\put(0,10){\line(1,0){10}}
\put(10,0){\line(0,1){10}}
\put(0,0){\line(1,0){10}}
\end{picture}}
\savebox5{\rule[-2\unitlength]{0pt}{10\unitlength}%
\begin{picture}(10,10)(0,2)
\put(0,10){\line(1,0){10}}
\put(10,0){\line(0,1){10}}
\put(0,10){\line(1,-1){10}}
\end{picture}}

\newlength\cellsizee 
\savebox6{%
\begin{picture}(18,18)
\put(0,0){\line(1,0){18}}
\put(0,0){\line(0,1){18}}
\put(18,0){\line(0,1){18}}
\put(0,18){\line(1,0){18}}
\end{picture}}
\newcommand\cellifyy[1]{\def\thearg{#1}\def\nothing{}%
\ifx\thearg\nothing
\vrule width0pt height\cellsize depth0pt\else
\hbox to 0pt{\usebox6\hss}\fi%
\vbox to 18\unitlength{
\vss
\hbox to 18\unitlength{\hss$#1$\hss}
\vss}}
\newcommand\stableau[1]{\vtop{\let\\=\cr
\setlength\baselineskip{-16000pt}
\setlength\lineskiplimit{16000pt}
\setlength\lineskip{0pt}
\halign{&\cellifyy{##}\cr#1\crcr}}}
\savebox7{%
\begin{picture}(15,15)
\put(0,0){\line(1,0){15}}
\put(0,0){\line(0,1){15}}
\put(15,0){\line(0,1){15}}
\put(0,15){\line(1,0){15}}
\end{picture}}
\newcommand\expathh[1]{%
\hbox to 0pt{\usebox3\hss}%
\vbox to 15\unitlength{
\vss
\hbox to 15\unitlength{\hss$#1$\hss}
\vss}}

\def\eqn#1{eq.~(\ref{#1})}
\def\Eqn#1{Eq.~(\ref{#1})}
\def\eqns#1#2{eqs.~(\ref{#1}) and~(\ref{#2})}

\def\fig#1{fig.~{\ref{#1}}}

\def\sec#1{section~{\ref{#1}}}

\def\app#1{appendix~\ref{#1}}

\newbox\charbox
\newbox\slabox
\def\s#1{{      
        \setbox\charbox=\hbox{$#1$}
        \setbox\slabox=\hbox{$/$}
        \dimen\charbox=\ht\slabox
        \advance\dimen\charbox by -\dp\slabox
        \advance\dimen\charbox by -\ht\charbox
        \advance\dimen\charbox by \dp\charbox
        \divide\dimen\charbox by 2
        \raise-\dimen\charbox\hbox to \wd\charbox{\hss/\hss}
        \llap{$#1$}
}}
\def\sandpm#1.#2.#3{%
  \left\langle\smash{#1}{\vphantom1}^{+}\right|{#2}%
  \left|\smash{#3}{\vphantom1}^{-}\right\rangle}
\def\spa#1.#2{\left\langle#1\,#2\right\rangle}
\def\spb#1.#2{\left[#1\,#2\right]}
\def\spab#1.#2.#3{\langle\mskip-1mu{#1}^- 
                  | #2 | {#3}^-\mskip-1mu\rangle}
\def\spba#1.#2.#3{\langle\mskip-1mu{#1}^+ 
                  | #2 | {#3}^+\mskip-1mu\rangle}
\def\spbb#1.#2.#3{\langle\mskip-1mu{#1}^+ 
                  | #2 | {#3}^-\mskip-1mu\rangle}
\def\lor#1.#2{\left(#1\,#2\right)}
\catcode`@=11  

\def\tr{\mathop{\rm tr}\nolimits}

\def\ph#1.#2{{\left[#1\,#2\right]} \over {\left\langle#1\,#2\right\rangle}}

\def\Tr{\, {\rm Tr}}
\def\Gr{\mathop{\rm Gr}\nolimits}

\def\eps{\epsilon}
\def\Ord{{\cal O}}

\def\dlips{d{\rm LIPS}}

\def\A{{\cal A}}
\def\M{{\cal M}}
\def\I{{\cal I}}
\def\Perm{{\cal P}}

\def\pol{\varepsilon}

\def\e{\epsilon}
\def\lr{\leftrightarrow}
\def\hf{{\textstyle {1\over2}}}

\def\la{\langle}
\def\ra{\rangle}
\def\tree{{\rm tree}}
\def\oneloop{{1 \mbox{-} \rm loop}}

\def\lsl{\s{\ell}}
\def\ksl{\s{k}}
\def\Ksl{\s{K}}

\def\Soft{\mathop {\cal S} \nolimits}
\def\SoftGrav{\mathop{{\cal S}^{\rm gravity}}\nolimits}
\def\Split{\mathop{\rm Split}\nolimits}

\def\SplitGravlam{\mathop{\rm Split^{\rm gravity}_{-\lambda}}\nolimits}
\def\SplitGravp{\mathop{\rm Split^{\rm gravity}_{+}}\nolimits}
\def\SplitGravm{\mathop{\rm Split^{\rm gravity}_{-}}\nolimits}
\def\cg{c_\Gamma}
\def\Li{\mathop{\rm Li}\nolimits}
\def\aX{\langle X \rangle}

\def\bX{[ X ]}
\def\btX{[ \tilde{X} ]}

\begin{document}
\begin{titlepage}
\begin{flushright}

hep-th/9811140 \hfill SLAC--PUB--7992\\
UCLA/98/TEP/28\\
UFIFT--HEP--98--33\\
November, 1998\\
\end{flushright}

\vskip 2.cm

\begin{center}
\begin{Large}
{\bf Multi-Leg One-Loop Gravity Amplitudes from Gauge Theory}
\end{Large}

\vskip 1.5cm

{\large Z. Bern$^{\star,1}$, L. Dixon$^{\dagger,2}$, 
M. Perelstein$^{\dagger,2}$ and J.S. Rozowsky$^{\ddagger,3}$}

\vskip 0.5cm

$^\star${\it Department of Physics,
University of California at Los Angeles,
Los Angeles,  CA 90095-1547}

\vskip .3cm

$^\dagger${\it Stanford Linear Accelerator Center,
Stanford University, Stanford, CA 94309}

\vskip .3cm
$^\ddagger${\it Institute for Fundamental Theory, Department of Physics, 
\vskip .01cm
University of Florida, Gainesville, FL 32611}

\vskip .3cm
\end{center}

\begin{abstract}
By exploiting relations between gravity and gauge theories, we present
two infinite sequences of one-loop $n$-graviton scattering amplitudes:
the `maximally helicity-violating' amplitudes in $N=8$ supergravity,
and the `all-plus' helicity amplitudes in gravity with any minimally
coupled massless matter content.  The all-plus amplitudes correspond
to self-dual field configurations and vanish in supersymmetric
theories. We make use of the tree-level Kawai-Lewellen-Tye (KLT)
relations between open and closed string theory amplitudes, which in
the low-energy limit imply relations between gravity and gauge theory
tree amplitudes.  For $n\le 6$, we determine the all-plus
amplitudes explicitly from their unitarity cuts.  The KLT relations,
applied to the cuts, allow us to extend to gravity a previously found
`dimension-shifting' relation between (the cuts of) the all-plus
amplitudes in gauge theory and the maximally helicity-violating amplitudes
in $N=4$ super-Yang-Mills theory.  The gravitational version of the 
relation lets us determine the $n\le6$ $N=8$ supergravity amplitudes 
from the all-plus gravity amplitudes.  We infer the two series of 
amplitudes for all $n$ from their soft and collinear properties, 
which can also be derived from gauge theory using the KLT relations.

\vskip .3 cm 
\noindent
{\it PACS:} 04.50.+h; 04.65+e;04.60.-m\\
{\it Keywords:} Unitarity; Cutting; Supersymmetry; Supergravity; One-loop
\end{abstract}

\vskip 1cm
\begin{center}
{\sl Submitted to Nuclear Physics B}
\end{center}

\vfill
\noindent\hrule width 3.6in\hfil\break
${}^{1}$Research supported by the US Department of Energy under grant 
DE-FG03-91ER40662.\hfil\break
${}^{2}$Research supported by the US Department of Energy under grant 
DE-AC03-76SF00515.\hfil\break
${}^{3}$Research supported by the US Department of Energy under grants 
DE-FG03-91ER40662 and DE-FG02-97ER41029.\hfil\break
\end{titlepage}

\baselineskip 16pt


\section{Introduction}
\label{IntroSection}

Of the four known forces, three are well-described by gauge theories,
and one by general relativity.  While the quantum-mechanical
properties of the former are reasonably well understood, the same 
cannot be said for the latter.  In particular, gravity contains a
dimensionful coupling parameter, $G_N = 1/M_{\rm Planck}^2$, and
therefore it is non-renormalizable at the level of power-counting.  On
the other hand, in perturbation theory certain relations have been
found expressing gravity amplitudes in terms of gauge theory
amplitudes.  Such relations may lead to a better understanding of the
properties of quantum gravity.  At the classical (tree) level, Kawai,
Lewellen and Tye (KLT)~\cite{KLT} found explicit general relations
between closed- and open-string amplitudes.  In the
infinite-string-tension limit the KLT relations provide a
representation of tree-level gravity amplitudes as the `square' of
tree-level gauge-theory amplitudes (where appropriate permutation
sums and kinematic prefactors also have to be applied to the
latter). These relationships, however, are not obvious from the point
of view of the Lagrangian and the associated Feynman diagrams.

As an example of how such relations can lead to more explicit
information about gravity amplitudes, one can consider special
helicity assignments for the external gravitons and gauge bosons
(`gluons').  The `maximally helicity-violating' (MHV) configurations are
those where exactly two particles have a helicity opposite to that of
the remaining $n-2$ particles.  (Tree amplitudes where all of the
external particles have the same helicity, or just one has the
opposite helicity, vanish trivially by a supersymmetry Ward identity
(SWI)~\cite{SWI}.)  In gauge theory, the MHV tree amplitudes are known for 
all $n$, and are remarkably compact~\cite{ParkeTaylor}.  
As we review in \sec{KnownRelationSection},
Berends, Giele and Kuijf (BGK)~\cite{BGK} used these gauge theory
results together with the KLT relations to find a closed-form
expression for the tree-level MHV gravity amplitudes.  They also
verified the universal behavior of these amplitudes as one of the
graviton momenta becomes soft~\cite{WeinbergSoftG}, providing a
non-trivial consistency check on their results.

At the quantum level gravity should presumably be regarded as only an
effective low-energy limit of some more fundamental theory, such as
string or $M$ theory.  It is nevertheless interesting to examine more
carefully the quantum (loop) behavior of gravity, and its connection
with gauge theory.  The amplitudes that we investigate in this paper
are ultraviolet finite in four dimensions, and should match the 
low-energy limit of any corresponding fundamental gravity theory.

The relation between gravity and gauge theory amplitudes has recently been
exploited at the loop level for four-point amplitudes in the maximally
supersymmetric versions of the two theories, namely $N=8$ supergravity and
$N=4$ super-Yang-Mills theory~\cite{BDDPR}.  It was shown that a squaring
relation exists for the coefficients of various multi-loop integrals
appearing in the two answers, similar to a relation found at one
loop~\cite{GSB,GSW}.  This result was obtained for all terms in the
two-loop amplitudes, and for some of the terms in higher-loop amplitudes
(those that can be reconstructed entirely from two-particle unitarity
cuts).  The multi-loop relations have led to an improved understanding of
the ultraviolet divergence structure of $N=8$ supergravity.  In
particular, at two loops the theory is finite for $D<7$, in contrast to a
previous expectation of divergences beginning at $D=5$, which was based on
an $N=4$ superspace power-counting~\cite{HSTHS}.  Although the entire
structure of the $N=8$ amplitudes is not yet known beyond two loops, the
entirely-two-particle-constructible terms suggest that $N=8$ supergravity
in $D=4$ is finite at three loops, and should only begin to diverge at
five loops~\cite{BDDPR}; this is again in contrast to previous
expectations~\cite{NEightAll,GSB,HSTHS}.

In this paper we continue the program of exploiting the relationship
between gravity and gauge theory amplitudes to obtain non-trivial
results for gravity.  We use this relationship to obtain
one-loop amplitudes with an arbitrary number $n$ of external
gravitons.  We consider two special types of helicity configurations:
(1) the same MHV configurations that were considered at tree level by
BGK, for the maximally supersymmetric theories of $N=8$ supergravity
and $N=4$ super-Yang-Mills; and (2) the `all-plus' configurations
where all the gravitons (or gauge bosons) have the same helicity, for
non-supersymmetric theories with arbitrary massless matter content.
(The all-plus amplitudes vanish for all supersymmetric theories.)  
For gauge theory, both infinite sequences (1) and (2) are 
known~\cite{SusyFour,AllPlus,MahlonAllPlus}, and are reviewed in 
\sec{OneLoopYMSection}. The all-plus gravity amplitudes have been
presented in ref.~\cite{GravAllPlus}; here we provide details of the 
construction, as well as new results for the $N=8$ amplitudes.

These two sequences of gravity (gauge theory) amplitudes are not as
different as one might expect, even though they involve different
helicity states and different matter content.  Indeed, in
ref.~\cite{DimShift} a `dimension-shifting' relation was exhibited
between the gauge theory amplitudes of types (1) and (2): The all-plus
gauge amplitudes can be obtained from the $N=4$ MHV amplitudes by
shifting the dimension of the loop integration upward by 4 units,
$\int d^D L \to \int d^{D+4} L$, and multiplying by an overall
prefactor.  This relation was explicitly verified for $n=4,5,6$.  In
\sec{GravDimShiftSubSection} we use a combination of the KLT relations
and unitarity to extend this relation from gauge theory to gravity,
where it implies that the all-plus gravity amplitudes can be obtained
from one-loop MHV $N=8$ supergravity amplitudes by shifting the
dimension upward by 8 units, $\int d^D L \to \int d^{D+8} L$.
(Ref.~\cite{DimShift} also speculated on this relation for gravity,
but only provided evidence for $n=4$.)

The dimension-shifting relation can be applied in either direction.  Here
we shall explicitly determine the all-plus $n$-graviton amplitudes for
$n=4,5,6$, by calculating their unitarity cuts in an arbitrary dimension
$D$ (see \sec{CutConstructionSection}).  The cut calculations may be
effectively performed by recycling the analogous gauge theory cut
calculations.  By working in an arbitrary dimension, we can determine the
complete amplitudes from the cuts, free of the subtraction ambiguities
frequently associated with dispersion relations \cite{Review}.  Then 
(in \sec{N8CutSubSection}) we use the dimension-shifting relation of 
\sec{GravDimShiftSubSection} to obtain the MHV $N=8$ supergravity
amplitudes.  This provides a non-trivial example, beyond those in
ref.~\cite{BDDPR}, where gauge theory properties may be used to 
derive analogous results for gravity theories.

The one-loop MHV $N=4$ amplitudes can be written as linear combinations of
a restricted class of scalar box (four-point) integrals.  The one-loop MHV
$N=8$ supergravity amplitudes can also be expressed in terms of the same
class of box integrals.  The relation between the coefficients of these
integrals for the $N=8$ and $N=4$ cases is reminiscent of the
tree-level KLT relations.

The unitarity-based results for gravity for $n \le 6$ provide the starting
point for constructing an ansatz for all $n$ which satisfies all known
analytic properties.  In order to go beyond the explicit graviton
amplitude calculations for $n\le 6$, we use the analytic properties of the
two series of amplitudes as the $n$-point kinematics approaches special
regions.  In particular, we study the soft limits noted above, as well as
the limits where two gravitons become collinear.  Both of these limits for
gravity can be understood at tree-level from the corresponding limits in
gauge theory, by exploiting the KLT relations, as we show in
\sec{SoftCollSection}.  We also show that there are no loop corrections to
the structure of these limits in the case of gravity (unlike gauge
theory).  In \sec{AnsatzSection} we obtain ansatze that satisfy the
appropriate limits for both the all-plus helicity (self-dual) gravity and
$N=8$ supergravity series of amplitudes.  Although we do not have a proof
that the $n>6$ amplitudes are the unique expressions with the proper
limits, we know of no counterexample with six- or higher-point kinematics
where this method of obtaining amplitudes has failed to produce the
correct expression.

The all-plus gravity and gauge amplitudes in $D=4$ are of interest in
part because of their connection with self-dual gravity
(SDG)~\cite{SDG} and self-dual Yang-Mills theory
(SDYM)~\cite{SDYMActions}, i.e. gravity and gauge theory restricted to
self-dual configurations of the respective field strengths,
$R_{\mu\nu\rho\sigma} = {i\over2}\eps_{\mu\nu}{}^{\alpha\beta}
R_{\alpha\beta \rho\sigma}$ and $F_{\mu\nu} = {i\over 2}
\eps_{\mu\nu}{}^{\alpha\beta} F_{\alpha\beta}$, with $\eps_{0123} =
+1$.  This connection is simple to see at the linearized (free) level
of superpositions of plane waves of identical helicity.  It has been
further studied at tree
level~\cite{DuffIshamEtc,Cangemi,GravPerturbiner} and
at the one-loop level~\cite{Cangemi,ChalmersSiegel}.
Chalmers and Siegel~\cite{ChalmersSiegel,CSUnpublished} have presented
self-dual actions for gravity and gauge theory which reproduce the
all-plus scattering amplitudes at both tree level and one loop.  Their
actions have no amplitudes beyond one loop, and the tree-level
amplitudes vanish on-shell.  Thus the one-loop all-plus amplitudes
constitute a complete perturbative solution to the theories defined by
the Chalmers-Siegel self-dual actions.  (See also
ref.~\cite{GravAllPlus}.)

In fact the one-loop gravity-gauge-theory relations can be stated in
terms of just the components of the gauge amplitudes that dominate in
the limit of a large number of colors, $N_c$.  (These components are the 
$A_{n;1}$ partial amplitudes
defined in \sec{OneLoopYMSection}.)  The large-$N_c$ limit of $N=4$
super-Yang-Mills theory has recently attracted much attention through
its connection to superstring configurations in anti-de-Sitter
space~\cite{AdS}.  In this context, gauge theories with $N<4$
supersymmetry have been constructed by an
orbifold-style~\cite{Orbifold} truncation of the $N=4$ spectrum, and
it has been argued that at large-$N_c$ their amplitudes actually
coincide (up to overall constants) with those of the $N=4$
theory~\cite{LessSUSY}.  These results may provide some additional
motivation for studying the relation between large-$N_c$ $N=4$
amplitudes and supergravity amplitudes, although we know of no direct
connection between the purely perturbative relations found here and
the anti-de-Sitter-space results, which are non-perturbative, involving
a weak $\lr$ strong coupling duality.


\section{Review of Tree-Level Properties}
\label{KnownRelationSection}

In this section we review the tree-level KLT relations~\cite{KLT}, and
the known analytic properties of tree-level amplitudes in gauge theory
and in gravity.  These properties, as well as additional ones derived
in \sec{SoftCollSection}, will be used in \sec{AnsatzSection} to
obtain an ansatz for the all-plus gravity and MHV $N=8$
supergravity amplitudes with an arbitrary number of external legs.


\subsection{KLT Relations}
\label{KLTSubSection}

The KLT relations are between tree-level amplitudes in closed and open
string theories, and arise from the representation of any closed-string
vertex operator as a product of open-string vertex operators,
$$
V^{\rm closed}(z_i,\bar{z}_i) = V_{\rm left}^{\rm open}(z_i)\, 
\overline{V}_{\rm right}^{\rm  open}(\bar{z}_i) \,.
\equn\label{ClosedVertex}
$$
The left and right string oscillators appearing in $V_{\rm left}$ and
$\overline{V}_{\rm right}$ are distinct, but the zero mode momentum is
shared.  In the open-string tree amplitude, the $z_i$ are real
variables, to be integrated over the boundary of the disk, while in
the closed-string tree amplitude the $z_i$ are complex and integrated
over the sphere.  The closed-string integrand is thus a product of two
open-string integrands.  This statement holds for any set of
closed-string states, since they can all be written as tensor products
of open-string states.  KLT evaluated the $(n-3)$ two-dimensional
closed-string world-sheet integrals, via a set of contour-integral
deformations, in terms of the $(n-3)$ open-string integrals, and
thereby related the two sets of string amplitudes.

After taking the field-theory limit~\cite{GSB,GSW}, 
$\alpha' k_i\cdot k_j \to 0$, the KLT relations for four-, five- and 
six-point amplitudes are~\cite{BGK}, 
$$
\eqalign{
M_4^\tree (1,2,3,4) & = - i s_{12} A_4^\tree(1,2,3,4) \,
   A_4^\tree(1,2,4,3)\,, \cr
M_5^\tree(1,2,3,4,5) 
& = i s_{12} s_{34}  A_5^\tree(1,2,3,4,5) \, A_5^\tree(2,1,4,3,5)  \cr
& \hskip 2 cm 
+ i s_{13}s_{24} A_5^\tree(1,3,2,4,5) \, A_5^\tree(3,1,4,2,5) \,, \cr
M_6^\tree(1,2,3,4,5,6) 
& = - i s_{12} s_{45}  A_6^\tree(1,2,3,4,5,6) 
    [ s_{35} A_6^\tree(2,1,5,3,4,6)  \cr
& \hskip 4 cm 
   + (s_{34} + s_{35}) A_6^\tree(2,1,5,4,3,6) ] \cr
& \hskip 2 cm +\ \Perm(2,3,4) \,. \cr}
\equn\label{GravYM}
$$
Here the $M_n$'s are the amplitudes in a gravity theory stripped of
couplings, the $A_n$'s are the color-ordered amplitudes in a gauge
theory~\cite{TreeColor,DecouplingIdent}, 
$s_{ij}\equiv (k_i+k_j)^2$, and $\Perm(2,3,4)$
instructs one to sum over all permutations of the labels 2, 3 and 4.
The $n$ arguments of $M_n$ and $A_n$ are the external states $j$,
which have momentum $k_j$.  The $n$-point generalization of \eqn{GravYM}
\cite{KLT,BGK} is presented in \app{UniversalAppendix}.

Each gravity state $j$ appearing in $M_n$ is the tensor
product of the corresponding two gauge theory states appearing in the
$A_n$'s on the right-hand side of the equation.  In particular, each
of the 256 states of the $N=8$ supergravity multiplet, consisting of 1
graviton, 8 gravitinos, 28 gauge bosons, 56 gauginos, and 70 real
scalars, can be interpreted as a tensor product of two sets of the 16
states of the $N=4$ super-Yang-Mills multiplet, consisting of 1 gluon,
4 gluinos and 6 real scalars.  (In string theory, this correspondence 
may be understood in terms of the factorization of the closed string vertex
operator for each $N=8$ state into a product of $N=4$ open
string vertex operators.)  Thus a sum over the $N=8$ supergravity
states can be interpreted as a double sum over a tensor product of
$N=4$ super-Yang-Mills states.

Full amplitudes are obtained from $M_n^\tree$ and $A_n^\tree$ via,
$$
\eqalign{
\M_n^\tree(1,2,\ldots,n) &= 
\left({  \kappa \over 2} \right)^{(n-2)} 
M_n^\tree(1,2,\ldots,n)\,,
\cr
\A_n^\tree(1,2,\ldots,n) &=  g^{(n-2)} \sum_{\sigma \in S_n/Z_n}
{\rm Tr}\left( T^{a_{\sigma(1)}} 
T^{a_{\sigma(2)} }\cdots T^{a_{\sigma(n)}} \right)
 A_n^\tree(\sigma(1), \sigma(2),\ldots, \sigma(n)) \,,
\cr}
\equn
$$
where $\kappa^2 = 32\pi G_N$, and $S_n/Z_n$ is the set of all
permutations, but with cyclic rotations removed. The $T^{a_i}$ are
fundamental representation matrices for the Yang-Mills gauge group
$SU(N_c)$, normalized so that $\Tr(T^aT^b) = \delta^{ab}$.

The relations~(\ref{GravYM}) hold for arbitrary external states.  For
external gravitons and gluons it is convenient to quote the results in
a helicity basis, using the spinor helicity
formalism~\cite{SpinorHelicity}.  At tree-level --- and to all orders
for supersymmetric theories --- helicity amplitudes where all, or all
but one, of the external particles have the same helicity vanish by a
SWI~\cite{SWI},
$$
 M_n(\pm,+,+,\ldots,+) = A_n(\pm,+,+,\ldots,+) = 0 \,,
\equn\label{SWIVanish}
$$
where the helicity assignments are for outgoing particles.
These relations hold for any states in the respective $N=8$ and $N=4$
multiplets.  (For scalar states, one interprets `helicity' as particle
vs. anti-particle.)

For a given number of external legs $n$, the simplest non-vanishing
tree amplitudes --- and supersymmetric loop amplitudes --- are the
maximally helicity-violating (MHV) amplitudes, where exactly two
helicities are opposite to the majority.  At tree-level, and to all loop
orders in $N=4$ super-Yang-Mills theory, the MHV $n$-gluon amplitudes are
all related to each other by the $N=4$ SWI~\cite{DimShift},
$$
{1\over\spa{i}.{j}^4}\,
 \A_n(1^+, 2^+, \ldots, i^-, \ldots, j^-, \ldots, n^+) = 
{1 \over \spa{a}.{b}^4} \,
\A_n(1^+, 2^+, \ldots, a^-, \ldots, b^-, \ldots, n^+) \,,
\equn\label{PermIdentityYM}
$$
where $i$ and $j$ are the only negative helicity legs on the left-hand
side and $a$ and $b$ are the only negative helicities on the
right-hand side.  (We will generally indicate the type of external
particle by a subscript; e.g. $1_g^-$ for a negative-helicity gluon.
However, for gluons in gauge theory and gravitons in gravity, we will
usually omit the subscript.)  Thus at tree-level it suffices to give
the formula~\cite{ParkeTaylor}
$$
A_n^\tree(1^-, 2^-, 3^+, 4^+, \ldots, n^+) 
= i \, {\spa1.2^4 \over \spa1.2\spa2.3\spa3.4\cdots \spa{n}.1}\,.
\equn\label{YMTreeMHV}
$$
We use the notation 
$\la k_i^- | k_j^+\ra = \spa{i}.j$ and $\la k_i^+| k_j^-\ra=\spb{i}.j$, 
where $|k_i^{\pm}\ra$ are massless Weyl spinors, labeled with the sign of
the helicity and normalized by 
$\spa{i}.j \spb{j}.i = s_{ij} = 2k_i\cdot k_j$.
For the case where both energies are positive the spinor inner
products are given by 
$$
\spa{i}.{j} = \sqrt{|s_{ij}|} e^{i\phi_{ij}} \,, \hskip 3 cm 
\spb{i}.{j} = \sqrt{|s_{ij}|} e^{-i(\phi_{ij}+\pi)} \,, 
\equn\label{SpinorExplicit} 
$$
where
$$
\cos\phi_{ij}\ =\ { k_i^1 k_j^+ - k_j^1 k_i^+ 
             \over \sqrt{|s_{ij}| k_i^+ k_j^+} }
\, , \hskip 2cm 
\sin\phi_{ij}\ =\ { k_i^2 k_j^+ - k_j^2 k_i^+ 
             \over \sqrt{|s_{ij}| k_i^+ k_j^+} }\,,
\equn\label{SpinorPhase}
$$
and $k_i^+ = k_i^0 + k_i^3$.  (The cases where one or both of the
energies are negative are similar, except for additional overall
phases.)  For later use, we also define the spinor strings
$$
\eqalign{
\spab{i}.{(l+m)}.{j} &\equiv 
\langle k_i^- \, |\, (\ksl_l+\ksl_m) \, | \, k_j^- \rangle \,, \cr
\spbb{i}.{lm\cdots}.{j} &\equiv 
\langle k_i^+ \, |\, \ksl_l\ksl_m\cdots \, | \, k_j^- \rangle \,, \cr
\spab{i}.{\ell_m}.{j} &\equiv 
\langle k_i^- \, |\, \lsl_m \, | \, k_j^- \rangle \,, \cr
}\equn\label{SpinorStringDef}
$$ 
etc., where $\ell_m$ is a loop momentum.

The MHV graviton tree (and $N=8$ loop) amplitudes satisfy an $N=8$ SWI 
analogous to \eqn{PermIdentityYM},
$$
{1\over\spa{i}.{j}^8}\,
 \M_n(1^+, 2^+, \ldots, i^-, \ldots, j^-, \ldots, n^+) = 
{1 \over \spa{a}.{b}^8} \,
\M_n(1^+, 2^+, \ldots, a^-, \ldots, b^-, \ldots, n^+) \,.
\equn\label{PermIdentityGrav}
$$
The MHV four- and five-graviton tree amplitudes,%
\footnote{Our overall phase conventions differ from those of
ref.~\cite{BGK} by a `$-i$'.}  
which satisfy \eqn{PermIdentityGrav} as well as the appropriate KLT
relations (\ref{GravYM}), are~\cite{BGK}
$$
\eqalign{
M_4^\tree(1^-, 2^-, 3^+, 4^+)  &=  
i \, \spa1.2^8 { \spb1.2 \over  \spa3.4 \, N(4) }  \,, \cr
M_5^\tree(1^-, 2^-, 3^+, 4^+, 5^+)  &=  
i \, \spa1.2^8 { \pol(1,2,3,4) \over N(5) } \,, \cr}
\equn\label{GravTreeFourFive}
$$
where 
$$
\pol(i,j,m,n) 
\equiv 4i\pol_{\mu\nu\rho\sigma} k_i^\mu k_j^\nu k_m^\rho k_n^\sigma
\ =\ \spb{i}.{j}\spa{j}.{m}\spb{m}.{n}\spa{n}.{i}
   - \spa{i}.{j}\spb{j}.{m}\spa{m}.{n}\spb{n}.{i} \,,
\equn\label{LeviCivitaDef}
$$
and
$$
 N(n) \equiv \prod_{i=1}^{n-1} \prod_{j=i+1}^n \spa{i}.{j} \,.
\equn\label{NnDef}
$$

For $n>4$, Berends, Giele and Kuijf~\cite{BGK} presented the expression,
$$
\eqalign{
M_n^\tree(1^-, 2^-, 3^+, \ldots, n^+) &= -i \, {\spa1.2}^8 \cr
& \hskip-3cm \times \Biggl[ 
 { \spb{1}.{2} \spb{n-2}.{\ n-1} \over \spa{1}.{\ n-1} \, N(n) }
 \biggl( \, \prod_{i=1}^{n-3} \prod_{j=i+2}^{n-1} \spa{i}.{j} \biggr)
   \prod_{l=3}^{n-3} \Bigl( - \spab{n}.{\Ksl_{l+1,n-1}}.{l} \Bigr)
\ +\ \Perm(2,3,\ldots,n-2) \Biggr] \,, \cr}
\equn\label{BGKMHV}
$$
where $K^\mu_{i,j} \equiv \sum_{s=i}^{j} k_s^\mu$, and $+\Perm(M)$
instructs one to sum the quantity inside the brackets over all
permutations of the set $M$.  They numerically verified its
correctness for $n \leq 11$.  The expression in brackets is totally
symmetric (although this is not manifest), as is required for consistency 
with \eqn{PermIdentityGrav}.


\subsection{Soft and Collinear Properties at Tree Level}
\label{SoftCollinearSubSection}

There are two important universal limits of color-ordered $n$-gluon tree
amplitudes.  In the limit that the gluon $s$ becomes soft, $A_n^\tree$ has
the universal behavior~\cite{SoftBG},
$$
A_n^\tree(\ldots,a,s^\pm,b,\ldots)\ \mathop{\longrightarrow}^{k_s\to0}\
    \Soft^\tree(a,s^\pm,b) \times
      A_{n-1}^\tree(\ldots,a,b,\ldots) \,,
\equn\label{YMTreeSoft}
$$
where the soft (eikonal) factors are
$$
\eqalign{
\Soft^\tree(a,s^+,b)
           \ &=\ {\spa{a}.{b}\over\spa{a}.{s}\spa{s}.{b}}\ ,\cr
\Soft^\tree(a,s^-,b)
           \ &=\ {-\spb{a}.{b}\over\spb{a}.{s}\spb{s}.{b}}\ .\cr}
\equn\label{YMTreeSoftFactor}
$$
In the collinear limit where two gluon momenta $k_a$ and $k_b$ become
parallel (denoted by $a \parallel b$), we have $k_a \approx z k_P$ and 
$k_b \approx (1-z) k_P$ for some $z \in [0,1]$, where $k_P \equiv k_a +
k_b$.  The behavior of a tree amplitude in this limit
is~\cite{ManganoReview}
$$
A_n^\tree(\ldots,a^{\lambda_a},b^{\lambda_b},\ldots)\ 
\mathop{\longrightarrow}^{a \parallel b}\
\sum_{\lambda=\pm}
 \Split^\tree_{-\lambda}(z,a^{\lambda_a},b^{\lambda_b}) \times
      A_{n-1}^\tree(\ldots,P^\lambda,\ldots)\ ,
\equn\label{YMTreeColl}
$$
where the gluon splitting amplitudes are 
$$
\eqalign{
 \Split^\tree_{+}(z,a^{+},b^{+})\ &=\ 0\,,\cr
 \Split^\tree_{-}(z,a^{+},b^{+})
           \ &=\ {1\over \sqrt{z (1-z)}\spa{a}.b}\, ,\cr
 \Split^\tree_{+}(z,a^{-},b^{+})
          \ &=\ \sqrt{{z^3 \over 1-z}}{1\over\spa{a}.b}\, ;\cr}
\equn\label{YMTreeSplit}
$$
the remaining ones may be obtained by parity.  Similar expressions
exist including fermions.  For example, for a gluon splitting
into two fermions, the color-ordered splitting amplitudes are
$$
\Split^\tree_{+}(z,a_{\bar{q}}^{-},b_q^{+})
            =  {z\over \spa{a}.{b}}\, , 
\hskip2cm
\Split^\tree_{-}(z,a_{\bar{q}}^{-},b_q^{+})
            =  {1-z\over \spb{a}.{b}}\, .
\equn\label{QuarkTreeSplit}
$$
A more complete discussion of splitting amplitudes may be found in 
reviews~\cite{ManganoReview,LDTASI,Review}.  (Our sign conventions
in the splitting functions are the ones used in ref.~\cite{Fermion}.)

Gravity amplitudes, like gauge amplitudes, are known to satisfy universal 
soft limits\cite{WeinbergSoftG,BGK}.  The gravitational soft limits have
the form,
$$
M_n^\tree(\ldots,a,s^\pm,b,\ldots)\ \mathop{\longrightarrow}^{k_s\to0}\
    \SoftGrav(s^\pm) \times
   M_{n-1}^\tree(\ldots,a,b,\ldots)\, .
\equn\label{GravTreeSoft}
$$
For the limit $k_n \rightarrow 0$ in $M_n^\tree(1,2,\ldots,n)$,
the gravitational soft factor (for positive helicity) is
$$
\eqalign{
\Soft_n\ \equiv\
\SoftGrav(n^+)
\ &=\ { -1 \over \spa{1}.{n} \spa{n,}.{n-1} }
  \sum_{i=2}^{n-2} { \spa{1}.{i} \spa{i,}.{n-1} \spb{i}.{n}
                                         \over \spa{i}.{n} }\, .\cr}
\equn\label{GravTreeSoftFactor}
$$
Although it is not manifest, $\Soft_n$ is also symmetric under the
interchange of legs $1$ and $n-1$ with the others.  For all $n$, BGK
verified that the MHV amplitudes~(\ref{BGKMHV}) have the correct
behavior as an external momentum becomes soft.  

In \sec{SoftCollSection} we will show that gravity tree amplitudes also
have universal behavior as two external momenta become collinear, and that
the splitting amplitudes are composed of products of pairs of the ones for
gauge theory.  We shall further demonstrate that the tree-level soft
factors and splitting amplitudes for gravity do not incur any higher loop 
corrections, in contrast to the situation for gauge theory.


\section{One-Loop MHV Amplitudes in Gauge Theory}
\label{OneLoopYMSection}

Before proceeding to gravity, it is useful to review the structure of the
one-loop maximally helicity-violating (MHV) amplitudes in gauge
theory~\cite{AllPlus,MahlonAllPlus,SusyFour,SusyOne}.  These
amplitudes were constructed by techniques similar to those used in the
following sections for the corresponding gravity amplitudes.  In
particular we shall use unitarity or cutting techniques
\cite{Cutting,SusyFour,SusyOne,Massive}, as well as the factorization
bootstrap approach~\cite{ParkeTaylor,AllPlus} of finding ansatze for
amplitudes based on their known kinematic poles.  In
\app{RecursiveAppendix} we outline an alternative
approach to the all-plus gravity amplitudes (as well as to various
tree-level gravity amplitudes), one based on recursive 
techniques~\cite{RecursiveBG,RecursiveK,MahlonAllPlus}.

The unitarity cuts in the $N=4$ MHV gauge case are simple enough that the
direct computation can be performed for all $n$ simultaneously.  We
have not been able to do that yet for the analogous $N=8$ MHV supergravity
computation, and so we shall resort to an ansatz for $n>6$, based on soft 
and collinear limits.


\subsection{General Properties of One-Loop Amplitudes}
\label{OneLoopSubSection}

We first define one-loop amplitudes $M_n$ for gravity and $A_{n;j}$ for
gauge theory, from which all couplings have been removed.  Color has
also been removed from the $A_{n;j}$, according to the one-loop color
decomposition~\cite{LoopColor}.  For the case where all states are in
the adjoint representation, the full amplitudes are given by,
$$
\eqalign{
\M_n^\oneloop(1,2,\ldots,n) &= 
\left({  \kappa \over 2} \right)^{n} 
M_n(1,2,\ldots,n)\,,
\cr
\A_n^\oneloop(1,2,\ldots,n) &=  g^n
\sum_{j=1}^{\lfloor{n/2}\rfloor+1} \sum_{\sigma \in S_n/S_{n;j}}
  \Gr_{n;j}(\sigma) \, A_{n;j}(\sigma(1),\ldots,\sigma(n)) \,,\cr}
\equn\label{OneLoopColorDecomp}
$$
where $\lfloor x\rfloor$ denotes the greatest integer less than or
equal to $x$, the (unpermuted) color trace structures are 
$\Gr_{n;1}(1) \equiv N_c \, \Tr\bigl(T^{a_1}\cdots T^{a_n}\bigr)$
and $\Gr_{n;j}(1) = \Tr\bigl(T^{a_1}\cdots T^{a_{j-1}}\bigr) 
\, \Tr\bigl(T^{a_j}\cdots T^{a_n}\bigr)$ for $j>1$,
and $S_{n;j}$ is the subset of permutations $S_n$ that leaves the
trace structure $\Gr_{n;j}$ invariant.  Similar color decompositions
exist for the cases with fundamental representation particles in the
loop.  In fact, the partial amplitudes $A_{n;j}$ for $j>1$ can be
expressed in terms of the $A_{n;1}$ through the
formula~\cite{SusyFour},
$$
 A_{n;j}(1,2,\ldots,j-1;j,j+1,\ldots,n)\ =\
 (-1)^{j-1} \sum_{\sigma\in COP\{\alpha\}\{\beta\}} A_{n;1}(\sigma)\, .
\equn\label{SublAnswer}
$$
Here $\alpha_i \in \{\alpha\} \equiv \{j-1,j-2,\ldots,2,1\}$,
$\beta_i \in \{\beta\} \equiv \{j,j+1,\ldots,n-1,n\}$,
and $COP\{\alpha\}\{\beta\}$ is the set of all
permutations of $\{1,2,\ldots,n\}$ with $n$ held fixed
that preserve the cyclic
ordering of the $\alpha_i$ within $\{\alpha\}$ and of the $\beta_i$
within $\{\beta\}$, while allowing for all possible relative orderings
of the $\alpha_i$ with respect to the $\beta_i$.

Thus the full gauge amplitude can be constructed just from the
$A_{n;1}$, which are {\it color-ordered} (i.e, they only receive
contributions from planar graphs with a fixed ordering of the external
legs), and therefore have simpler analytic properties than the
remaining $A_{n;j}$.  For this reason we need only explicitly discuss
the case of $A_{n;1}$.  The $A_{n;1}$ contributions are the ones which
dominate the amplitude for a large number of colors $N_c$.

We consider one-loop amplitudes where the external momenta are taken
to lie in four dimensions, but the number of dimensions $D$ appearing
in the loop-momentum integration measure $d^D L$ remains arbitrary
(for the time being).  (To maintain supersymmetry we leave the number
of states at their four-dimensional values.)  In general, the
$m$-point loop integrals with $m\geq5$ which appear in such one-loop
amplitudes can be reduced down to at most box (four-point) integrals
and pentagon (five-point) integrals, where the pentagon integrals are
scalar integrals (i.e., they contain no loop momenta in the numerator
of the integrand) and are evaluated in $D+2$ dimensions~\cite{Integrals}.
Furthermore, if we now set $D=4-2\e$, then to $\Ord(\e^0)$ the pentagon 
contributions may be neglected, because the scalar pentagon integrals in 
$D=6-2\e$ have no poles as $\e \to 0$ (they are infrared and ultraviolet 
finite in $D=6$), and because they are generated in the integral 
reduction procedure with a manifest $\e$ prefactor~\cite{Integrals}.

For an amplitude in a generic theory, after applying these reductions
the box integrals may have powers of the loop momentum $L$ inserted
in the numerator of the integrand, in addition to the four scalar
propagators which make up the denominator.  The amplitude may also
contain triangle and bubble integrals arising from the corresponding
Feynman diagrams.  The all-plus helicity and $N=4$ supersymmetric
cases which we discuss below are special, however, and do not contain
the full set of possible scalar integrals.


\subsection{Review of One-Loop Soft and Collinear Properties in 
Gauge Theory}

The collinear limits for the leading color-ordered one-loop amplitudes, 
$A_{n;1}$, are similar to the tree-level case and have the form
$$
\eqalign{
A_{n;1}(\ldots,a^{\lambda_a},b^{\lambda_b},\ldots)\ 
\mathop{\longrightarrow}^{a \parallel b}\
\sum_{\lambda=\pm}  \biggl(
 & \Split^\tree_{-\lambda}(z, a^{\lambda_a},b^{\lambda_b})\,
      A_{n-1;1}(\ldots,P^\lambda,\ldots) \cr
& + \Split^\oneloop_{-\lambda}(z,a^{\lambda_a},b^{\lambda_b})\,
      A_{n-1}^\tree(\ldots,P^\lambda,\ldots) \biggr) \,.
\cr}
\equn\label{OneLoopSplit}
$$
In addition to the tree-level splitting amplitudes,
$\Split^\tree_{-\lambda}(z,a^{\lambda_a},b^{\lambda_b})$,
one-loop corrections now also appear,
$\Split^\oneloop_{-\lambda}(z,a^{\lambda_a},b^{\lambda_b})$.
Both quantities are universal, depending only on the two momenta becoming 
collinear, and not upon the specific amplitude under
consideration~\cite{Factorization}.  The explicit values of the
$\Split^\oneloop_{-\lambda}(z,a^{\lambda_a},b^{\lambda_b})$ (which we shall
not need here) were originally determined~\cite{SusyFour} from the 
four-~\cite{FourParton} and five-point~\cite{FiveParton,Fermion} one-loop 
gauge amplitudes.  Their universality for an arbitrary number of external 
legs was demonstrated in ref.~\cite{Factorization}.

Similarly, as the momentum of an external leg becomes soft the
color-ordered one-loop amplitudes behave as,
$$
\eqalign{
A_{n;1}(\ldots,a,s^\pm,b,\ldots)\ &\mathop{\longrightarrow}^{k_s\to0}\ 
    \Soft^\tree(a,s^\pm,b)\,
      A_{n-1;1}(\ldots,a,b,\ldots) \cr
& \hskip 2 cm 
  + \Soft^\oneloop(a,s^\pm,b)\,
      A_{n-1}^\tree(\ldots,a,b,\ldots) \,, \cr}
\equn\label{OneLoopSoft}
$$
where $\Soft^\oneloop(a,s^\pm,b)$ is universal.

In the application of \eqns{OneLoopSplit}{OneLoopSoft} to the collinear
and soft limits of the one-loop all-plus gauge amplitudes, the second
term always drops out, because of the vanishing of the tree-level 
amplitudes with all plus helicities, or all but one plus.
In the $N=4$ super-Yang-Mills MHV case, however, the contributions of
$\Split^\oneloop_{-\lambda}(z,a^{\lambda_a},b^{\lambda_b})$ and
$\Soft^\oneloop(a,s^\pm,b)$ survive.


\subsection{All-Plus Amplitudes}
\label{AllPlusGaugeSubSection}

The analytic properties of the one-loop all-plus amplitudes in
gauge theory~\cite{AllPlus,MahlonAllPlus} are remarkably simple.  First of
all, the unitarity cuts vanish in four dimensions, since \eqn{SWIVanish}
implies that at least one of the two tree amplitudes on either side of a
unitarity cut vanishes, for every possible helicity assignment for the two
gluons crossing the cut.  Similarly, by considering their factorization on
particle poles, one finds that the one-loop all-plus amplitudes cannot
contain multi-particle poles, i.e., factors of the form
$1/(k_{i_1}+k_{i_2}+\cdots+k_{i_m})^2$ with $m>2$.  The only permitted
kinematic singularities are the ones where one external momentum becomes
soft, or two external momenta become collinear.  Finally, the loop-momentum
integration does not generate any infrared nor ultraviolet divergences or
associated logarithms.  In summary, $A_{n;1}(1^+,2^+,\ldots,n^+)$ is a
finite rational function of the momenta, cyclically symmetric in its $n$
arguments, with singularities only in the regions where a momentum is soft
or two cyclically (color) adjacent momenta are collinear.

These analytic properties were crucial in obtaining an ansatz for the
explicit form of the amplitudes~\cite{AllPlus}, which was verified by
Mahlon via recursive techniques~\cite{MahlonAllPlus}.  The one-loop 
amplitudes for $n$ identical-helicity gluons in pure Yang-Mills
theory are,
$$
A_{n;1}(1^+,2^+,\ldots,n^+) = -{i \over 48\pi^2}\,
\sum_{1\leq i_1 < i_2 < i_3 < i_4 \leq n}
{ {\rm tr}_-[i_1 i_2 i_3 i_4] \over \spa1.2 \spa2.3 \cdots \spa{n}.1}\,,
\equn\label{YMAllnPlus}
$$
where $\tr_\pm[i_1 i_2 i_3 i_4] \equiv {1\over 2}\tr[(1\pm\gamma_5)
\ksl_{i_1} \ksl_{i_2} \ksl_{i_3} \ksl_{i_4}]$.
These amplitudes are generated by actions for self-dual 
Yang-Mills theory~\cite{Cangemi,ChalmersSiegel} as well as ordinary
gauge theory.

The all-plus amplitudes vanish in any supersymmetric theory by the
SWI~(\ref{SWIVanish}).  Thus the contribution from a gluon circulating
around the loop is the negative of that from an adjoint fermion in the loop,
and equal to that from an adjoint scalar.   For a fundamental 
representation fermion one must divide the contribution to $A_{n;1}$ by 
an additional factor of $N_c$. In particular, for QCD with $n_{\! f}$
flavors of quarks we have
$$
A_{n;1}^{\rm QCD} (1^+,2^+,\ldots,n^+) = 
\Bigl(1 - {n_{\! f} \over N_c} \Bigr) A_{n;1}(1^+,2^+,\ldots,n^+) \,.
\equn
$$

For $n\le 6$, the all-plus amplitudes have also been computed
to all orders in the dimensional regularization parameter $\eps$ (but
with four-dimensional external momenta) via their unitarity
cuts~\cite{DimShift}.  These amplitudes may be compactly expressed as 
$$
\eqalign{
A_{4;1}(1^+,2^+,3^+,4^+) &=
{ -2s_{12}s_{23} \over \spa1.2\spa2.3\spa3.4\spa4.1}
  \I_4^{1234}[\mu^4] \,, \cr
A_{5;1}(1^+,2^+,3^+,4^+,5^+) &=
 {1\over \spa1.2\spa2.3\spa3.4\spa4.5\spa5.1}
\Bigl\{
\Bigl[ -s_{12}s_{23} \I_4^{123(45)}[\mu^4] +\hbox{cyclic} \Bigr] \cr
& \hskip 5 cm 
+ 2 \, \pol(1,2,3,4) \I_5^{12345}[\mu^6] \Bigr\} \,, \cr
A_{6;1}(1^+,2^+,3^+,4^+,5^+,6^+) &=
 {1\over  \spa1.2\spa2.3\spa3.4\spa4.5\spa5.6\spa6.1} \Biggl\{ 
\biggl[ -s_{12} s_{23}  \, \I_4^{123(456)}[\mu^4]  \cr
& \hskip0.3cm 
- {1\over2} (t_{123} t_{234} - s_{23} s_{56}) \, \I_4^{1(23)4(56)}[\mu^4] 
+ \pol(1,2,3,4)\, \I_5^{1234(56)}[\mu^6] + \hbox{cyclic} \biggr] \cr
& \hskip0.3cm 
-  \tr[123456]  \, 
 \I_6^{123456}[\mu^6] \Biggr\} \,, \cr}
\equn\label{ExactYM}
$$
where 
$s_{ij} = (k_i + k_j)^2$, $t_{ijl} = (k_i + k_j + k_l)^2$, and `$+$~cyclic'
implies a sum over the $n$ cyclic permutations (for $A_{n;1}$) of the 
quantity within the brackets $([\ ])$ in which the phrase appears.

In \eqn{ExactYM}, $\I_5^{12345}$ and $\I_6^{123456}$ are pentagon and
hexagon integrals where all the external legs are massless.  The integral 
$\I_5^{1234(56)}$ is a one-mass pentagon integral, where legs $5$ and $6$ 
form the one external mass.  The parentheses in the arguments
of the one- and two-mass box integrals $\I_4$ similarly indicate the 
grouping of massless external legs for the amplitude into massive legs 
for the integral.  (See \app{IntegralAppendix} for
further exposition of our notation for the integrals.)
In the integration measure $d^D L$ for the integrals, the 
$(D=4-2\e)$-dimensional loop momentum $L$ can be decomposed into a 
4-dimensional part $\ell$ and a $(-2\eps)$-dimensional part $\mu$,
as $L=\ell+\mu$.  Following the prescriptions of 't Hooft and
Veltman~\cite{DimensionalRegularization}, we take the four- and
$(-2\eps)$-dimensional parts of the loop momenta to be orthogonal, so
that $L^2 = \ell^2 - \mu^2$.  The symbol `$[\mu^{2r}]$' instructs one 
to insert an extra factor of $\mu^{2r} \equiv (\mu^2)^r$ into the loop 
integrand before performing the integral, i.e.,
$$
\I_m[\mu^{2r}] 
\equiv \int {d^{4}\ell \over (2\pi)^{4} }
    {d^{-2\e} \mu \over (2\pi)^{-2\e} }
  { (\mu^2)^r \over (\ell^2 - \mu^2)((\ell-K_1)^2 - \mu^2) \ldots (
    (\ell-\sum_{i=1}^{m-1} K_i )^2 - \mu^2) } \,.
\equn\label{MuIntegralDef}
$$
Carrying out the $\mu$ integration explicitly leads to the 
formula~\cite{MahlonAllPlus,Massive,DimShift}
$$
\I_m[\mu^{2r}] =  -\e (1-\e) \cdots (r-1-\e) \, (4\pi)^r
\, \I_m^{D=4+2r-2\e} \,,
\equn\label{DimShiftIntegral}
$$
where $\I_m^{D=4+2r-2\e}$ is $\I_m$ with the number of dimensions
in the loop-momentum integration shifted upward by $2r$; i.e., one
replaces $D\to D+2r$ in \eqn{IntegralDef}.

The fact that the all-orders-in-$\e$ formulas~(\ref{ExactYM}) for the 
all-plus amplitudes contain insertions of the $(-2\eps)$-dimensional 
components of the loop-momentum is just a reflection of the vanishing 
of the amplitudes' unitarity cuts for $D=4$ ($\e\to0$).
It is straightforward to show that \eqn{ExactYM} reduces to the $n\le 6$
cases of \eqn{YMAllnPlus} as $\e \to 0$.  The explicit $\e$ in 
the prefactor of $\I_m^{D=4+2r-2\e}$ in \eqn{DimShiftIntegral} means
that only the $1/\e$ pole coming from the ultraviolet divergence of 
$\I_m^{D=4+2r-2\e}$ will contribute.  These contributions, which are pure
numbers, are given in \app{IntegralAppendix}.


\subsection{$N=4$ Super Yang-Mills Amplitudes}
\label{N4SubSection}

The one-loop MHV amplitudes of $N=4$ super-Yang-Mills theory provide
another example of amplitudes that may be evaluated for an arbitrary
number of external legs. The higher degree of supersymmetry present in $N=4$
super-Yang-Mills theory considerably simplifies the analytic properties of
its loop amplitudes.  (Infinite sequences of MHV amplitudes have also 
been determined for $N=1$ supersymmetric theories, but their analytic 
structure is more complicated~\cite{SusyOne}.)   

In particular, supersymmetry cancellations forbid
all triangle and bubble integrals, and only scalar box integrals 
(with no loop momenta in the numerator) may appear~\cite{SusyFour}.  These
supersymmetry cancellations, which may be seen in $N=1$
superspace~\cite{Superspace}, or in components by using a string-based
approach~\cite{ZBTASI,BernMorgan}, imply a maximum of $m-4$ powers of
loop momentum in the numerator of an $m$-point integral.  The integral
reduction procedure mentioned in \sec{OneLoopSubSection} uses equations 
such as $L\cdot k_i = -\hf( (L-k_i)^2 - L^2 - k_i^2 )$, where $L$ is 
the loop momentum and $k_i$ is an external momentum.  The factors $L^2$ and 
$(L-k_i)^2$ cancel denominator factors from scalar propagators and reduce
the number of external legs for the integral by one~\cite{PassarinoVeltman}.  
Thus the degree of the loop-momentum polynomial in the numerator of the 
integral is reduced by one whenever the number of legs
for the integral is reduced by one.  As a consequence, the $m$-point
integrals in $N=4$ super-Yang-Mills theory can lead, after reduction,
to at most scalar box integrals.  Later, in \sec{N8PowerSubSection},
we shall compare this loop-momentum power counting to what we find
from inspecting the $N=8$ MHV supergravity amplitudes.

\begin{figure}[ht]
\centerline{\epsfxsize 4. truein \epsfbox{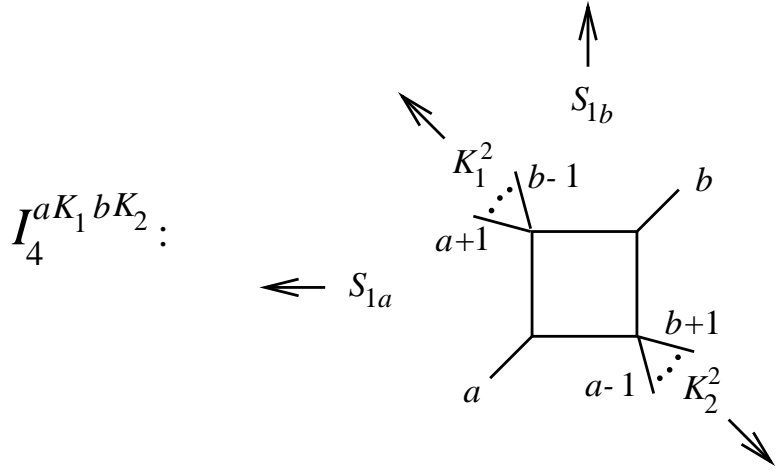}}
\vskip -.2 cm
\caption[]{
\label{EasyTwoMassFigure}
\small Kinematics of the two-mass box integrals $\I_4^{a K_1 b K_2}$ that
enter $n$-point MHV amplitudes in both $N=4$ super-Yang-Mills theory and
$N=8$ supergravity.  Here $a$ and $b$ label the external massless legs for
the integral, which coincide with two external momenta for the amplitude,
$k_a$ and $k_b$.  The massive legs carry momenta $K_1$ and $K_2$, which
are sums of the remaining external momenta.  The four Lorentz invariants
are the masses $K_1^2$ and $K_2^2$, and the Mandelstam invariants
$S_{1a} = (K_1+k_a)^2$ and $S_{1b} = (K_1+k_b)^2$.
}
\end{figure}

Although the $N=4$ power-counting allows any scalar box integral to
appear at one loop, for the MHV helicity configurations one finds only
the two-mass scalar box integrals where the two massive legs are
diagonally opposite~\cite{SusyFour}.  (Massless legs of the integral
correspond directly to external momenta of the amplitude, while
massive legs correspond to sums of external momenta.)  Denoting the
massless legs by $a$ and $b$, we define
$$
\I_4^{a K_1 b K_2} \equiv
\I_4^{a\, (a+1, \ldots, b-1)\, b\, (b+1, \ldots, a-1)}
 = \int { d^D L \over (2\pi)^D }
 { 1 \over L^2 (L-k_a)^2 (L-k_a-K_1)^2 (L+K_2)^2 }\ ,
\equn\label{EasyTwoMassBoxDef}
$$
where $K_1 = k_{a+1} + k_{a+2} + \cdots + k_{b-1}$ is the sum of the 
adjacent momenta between $a$ and $b$ (in the cyclic sense) and 
$K_2 = -k_a-K_1-k_b$.  (See \fig{EasyTwoMassFigure}.) 
In general we label the one-loop integrals by their external legs, 
following the cyclic ordering around the loop.  The parentheses group 
together those legs of the amplitude which combine together to form a 
massive leg of the integral.  As a more compact notation, 
we sometimes label the massive legs just by their total momentum (e.g.,
$K_1$ or $K_2$).  See \app{IntegralAppendix} for more details.
We use the same labeling for the coefficients of the integrals.

%

The explicit form for the integrals~(\ref{EasyTwoMassBoxDef}) near
$D=4$~\cite{Integrals} is given in \eqn{EasyTwoMassAnswer}.  In terms
of these integrals, the $N=4$ MHV amplitudes are given by~\cite{SusyFour}
$$
A_{n;1}(1^-,2^-,3^+,\ldots,n^+) 
= {1\over2} \spa1.2^4 
\sum_{{a,b \atop \rm cyclic}}\alpha_{a K_1 b K_2} 
                 \I_4^{a K_1 b K_2} +\ \Ord(\e),
\equn\label{N4YMAlphaDecomp}
$$
where the sum is over all integrals with the {\it standard} $123\ldots n$
cyclic ordering of external legs, and over all distinct non-adjacent 
pairs of massless legs $a,b$.  The coefficients of the box integrals are
$$
\alpha_{a K_1 b K_2} 
= - { (K_1+k_a)^2 (K_1+k_b)^2 - K_1^2 K_2^2 
   \over \spa1.2\spa2.3\spa3.4\cdots \spa{n}.1}
= { \spab{a}.{K_1}.{b} \spab{b}.{K_2}.{a}
   \over \spa1.2\spa2.3\spa3.4\cdots \spa{n}.1} 
= {1\over2} { \tr[a \, K_1 \, b \, K_2]
   \over \spa1.2\spa2.3\spa3.4\cdots \spa{n}.1} \,,
\equn\label{YMcoeffs}
$$
where $\spab{i}.{K}.{j} \equiv \langle i^- | \hskip-1mm \Ksl | j^-\rangle$
and $\tr[a \, K_1 \, b \, K_2] \equiv
\tr[\ksl_a \, \Ksl_1 \,\ksl_ b \, \Ksl_2]$.

For the purpose of facilitating comparisons to the gravity 
results, an alternative representation for the coefficients is
$$
\alpha_{a K_1 b K_2} = 
 {1\over2} \, g(a, K_1, b) \, g(b, K_2, a) \, \tr[a \, K_1 \, b \, K_2]  \,, 
\equn\label{YMcoeffsAlt}
$$
where the functions $g$ are 
$$
g(a, K_1, b) \equiv g(a, \{ a+1, a+2, \ldots, b-1 \}, b)
= {1\over \spa{a,}.{a+1} \spa{a+1,}.{a+2} \cdots \spa{b-1,}.b} \,.
\equn\label{gDef}
$$

For $n=4$, \eqn{N4YMAlphaDecomp} is the exact answer, to all orders in
$\e$.  For $n>4$, the $\Ord(\e)$ terms contain pentagon and higher-point 
integrals evaluated in $D=6-2\e$, with a manifest $\e$ prefactor.  
These terms are currently known only for $n=5,6$~\cite{DimShift} 
--- see below.


\subsection{`Dimension-Shifting' Relations}
\label{DimShiftSubSection}

The one-loop MHV amplitudes (\ref{N4YMAlphaDecomp}) in $N=4$
super-Yang-Mills theory bear a curious `dimension-shifting' relation
to the sequence of one-loop `all-plus' amplitudes (\ref{YMAllnPlus}).
The relation between $N=4$ MHV and all-plus amplitudes
may be expressed in two different ways~\cite{DimShift},
$$
\eqalign{
A_{n;1} (1^+, 2^+, \ldots, n^+) &= 
- 2 \e (1-\e) (4 \pi)^2 
\left[  {A_{n;1}^{N=4} (1^+, 2^+, \ldots, i^-, \ldots, j^-, \ldots,
  n^+) \over \spa{i}.j^4} 
    \biggr|_{D \rightarrow D+4} \right] \cr
&= 2 {A_{n;1}^{N=4} (1^+, 2^+, \ldots, i^-, \ldots, j^-, \ldots,  n^+)
[\mu^4]  \over \spa{i}.j^4} \,. \cr}
\equn\label{YMDimShiftRel}
$$
Here `$D \rightarrow D+4$' instructs one to replace the $D$ appearing in the
loop-momentum integration measure $d^D L$ by $D+4$ (where $D=4-2\e$)
in all integrals appearing in $A_{n;1}$.  The notation `$[\mu^4]$' 
means that an extra factor of $\mu^4$ should be inserted into 
the numerator of every loop integral in the amplitude. The 
equivalence of the two forms follows from \eqn{DimShiftIntegral}.
This relation (\ref{YMDimShiftRel}) has been established for $n=4,5,6$ 
but remains a conjecture for $n\ge 7$. 

A few comments about the dimension-shifting relation~(\ref{YMDimShiftRel})
are in order.  First of all, the manifest symmetry of the all-plus 
amplitudes on the left-hand side of the relation under the cyclic 
substitution $1\to 2 \to \ldots \to n \to 1$ is also present on the 
right-hand side, as a consequence of the SWI~(\ref{PermIdentityYM}).
Secondly, in four dimensions the all-plus amplitudes have no unitarity 
cuts, and hence are pure rational functions~\cite{AllPlus,MahlonAllPlus},
while the $N=4$ MHV amplitudes have cuts in all channels.
These facts are consistent with \eqn{YMDimShiftRel} because of the
manifest $\e$ on the right-hand side (in the first form of the relation):
Only the ultraviolet $1/\e$ poles in the higher-dimensional box,
pentagon, etc., integrals for the $N=4$ amplitudes contribute as $D\to4$
($\e\to0$), and these have rational-function coefficients.

The dimension-shifting relation~(\ref{YMDimShiftRel}) is a statement about
all orders in $\e$, so its complete verification requires all-orders
evaluation of both sides.  The all-$n$ formulae~(\ref{N4YMAlphaDecomp})
and (\ref{YMAllnPlus}), which are only valid through $\Ord(\e^0)$, are not
sufficient for this purpose.  However, for $n=4,5,6$ the relation was
verified in ref.~\cite{DimShift}, with the all-plus amplitudes given in
\eqn{ExactYM}.  The $N=4$ amplitudes may be obtained from this equation by
applying the dimension-shifting formula; the net effect on the integrals
is to remove four powers of $\mu$ from their arguments.

For the all-plus amplitudes, the pentagon terms in \eqn{ExactYM} 
do contribute as $\e\to 0$, since these integrals are ultraviolet 
divergent at $D=10$, canceling the $\eps$ prefactor in 
\eqn{DimShiftIntegral}.  On the other hand, for the $N=4$ amplitudes
the pentagon and hexagon terms are finite and do not cancel the 
overall $\eps$ implied by the remaining $\mu^2$ arguments.  
This leaves only box integrals in the expression~(\ref{N4YMAlphaDecomp}) 
for the $N=4$ MHV amplitudes in $D=4-2\e$ as $\e\to0$.

Eqs.~(\ref{ExactYM}) --- or rather the cuts of 
eqs.~(\ref{ExactYM}) in various channels --- will play a role in 
section~\ref{CutConstructionSection} as we construct the analogous 
amplitudes in $N=8$ supergravity and pure (or self-dual) gravity, 
using in part the KLT relations between tree amplitudes on either side 
of the cuts.

In direct analogy to \eqn{YMDimShiftRel}, one may conjecture a
relation between one-loop MHV amplitudes in $N=8$ supergravity, 
and all-plus amplitudes in pure gravity~\cite{DimShift}:
$$
\eqalign{
M_n (1^+, 2^+, \ldots, n^+) &= 
- 2 \e (1-\e)(2-\e)(3-\e) (4 \pi)^4 \left[ 
{M_n^{N=8}(1^+,2^+,\ldots,i^-,\ldots,j^-,\ldots,n^+) \over \spa{i}.j^8} 
    \biggr|_{D \rightarrow D+8} \right] \cr
&= 2 {M_n^{N=8}(1^+, 2^+, \ldots, i^-, \ldots, j^-, \ldots,  n^+)
[\mu^8]  \over \spa{i}.j^8} \,. \cr}
\equn\label{GravDimShiftRel}
$$
This equation respects the heuristic relation 
`gravity $\sim$ (gauge theory)$^2$',
since the $\mu^4$ gauge theory factor becomes a $\mu^8$ factor in the
gravity case.  Here $M_n^{N=8}$ refers to an entire $N=8$ multiplet 
circulating in the loop, while $M_n(1^+, 2^+, \ldots, n^+)$ gives the
contribution of a graviton in the loop.  For the same reasons as in
the gauge case, this $M_n$ could equally well be calculated with a 
massless scalar in the loop instead~\cite{GZ}.  In ref.~\cite{DimShift}
\eqn{GravDimShiftRel} was only verified for the simplest case, $n=4$.
In the following section we shall see that it holds for $n=5$ and 6 as
well, thus strengthening the all-$n$ conjecture.


\section{Cut Construction of One-Loop MHV $N=8$ Supergravity and All-Plus 
Gravity Amplitudes}
\label{CutConstructionSection}

In this section we construct the one-loop $n$-point all-plus gravity
amplitudes from their unitarity cuts, for $n\leq6$.  (The case of $n=4$
has been computed previously in refs.~\cite{GZ,DN}.)  Then we exploit the
gauge dimension-shifting relations of \sec{DimShiftSubSection} to obtain
the $N=8$ MHV supergravity amplitudes.  These calculations will provide a
firm basis from which we shall construct ansatze for an arbitrary number
of external legs in \sec{AnsatzSection}, using the soft and collinear
behavior of gravity amplitudes to be discussed in \sec{SoftCollSection}.


\subsection{Brief Review of Cutting Method}
\label{CuttingReviewSubSection}

The cutting method that we use has been discussed extensively for the
case of gauge theory amplitudes, and reviewed in ref.~\cite{Review},
so we only briefly describe it.  This is a proven technology, having
been used in the calculation of analytic expressions for the QCD
one-loop helicity amplitudes for $Z \rightarrow 4$
partons~\cite{Zjets}, in the construction of infinite
sequences~\cite{AllPlus,SusyFour,SusyOne} of one-loop MHV amplitudes
and for two-loop four-point amplitudes in maximally supersymmetric
cases~\cite{BRY,BDDPR}.  This technique allows for a complete
reconstruction of the amplitudes from the cuts, provided that all cuts
are known in arbitrary dimension.  Because on-shell expressions are
used throughout, gauge invariance, Lorentz covariance and unitarity
are manifest.

The unitarity cuts of one-loop amplitudes are given simply by
phase-space integrals of products of tree amplitudes, summing over all
intermediate states that can cross the cut. For example, the cut in
the channel carrying momentum $k_{m_1}+\cdots+k_{m_2}$ for 
$M_n(1,2,\ldots,n)$ is given by
$$
\eqalign{
C_{m_1\ldots m_2} 
= i \sum_{\lambda_1,\lambda_2} \int \dlips(-L_1,L_2)\ 
  & M_{m_2-m_1+3}^\tree
       ((-L_1)^{-\lambda_1},m_1,\ldots,m_2,L_2^{\lambda_2}) \cr
\times \,  
  & M_{n+m_1-m_2+1}^\tree
       ((-L_2)^{-\lambda_2},m_2+1,\ldots,m_1-1,L_1^{\lambda_1}) \,, 
\cr}
\equn\label{BasicCutEquation}
$$
where the integration is over the two-particle $D$-dimensional 
Lorentz-invariant phase-space, and $\lambda_{1,2}$ denote the 
helicity/particle-type of the states crossing the cut.  
(Polarization labels for the external graviton states have been suppressed.)
One can replace~\cite{Review} the phase-space integral with an
unrestricted loop momentum integral $\int d^D L$, yet continue
to apply the on-shell conditions $L_1^2 = L_2^2 = 0$, so long as 
one remembers that only functions with a cut in the given channel
are reliably computed in this way.  (The positive energy conditions
are automatically imposed with the use of Feynman propagators.)

A principal advantage of the cutting approach for gauge theory
calculations is that the tree amplitudes on either side of the cut can be
simplified {\it before} attempting to evaluate the cut
integral~\cite{Review}.  In the case of gravity, the KLT relations 
provide convenient representations of the tree amplitudes.  
In the supersymmetric case, on-shell supersymmetry Ward
identities can also be used to reduce the amount of work required.  (To
maintain the supersymmetry cancellations the dimensional regularization
scheme should not alter the number of states from their four-dimensional
values~\cite{Siegel}.)

For the non-supersymmetric all-plus calculation, the
SWI~(\ref{SWIVanish}) allow us to replace gravitons or any other
massless particles in the loop with massless scalars~\cite{GZ}.  
That is, at one loop we have,
$$
M_n^{\rm any\ states}(1^+, 2^+, \ldots, n^+) = N_s
M_n^{\rm scalar}(1^+, 2^+, \ldots, n^+)\, ,
\equn\label{ScalarInLoop}
$$
where $N_s$ is the number of bosonic states minus fermionic states
circulating in the loop in $M_n^{\rm any\ states}$.  (We have taken the
normalization of the `scalar' amplitude to be that for a single real
scalar state.)  For scalars (or fermions) crossing the cuts, a detailed
study of the effect of the $D$-dimensional loop momentum has previously
been presented~\cite{Massive,Review,DimShift}, and the requisite gauge
theory tree amplitudes, where the scalar carries non-zero momenta in the
extra $(-2\e)$ dimensions, have been computed for $n\le 6$
\cite{DimShift}.  We shall obtain the scalar$+$graviton tree amplitudes
from the scalar$+$gluon tree amplitudes using the KLT relations.  Thus we
shall be able to directly evaluate the cuts for the all-plus amplitudes to
all orders in $\e$, which in turn gives the full amplitudes to all orders
in $\e$~\cite{TwoLoopUnitarity,Review,Massive}.

One good way to obtain the $N=8$ MHV amplitudes with up to six legs, through
$\Ord(\e^0)$, is to make use of a `cut constructible' criterion that
allows one to use four-dimensional momenta in the cuts without introducing
any errors in $\Ord(\e^0)$ rational functions, assuming that certain power
counting criteria are satisfied~\cite{SusyOne}.  In the present case,
however, we can avoid explicit computations of the $N=8$ MHV cuts by
instead obtaining the $N=8$ amplitudes from the all-plus 
amplitudes using the gravitational version~(\ref{GravDimShiftRel}) 
of the gauge theory `dimension-shifting' results presented in 
\sec{DimShiftSubSection}.  We now explain how the gravitational 
relation can be derived from the gauge theory one.


\subsection{Dimension-Shifting Relations between Gravity Cuts}
\label{GravDimShiftSubSection}

In ref.~\cite{DimShift} the dimension-shifting formula
(\ref{GravDimShiftRel}) was shown to hold for the four-graviton amplitudes,
and was conjectured to hold for $n$-point amplitudes.  Here we demonstrate
that it does hold at $n$-points, if the gauge theory 
relation~(\ref{YMDimShiftRel}) holds at $n$-points. 
Since the latter relation has been proven for $n\le6$, this establishes 
the gravity dimension-shifting relation (\ref{GravDimShiftRel}) up to 
six points, but leaves the $n\ge7$ cases as a conjecture.

We begin with the cuts of the gauge theory dimension-shifting 
formula (\ref{YMDimShiftRel}),
$$
\eqalign{
 A^\tree_{m+2} & (-L_1^s,i_1^+, i_2^+, \ldots, L_2^s, \ldots, i_m^+) \, 
 \times A^\tree_{n-m+2}(-L_2^s,i_{m+1}^+, i_{m+2}^+, \ldots, 
                     L_1^s, \ldots, i_n^+) \cr
&\hskip-.9cm
 = {\mu^4 \over \spa{i}.j^4} \sum_{N=4\ \rm states} 
  A^\tree_{m+2}(-L_1,i_1^+, i_2^+,\ldots, L_2,\ldots, i_m^+) 
\times A^\tree_{n-m+2}(-L_2,i_{m+1}^+,i_{m+2}^+,\ldots,L_1,\ldots,i_n^+)\,,
\cr}
\equn\label{YMCutDimShift}
$$
where the superscript $s$ denotes a scalar line, the sum on the right-hand
side runs over all $N=4$ super-Yang-Mills states that can cross
the cut, and we have suppressed the $\lambda_{1,2}$ helicity/state labels
for these states.  (The reason the `2' in \eqn{YMDimShiftRel} has
disappeared from \eqn{YMCutDimShift} is that $A_{n;1}$ in the former
equation corresponds to 2 real scalars circulating in the loop.)
\Eqn{YMDimShiftRel} was originally derived in its loop-momentum integrated
version.  Nevertheless, it turns out that the manipulations
used in ref.~\cite{DimShift} to verify the relations between the
amplitudes can be arranged so as not to introduce any total derivatives.
This means that \eqn{YMDimShiftRel} holds point-by-point in the
integrands.

Also, the cuts of \eqn{YMDimShiftRel} (a leading-in-$N_c$ equation) only
correspond directly to the configurations where the cut loop momenta $L_1$
and $L_2$ are adjacent, i.e.  
$A^\tree_{m+2}(-L_1^s,i_1^+,\ldots,i_m^+,L_2^s) \, \times A^{\rm
tree}_{n-m+2}(-L_2^s,i_{m+1}^+,\ldots,i_n^+,L_1^s)$ on the left-hand side
of \eqn{YMCutDimShift}.  However, it is possible to obtain all the other
permutations of this equation, by using the following relation among
color-ordered tree amplitudes~\cite{GenDecoupling},
$$
A^\tree(1,\{\alpha\},2,\{\beta\}) 
= (-1)^{n_\beta} \sum_{\sigma\in OP\{\alpha\}\{\beta^T\}}
     A^\tree(1,\sigma(\{\alpha\}\{\beta^T\}),2) \,,
\equn\label{TreeSublAnswer}
$$
where $n_\beta$ is the number of elements in $\{\beta\}$, the set
$\beta^T$ is $\beta$ with the ordering reversed, and
$OP\{\alpha\}\{\beta^T\}$ is the set of all permutations of $\{\alpha\}
\cup \{\beta^T\}$ that preserve the ordering of elements within each of
the two sets.  \Eqn{TreeSublAnswer} can be
inserted twice each into the left- and right-hand sides of
\eqn{YMCutDimShift}, in order to reduce the general case to the case where
$L_1$ and $L_2$ are adjacent.

We may now use the $n$-point KLT equation (\ref{KLTGeneral}) 
to rewrite the cuts of the all-plus gravity amplitudes in terms of 
gauge theory cuts,  
$$
\eqalign{
&M_{m+2}^\tree(-L_1^s, 1^+,2^+, \ldots, m^+, L_2^s) \times 
 M_{n-m+2}^\tree(-L_2^s, (m+1)^+,\ldots, n^+, L_1^s) \cr
 & \hskip .2 cm
 = \biggl( \sum_{\rm perms} f \bar f \,  
  A_{m+2}^\tree(-L_1^s, 1^+,2^+, \ldots, m^+, L_2^s)
\ A_{m+2}^\tree(i_1^+,\ldots,-L_1^s, m^+, i_{m-1}^+, \ldots, L_2^s) 
  \biggr) \cr
 & \hskip 1.2 cm
 \times 
 \biggl( \sum_{{\rm perms}'} f' \bar{f}' \, 
  A_{n-m+2}^\tree(-L_2^s, (m+1)^+, \ldots, n^+, L_1^s) 
\ A_{n-m+2}^\tree(i_{m+1}^+,\ldots,-L_2^s, n^+, i_{n-1}^+, \ldots, L_1^s) 
  \biggr) \,, \cr}
\equn\label{GravDimShift1}
$$
where `perms' and `${\rm perms}'$' stand for the full sum over KLT
permutations in \eqn{KLTGeneral}, and $f$, $\bar f$, $f'$ and
$\bar{f}'$ are the functions $f$, $\bar{f}$ defined in \eqn{fDef}, for
the appropriate sets of arguments. (The KLT equations hold in any
dimension $D\le 10$ where string constructions exist and may be
analytically continued to arbitrary dimensions.) After rearranging the
right-hand side of \eqn{GravDimShift1} and applying the cut version of
the gauge theory dimension-shifting formula~(\ref{YMCutDimShift}), we
obtain
$$
\eqalign{
&M_{m+2}^\tree(-L_1^s, 1^+,2^+, \ldots, m^+, L_2^s) \times 
 M_{n-m+2}^\tree(-L_2^s, (m+1)^+,\ldots, n^+, L_1^s) \cr
 & \hskip .2 cm
\vphantom{\biggl[}
= \sum_{\rm perms} \sum_{{\rm perms}'} f \bar f \, f' \bar{f}' 
[ A_{m+2}^\tree(-L_1^s, 1^+,2^+, \ldots, m^+, L_2^s)
\ A_{n-m+2}^\tree(-L_2^s, (m+1)^+, \ldots, n^+, L_1^s) ] \, \cr
& \hskip 1.2 cm \times
 [ A_{m+2}^\tree(i_1^+,\ldots,-L_1^s, m^+, i_{m-1}^+, \ldots, L_2^s) 
 \ A_{n-m+2}^\tree(i_{m+1}^+,\ldots,-L_2^s, n^+, i_{n-1}^+, \ldots, L_1^s)
] \cr
& \hskip .2 cm \vphantom{\biggr]}
= \sum_{\rm perms} \sum_{{\rm perms}'} f \bar f \, f' \bar{f}' \cr
& \hskip .4 cm \times
\sum_{N=4\ \rm states}  \biggl[ {\mu^4 \over \spa{i}.j^4} 
  A_{m+2}^\tree(-L_1, 1^+,2^+, \ldots, m^+, L_2)
\ A_{n-m+2}^\tree(-L_2, (m+1)^+, \ldots, n^+, L_1) \biggr] \, \cr
& \hskip .4 cm \times
\sum_{N=4\ \rm states} \biggl[ {\mu^4 \over \spa{i}.j^4} 
   A_{m+2}^\tree(i_1^+,\ldots,-L_1, m^+, i_{m-1}^+, \ldots, L_2) 
 \ A_{n-m+2}^\tree(i_{m+1}^+,\ldots,-L_2, n^+, i_{n-1}^+, \ldots, L_1)
 \biggr] \,. \cr}
\equn
$$
As discussed in \sec{KLTSubSection}, in terms of a $D=4$ decomposition
of states the double sum over the 16 $N=4$ super-Yang-Mills states may
be reassembled as a single sum over the 256 $N=8$ supergravity
states. (In higher dimensions up to $D=10$, the sum over $N=4$ states
can be reassembled into a sum over the appropriate multiplet in the
higher dimensional theory.) This yields,
$$
\eqalign{
&M_{m+2}^\tree(-L_1^s, 1^+,\ldots,m^+, L_2^s) \times 
 M_{n-m+2}^\tree(-L_2^s, (m+1)^+,\ldots, n^+, L_1^s) \cr
& \hskip 1 cm
= {\mu^8 \over \spa{i}.j^8} \sum_{N=8\ \rm states}
M_{m+2}^\tree(-L_1, 1^+,2^+, \ldots, m^+, L_2) \times 
 M_{n-m+2}^\tree(-L_2, (m+1)^+,\ldots, n^+, L_1)\,. \cr}
\equn\label{GravCutDimShift}
$$
Note that the precise details of which permutation sums are included, or
what exactly the $f$ functions are, are unimportant in the derivation of
\eqn{GravCutDimShift}, because the same permutations and $f$ functions appear
in the KLT expressions for any of the tree amplitudes appearing in the
cuts, regardless of the particle type of the states crossing the cut.
Since \eqn{GravCutDimShift} can be used to obtain all cuts of the
amplitudes and is valid to all orders in $\eps$, the gravity
dimension-shifting relation (\ref{GravDimShiftRel}) is established
for all values of $n$ for which \eqn{YMCutDimShift} has been proven 
(currently this is for $n\le 6$).

\begin{figure}[ht]
\centerline{\epsfxsize 4 truein \epsfbox{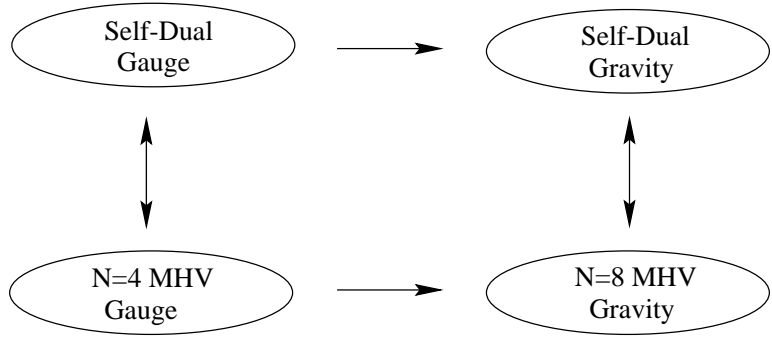}}
\vskip -.2 cm
\caption[]{
\label{SquareOfRelationsFigure}
\small Relations between infinite sequences of one-loop
amplitudes in four different theories.  The vertical arrows correspond
to the `dimension-shifting' relations of ref.~\cite{DimShift}, within
(super-) Yang-Mills theory and (super-) gravity.  (These remain a
conjecture for $n\ge 7$ legs.)  The horizontal 
arrows correspond to the gauge-gravity relations which follow from 
the KLT equations.}
\end{figure}

The relationships between the infinite sequences of one-loop all-plus 
(or self-dual) amplitudes and MHV amplitudes in maximally
supersymmetric theories, and between gravity and gauge theory, are
summarized in \fig{SquareOfRelationsFigure}.  The horizontal arrows
correspond to the gauge-gravity relations that follow from the KLT
equations, and the vertical arrows represent the dimension-shifting 
relations.

%

Using \eqn{GravDimShiftRel} it is then sufficient to calculate 
the all-plus graviton amplitudes as a function of $\e$ (i.e., for
arbitrary $D=4-2\e$), in order to obtain the $N=8$ supergravity 
MHV amplitudes.


\subsection{All-Plus Amplitudes for $n\le 6$}
\label{AllPlusSubSection}

As a warmup we first calculate the four-graviton all-plus amplitude,
before proceeding to the five- and six-graviton cases.  Using
\eqn{ScalarInLoop} we can replace the graviton in the loop with two real
scalars.  Thus, the cut in the $s_{12}$ channel is
$$
M_4(1^+, 2^+, 3^+, 4^+) \Bigr|_{s_{12}\rm -cut} 
= \int {d^D L_1 \over (2\pi)^D} \; {i \over L_1^2} 
 M_4^\tree(-L_1^s, 1^+, 2^+, L_3^s) \, {i \over L_3^2} 
  \, M_4^\tree(-L_3^s, 3^+, 4^+, L_1^s) \Bigr|_{s_{12}\rm-cut}\,, 
\equn\label{FourPtCut}
$$
where the superscript $s$ denotes that the cut lines are scalars, and
$L_3 = L_1 - k_1 - k_2$.  An overall 2, from two real scalars propagating
in the loop, is canceled by an identical-particle phase-space factor of 
1/2.  Using the KLT expressions
(\ref{GravYM}) we may replace the gravity tree amplitudes appearing in
the cuts with products of gauge theory amplitudes.  The required gauge 
theory tree amplitudes, with two external scalar legs and two gluons, 
are relatively simple to obtain~\cite{Massive,Review},
$$
\eqalign{
A_4^\tree(-L_1^s,1^+,2^+,L_3^s) & =  i
{\mu^2\spb1.2\over\spa1.2 [( \ell_1 -k_1)^2 -\mu^2] }\,, \cr
A_4^\tree(-L_1^s, 1^+, L_3^s, 2^+) 
& = - i {\mu^2\spb1.2\over\spa1.2}
 \biggl[{1\over (\ell_1 -k_1)^2 -\mu^2}
 + {1\over (\ell_1 -k_2)^2 -\mu^2}\biggr] \,, \cr}
\equn
$$
where the gluon momenta are four-dimensional, but the scalar momenta
are allowed to have a $(-2\e)$-dimensional component $\vec{\mu}$,
with $\vec{\mu}\cdot\vec{\mu} = \mu^2 > 0$.  The overall factor 
of $\mu^2$ appearing in these tree amplitudes means that they vanish 
in the four-dimensional limit, in accord with the SWI~(\ref{SWIVanish}).
In the KLT relation (\ref{GravYM}), one of the propagators
cancels, leaving
$$
M_4(-L_1^s, 1^+, 2^+, L_3^s) = 
- i \biggl({\mu^2\spb1.2\over\spa1.2}\biggr)^2
 \biggl[{1\over (\ell_1 -k_1)^2-\mu^2}
 + {1\over (\ell_1 -k_2)^2-\mu^2}\biggr] \,. 
\equn\label{GravityScalarTreeFour}
$$
Inserting \eqn{GravityScalarTreeFour} and the permuted formula for 
$M_4(-L_3^s, 3^+, 4^+, L_1^s)$ into the cut~(\ref{FourPtCut}) yields
$$
\eqalign{
&M_4(1^+, 2^+, 3^+, 4^+) \bigl|_{s_{12}\rm -cut} \cr
&\hskip 2 cm 
= {\spb1.2^2\spb3.4^2\over\spa1.2^2\spa3.4^2} 
\int {d^4 \ell \over (2\pi)^4} 
\int {d^{-2\eps} \mu \over (2\pi)^{-2\eps} } 
\, \mu^8 \; {1 \over \ell^2 -\mu^2}\biggl[{1\over (\ell -k_1)^2 - \mu^2}
 + {1\over (\ell -k_2)^2 -\mu^2}\biggr] \cr
& \hskip 5 cm \times
   {1 \over (\ell - k_1 - k_2 )^2 -\mu^2}
 \biggl[{1\over (\ell + k_4)^2 -\mu^2} + {1\over (\ell +k_3)^2 -\mu^2}\biggr] 
 \biggr|_{s_{12}\rm -cut} \ , \cr}
\equn
$$
which corresponds to a sum of $(12-2\eps)$-dimensional scalar
integrals, using (\ref{DimShiftIntegral}).  By symmetry, the other cuts
are the same up to relabelings.  Combining all three cuts into a
single function that has the correct cuts in all channels yields
$$
\eqalign{
M_4(1^+, 2^+, 3^+, 4^+) & = 
    2 {\spb1.2^2\spb3.4^2\over\spa1.2^2\spa3.4^2}
\Bigl(\I_4^{1234}[\mu^8] + 
      \I_4^{3124}[\mu^8] +
      \I_4^{2314}[\mu^8] \Bigr)\,, \cr }
\equn\label{FourGravAllPlus}
$$
where 
$$
\I_4^{1234}[\mu^8] = 
\int {d^4 \ell \over (2\pi)^4} 
\int {d^{-2\eps} \mu \over (2\pi)^{-2\eps} } 
\, \mu^8 \; {1 \over [\ell^2 -\mu^2]  [(\ell - k_1)^2 - \mu^2]
       [(\ell - k_1-k_2)^2 - \mu^2]
       [(\ell + k_4)^2 - \mu^2] }
\equn
$$
is the scalar box integral with the external legs arranged in the 
order 1234.  The two other scalar integrals that appear correspond 
to the two other distinct orderings of the four external legs.
(See \app{IntegralAppendix} for our notation for general one-loop
integrals.)  Using \eqn{ScalarInLoop}, this result can be applied to 
any set of massless fields circulating in the loop.

The spinor factor $\spb1.2^2\spb3.4^2/(\spa1.2^2\spa3.4^2)$ in 
\eqn{FourGravAllPlus} is actually completely symmetric, although not
manifestly so.  By rewriting this factor and using \eqn{I4D12} for the box
integral, the final one-loop result in $D=4$ is
$$
M_4(1^+, 2^+, 3^+, 4^+) = - \, {i \over (4 \pi)^2} \,
\biggl( {st \over \spa1.2 \spa2.3 \spa3.4 \spa4.1} \biggr)^2 \  
{s^2 + t^2 + u^2 \over 120}\ +\ \Ord(\eps)\,,
\equn\label{FourGravAllPlusDFour1}
$$
in agreement with a previous calculation~\cite{DN}.

For the purpose of constructing an ansatz for $n\ge 7$, 
it is useful to write the $n=4$ amplitude as
$$
\eqalign{
M_4(1^+, 2^+, 3^+, 4^+) 
& = - {i \over (4 \pi)^2} {1\over480} \Bigl[
  h(1,\{3\},2) h(2,\{4\},1) \, \tr^3[1 3 2 4] \cr
& \hskip .5 cm 
+ h(1,\{2\},3) h(3,\{4\},1) \, \tr^3[1 2 3 4] 
+ h(1,\{2\},4) h(4,\{3\},1) \, \tr^3[1 2 4 3] \Bigr]
+ \Ord(\eps)\,, \cr}
\equn\label{FourGravAllPlusDFour}
$$
where $\tr[i_1 i_2 i_3 i_4] \equiv \tr[
\ksl_{i_1} \ksl_{i_2} \ksl_{i_3} \ksl_{i_4}]$ and 
$$
h(a,\{1\},b) = {1\over \spa{a}.{1}^2 \spa{1}.{b}^2} \,.
\equn\label{Half1}
$$

Next we compute $M_5(1^+,2^+,3^+,4^+,5^+)$ to all orders in
$\eps$.  Its total symmetry implies that the
$s_{12}$ cut again suffices for its complete reconstruction.
Thus we require the tree amplitudes for two scalars and either two or 
three gravitons,
$M_4^\tree(-L_1^s,1^+,2^+,L_3^s)$ from \eqn{GravityScalarTreeFour}, and
$M_5^\tree(-L_3^s,3^+,4^+,5^+,L_1^s)$, which may be constructed from the 
gauge amplitudes for two scalars and three gluons~\cite{DimShift}, 
$$
\eqalign{ 
A_5^\tree(L_1^s,-L_3^s,3^+,4^+,5^+) &= 
i \, \mu^2 \, {\langle 5^+ | (3+4) \ell_3 | 3^- \rangle  
 \over (L_3 -k_3)^2 \, \spa3.4\spa4.5\, (L_1 +k_5)^2 } \,, \cr
A_5^\tree(L_1^s, 3^+, -L_3^s, 4^+, 5^+) 
& = - A_5^\tree(L_1^s, -L_3^s, 3^+, 4^+, 5^+)
    - A_5^\tree(L_1^s, -L_3^s, 4^+, 3^+, 5^+) \cr
& \hskip 2 cm 
    - A_5^\tree(L_1^s,  -L_3^s, 4^+, 5^+, 3^+)\,, \cr}
\equn\label{YMScalarGluon}
$$
using the five-point KLT relation~(\ref{GravYM}).
The second equation (a special case of \eqn{TreeSublAnswer}) 
follows from the $U(1)$ decoupling 
identity~\cite{DecouplingIdent,RecursiveBG,ManganoReview}.

After applying several spinor-product identities to the right-hand side of
the five-point KLT relation, we obtain the manifestly symmetric form
$$
\eqalign{
M_5^\tree(-L_3^s,3^+,4^+,5^+,L_1^s) &= 
 - i {\mu^4 \over \spa3.4^2\spa3.5^2\spa4.5^2} \biggl[ 
 {s_{34} s_{45} \over (L_3 - k_3)^2 (L_1 + k_5)^2 }
 \Bigl( \langle 5^- | 3 (4+5) \ell_1 | 5^- \rangle + \mu^2 s_{35} \Bigr) \cr
& \hskip 3 cm 
 + \Perm(3,4,5) \biggr]\,. \cr}
\equn\label{GravScalarFive}
$$

Inserting \eqns{GravityScalarTreeFour}{GravScalarFive} into the 
$s_{12}$-channel cut of the five-point all-plus amplitude gives
$$
\eqalign{
M_5(1^+,2^+,\ldots,5^+) \bigr|_{s_{12}\hbox{-} \rm cut}
&= { \spb1.2^2 \over \spa1.2^2 \, \spa3.4^2\spa3.5^2\spa4.5^2} 
 \int {d^D L \over (2\pi)^D} \, \mu^8 \cr
& \hskip 1 cm \times
\Biggl\{ {1\over L_1^2 L_2^2 L_3^2} \biggl[ {s_{34} s_{45} \over L_4^2 L_5^2}
 \Bigl( \spab{5}.{3(4+5)\ell_1}.{5} + \mu^2 s_{35} \Bigr) \cr
& \hskip 6 cm  
 + \Perm(3,4,5) \biggr] + \Perm(1,2) \Biggr\} 
   \Biggr|_{s_{12}\hbox{-} \rm cut}\,,
\cr}
\equn\label{CutScalarOne}
$$
where $L_i = L - k_1 - \cdots - k_{i-1}$ and $L_i^2 = \ell_i^2 - \mu^2$.
The $L_1^2 L_2^2 L_3^2 L_4^2 L_5^2$ denominator factors signal the 
presence of a pentagon integral with a 12345 cyclic ordering of external 
legs. All other orderings appearing in the $s_{12}$ cut are generated by 
the permutation sums.

By using standard integration formulas, and combining all cuts
into a single function, we find that the five-point all-plus gravity 
amplitude may be put into a form similar to the all-plus gauge amplitude
(\eqn{YMAllnPlus}),
$$
M_5(1^+,2^+,\ldots,5^+) 
= \beta_{123(45)} \, \I_4^{123(45)}[\mu^8]
-  2 \, {\spb1.2 \spb2.3 \spb3.4 \spb4.5 \spb5.1 \over
          \spa1.2 \spa2.3 \spa3.4 \spa4.5 \spa5.1} \I_5^{12345}[\mu^{10}] 
       + \hbox{perms}, 
\equn\label{FiveGravAllPlus}
$$
where the permutation sum is over all distinct one-mass box integrals and
massless pentagon integrals (no cyclic ordering is imposed, in contrast to
the gauge case).  There are 30 different box-integral terms and 12
pentagons in the sum.  The box coefficient is
$$
\beta_{123(45)} = 
- {\spb1.2^2 \spb2.3^2 \spb4.5 \over \spa1.4\spa1.5\spa3.4\spa3.5\spa4.5}\,.
\equn\label{Mu13Coeff}
$$
We can rewrite this coefficient in terms of the one given for the five-point
all-plus gauge amplitude in \eqn{ExactYM} (or equivalently, for the MHV 
amplitude in $N=4$ super-Yang-Mills theory in \eqn{YMcoeffs}): 
$$
\eqalign{
\beta_{123(45)} &= - s_{45} \,
{ s_{12}\, s_{23} \over \spa1.2\spa2.3\spa3.4\spa4.5\spa5.1 }
{ s_{12}\, s_{23} \over \spa1.2\spa2.3\spa3.5\spa5.4\spa4.1 } \cr
&=\ - s_{45} \, \alpha_{123(45)} \alpha_{123(54)} \,. \cr}
\equn\label{Mu13Rewrite}
$$
This relation is reminiscent of the tree-level four-point KLT 
relation in \eqn{GravYM}, in that 
\par\noindent
(1) a quantity in gravity is expressed as a product of gauge quantities,
\par\noindent
(2) one $s_{ij}$ appears as a prefactor on the gauge side of the relation,
and 
\par\noindent
(3) the indices on the $s_{ij}$ coincide with the arguments which are
permuted between the two gauge-theory factors.

Through $\Ord(\eps^0)$ the expression for the amplitude can be 
simplified considerably by non-trivial rearrangements to yield,
$$
\eqalign{
&M_5(1^+, 2^+, 3^+, 4^+, 5^+ ) \cr
& \hskip1cm =  
 {i \over (4 \pi)^2} \, {1 \over 960} \,  
 h(1,\{2\},3) \, h(3,\{4,5\},1) \tr^3[123(4+5)]\ +\ \hbox{perms} 
 + \Ord(\eps), \cr}
\equn\label{FiveGravAllPlusDFour}
$$
where $\tr[\cdots (i +j) \cdots] \equiv \tr[\cdots (\ksl_i + \ksl_j)
\cdots]$, $h(a,\{1\},b)$ is defined in \eqn{Half1},
$$
h(a,\{1,2\},b) =  
{ \spb1.2 \over 
  \spa1.2 \spa{a}.{1}\spa{1}.{b}\spa{a}.{2}\spa{2}.{b}} \,,
\equn\label{Half2}
$$
and the sum is over $10 \times 3 = 30$ distinct permutations.
(There are $({5\atop2}) = 10$ possible choices for the pair of arguments
in braces in the second $h$ function, and for each of these there are 
3 more choices for the argument in braces in the first $h$ function.)

We have also obtained the six-point all-plus amplitude to 
all orders in $\eps$ from the cuts.  The result of this computation is 
$$
M_6(1^+, \ldots, 6^+) = \beta_{1(23)4(56)} \, \I_4^{1(23)4(56)}[\mu^8]
               + \beta_{123(456)} \, \I_4^{123(456)}[\mu^8] 
               + \rho_{1234(56)} \,  \I_5^{1234(56)}[\mu^{10}] 
               + \hbox{perms},
\equn\label{SixGravAllPlus}
$$
where the sum over permutations again runs over all distinct integral 
functions.  The coefficients of the integrals are 
$$
\eqalign{
\beta_{1(23)4(56)} & = {\spb2.3\over\spa2.3} {\spb5.6\over\spa5.6} 
{\spab1.{(2+3)}.4^2 \, \spab4.{(5+6)}.1^2 \over 
  \spa1.2\spa2.4\spa1.3\spa3.4\spa4.5\spa5.1\spa4.6\spa6.1} \,, \cr
\beta_{123(456)} &=  {\spb1.2^2\spb2.3^2
  \over \spa1.4\spa4.3\spa1.5\spa5.3\spa1.6\spa6.3}   \cr
&\hskip1cm
 \times \left( \spa1.4\spa4.3 {\spb4.5\spb4.6\over\spa4.5\spa4.6} 
         + \spa1.5\spa5.3 {\spb4.5\spb5.6\over\spa4.5\spa5.6} 
         + \spa1.6\spa6.3 {\spb4.6\spb5.6\over\spa4.6\spa5.6} \right)\,, \cr
\rho_{1234(56)} & = 2 \, 
{ s_{12}\, s_{23} \, s_{34} \spab1.{(2+3)}.4 \spab4.{(5+6)}.1 
  \over \tr_5[1234] }
\Bigl[ c_1 + c_2 + c_2 \vert_{5\lr6} \Bigr] \,, \cr}
\equn\label{SixPtCoeffs}
$$
where $\tr_5[\cdots] \equiv \tr[\gamma_5 \cdots]$ and 
$$
\eqalign{
c_1 &= { \tr_+[1234]\ \spb5.6 
   \over \spa1.2^2\spa2.3^2\spa3.4^2
      \spa4.5\spa5.1\spa4.6\spa6.1\spa5.6 } \,, \cr
c_2 &= { 1\over \tr_5[123456] }
  { \spb1.2\spb2.3\spb3.4\spb4.5\spb5.6\spb6.1 
   \over \spa1.2\spa2.3\spa3.4\spa4.5\spa5.6\spa6.1 } \,. \cr
}\equn\label{AlphaDef}
$$

Again the expression for the amplitude through $\Ord(\eps^0)$ can
be simplified, to 
$$
\eqalign{
&M_6(1^+, 2^+, 3^+, 4^+, 5^+, 6^+) 
=  
- {i \over (4 \pi)^2}\,  {1\over 960} \, \Bigl\{
 h(1,\{2,3\},4) h(4,\{5,6\},1) \tr^3[1(2+3)4(5+6)] \cr
& \hskip 2 cm 
 + h(1,\{2\},3) h(3,\{4,5,6\},1) \tr^3[123(4+5+6)]  
 + \hbox{perms} \Bigr\} \,, \cr}
\equn\label{SixGravAllPlusDFour}
$$
where the permutation sum is over distinct terms and the new $h$ function
that appears is
$$
\eqalign{
h(a,\{1,2,3\},b) &=
{ \spb1.2\spb2.3\over
  \spa1.2\spa2.3 \spa{a}.{1}\spa{1}.{b}\spa{a}.{3}\spa{3}.{b} }
+ { \spb2.3\spb3.1\over
    \spa2.3\spa3.1 \spa{a}.{2}\spa{2}.{b}\spa{a}.{1}\spa{1}.{b} } \cr
&\hskip1cm
+ { \spb3.1\spb1.2\over
    \spa3.1\spa1.2 \spa{a}.{3}\spa{3}.{b}\spa{a}.{2}\spa{2}.{b}} \,. \cr}
\equn\label{Half3}
$$


\subsection{$N=8$ MHV Amplitudes from Dimension-Shifting}
\label{N8CutSubSection}

By using the gravitational dimension-shifting
relation~(\ref{GravDimShiftRel}), we may obtain the $N=8$ four-, five-, 
and six-point MHV amplitudes from the all-plus (self-dual) amplitudes
(\ref{FourGravAllPlus}), (\ref{FiveGravAllPlus}), and
(\ref{SixGravAllPlus}), by dividing out a factor of $\mu^8$ from each
integrand and multiplying by an overall factor of $\spa{i}.j^8/2$, where
$i$ and $j$ are the two negative helicity legs.  After removing
a factor of $\mu^8$, the pentagon integrals are no longer ultraviolet
divergent and are suppressed by an overall power of $\eps$ near $D=4$,
since a power of $\mu^2$ remains.  Hence, the $N=8$ amplitudes
through $\Ord(\eps^0)$ are given just by the box integral contributions.

A representation of the four-, five- and six-point amplitudes which is
convenient for extending the result to an arbitrary number of external 
legs is 
$$
\eqalign{
 &M_4^{N=8}(1^-, 2^-, 3^+, 4^+)  = {1\over 4} \spa1.2^8\,
\Bigl[  h(1,\{2\},3) h(3,\{4\},1) \tr^2[1234] \, \I_4^{1234}
 + \hbox{perms}\Bigr]  + \Ord(\eps)\,, \cr
& M_5^{N=8}(1^-, 2^-, 3^+, 4^+, 5^+) 
= -{1\over 8} \spa1.2^8\,
\Bigl[
 h(1,\{2\},3) h(3,\{4,5\},1) \tr^2[123(4+5)] \, \I_4^{123(45)} \cr
& \hskip 6 cm 
       + \hbox{perms}\Bigr] + \Ord(\eps)\,, \cr
& M_6^{N=8}(1^-, 2^-, 3^+, 4^+, 5^+,6^+)  = 
 {1\over 8} \spa1.2^8\, \Bigl[
      h(1,\{2\},3) h(3,\{4,5,6\},1) \tr^2[123(4+5+6)] 
      \, \I_4^{123(456)} \cr
& \hskip 6 cm 
    + h(1,\{2,3\},4) h(4,\{5,6\},1) \tr^2[1(2+3)4(5+6)] 
   \, \I_4^{1(23)4(56)} \cr
& \hskip 6 cm 
  + \hbox{perms} \Bigr] + \Ord(\eps)\,, \cr}
\equn\label{N8Explicit}
$$
where the permutation sums are over all distinct permutations.

As a check, we have explicitly calculated the cuts of the $N=8$ MHV
supergravity amplitudes up to six legs in $D=4$ (i.e., through
$\Ord(\e^0)$).  We find complete agreement with the results
(\ref{N8Explicit}) obtained via \eqn{GravDimShiftRel}.  This cut
calculation is similar to the one performed for the all-plus
amplitudes in \sec{AllPlusSubSection}, and makes use of the KLT relations
(\ref{GravYM}) to express the the gravity tree amplitudes appearing in the
cuts in terms of gauge theory amplitudes.


\subsection{Power-Counting for $N=8$ MHV Amplitudes vs. $N=4$}
\label{N8PowerSubSection}

Let us compare the structure of the $N=8$ MHV results~(\ref{N8Explicit})
with general expectations from loop-momentum power-counting.
First recall from \sec{N4SubSection} that in a one-loop amplitude in $N=4$
super-Yang-Mills theory, a maximum of $m-4$ powers of loop momentum can 
appear in the numerator of each $m$-point integral.
In a string-based approach~\cite{BDS,DN}, the loop-momentum integrand 
for $N=8$ supergravity is just the product of two $N=4$ integrands.
Therefore one expects a maximum of $2(m-4)$ powers of loop momentum 
to appear in the numerator of an $m$-point integral for $N=8$
supergravity.  After carrying out the same integral reductions sketched 
in \sec{N4SubSection}, this power-counting allows for box integrals with 
up to $n-4$ powers of loop momentum in the numerator~\cite{DunbarPrivate},
for an $n$-point amplitude.  Such integrals can be reduced
to scalar box integrals, but (for $n\geq5$) only at the expense of 
introducing scalar triangle and perhaps bubble integrals.  
On the other hand, we find no such integrals in \eqn{N8Explicit},
only scalar box integrals.  Nor will we find any need for
integrals besides scalar boxes in the all-$n$ ansatz in \sec{AnsatzSection}.  

In other words, all the $N=8$ MHV amplitudes are consistent with having at
most $(m-4)$ powers of loop momentum for each $m$-point integral, instead
of the $2(m-4)$ powers expected from `squaring' gauge theory.  This
better-than-expected ultraviolet behavior can be contrasted with the
recent analysis of multi-loop four-point amplitudes~\cite{BDDPR}, in which
the multi-loop $N=8$ amplitudes had exactly the same number of powers 
of loop momentum as expected from $N=4$ gauge theory.  The additional
cancellations we find in one-loop MHV amplitudes in $N=8$ 
supergravity amplitudes for $n\geq5$ legs presumably arise from sums over 
different orderings of external legs.  It would be interesting to
know whether they can be understood at the Lagrangian level, or in a 
string-based framework, and whether they might extend to non-MHV helicity 
configurations as well, or to theories with less supersymmetry.


\section{Soft and Collinear Behavior of Gravity Amplitudes}
\label{SoftCollSection}

In order to extend the results of the previous section beyond the
six-point level we will make use of the analytic behavior of gravity
amplitudes as momenta become soft or collinear.  A feature that the
all-plus gravity and $N=8$ MHV amplitudes have in common with the
all-plus gauge and $N=4$ MHV amplitudes is the absence of multi-particle 
kinematic poles.  (This may be demonstrated using the SWI
(\ref{SWIVanish}), which implies the vanishing of each product of
amplitudes, tree$\times$loop, that forms the residue of a multi-particle 
pole.)  It is therefore sufficient to focus on the soft and collinear limits,
which determine the two-particle poles.  We perform our analysis in 
Minkowski space-time with signature (1,3).

The behavior of tree-level gravity amplitudes as momenta become soft 
is well known~\cite{WeinbergSoftG,BGK}, and was reviewed in 
\sec{SoftCollinearSubSection}.  
However, the behavior as momenta become collinear --- outlined 
previously in ref.~\cite{GravAllPlus} --- is more subtle.%
\footnote{The suggestion that collinear limits in gravity are universal
was made by Chalmers and Siegel~\cite{CSUnpublished}.}  In the following
subsection we obtain the graviton collinear splitting amplitudes from the
gauge theory ones using the KLT expressions (\ref{GravYM}).  We then show
that these splitting amplitudes are universal: they apply to collinear
limits of amplitudes with an arbitrary number of external legs.
Furthermore, we shall argue in \sec{LoopSoftSplitSubSection} that the
tree-level soft functions and collinear splitting amplitudes suffer no 
higher loop corrections.  That is, we shall show that at {\it any} loop 
order (including tree level) a gravity amplitude behaves as
$$
M_n^{\rm loop}(\ldots,a^{\lambda_a},b^{\lambda_b},\ldots)
  \ \mathop{\longrightarrow}^{a \parallel b}\  
\sum_{\lambda} \SplitGravlam(z,a^{\lambda_a},b^{\lambda_b}) \times
         M_{n-1}^{\rm loop}(\ldots,P^\lambda,\ldots) \,,
\equn\label{GravLoopColl}
$$
when $k_a$ and $k_b$ are collinear, and as
$$
M_n^{\rm loop}(\ldots,a,s^\pm,b,\ldots)\ \mathop{\longrightarrow}^{k_s\to0}\
    \SoftGrav(s^\pm) \times
   M_{n-1}^{\rm loop}(\ldots,a,b,\ldots)\,,
\equn\label{GravLoopSoft}
$$
when $k_s$ becomes soft.


\subsection{Collinear Behavior of Gravity from Gauge Theory}
\label{CollinearGravSubSection}

Assuming that the collinear behavior of graviton amplitudes is universal,
the splitting amplitudes in \eqn{GravLoopColl} can be computed
in terms of the gauge splitting amplitudes, using the
four- and five-point KLT relations.  Taking 
$1 \parallel 2$ in the five-point gravity amplitude~(\ref{GravYM}) 
and applying \eqn{YMTreeColl} we have,
$$
\eqalign{
\Split^{\rm gravity}_{-(\lambda+\tilde\lambda)}
(z,1^{\lambda_1+\tilde\lambda_1},2^{\lambda_2+\tilde\lambda_2}) =  
-s_{12} &\times \Split^\tree_{-\lambda}
          (z,1^{\lambda_1},2^{\lambda_2}) \cr
        &\times \Split^\tree_{-\tilde\lambda}
          (z,2^{\tilde\lambda_2},1^{\tilde\lambda_1})\,, \cr}
\equn\label{GravYMSplit}
$$
where $\SplitGravlam$ is a tree-level gravity splitting
amplitude and the $\Split_{-\lambda}^\tree$ are gauge theory splitting
amplitudes, such as those given in \eqns{YMTreeSplit}{QuarkTreeSplit}.

Equation~(\ref{GravYMSplit}) may be applied to arbitrary $N=8$
supergravity states by factorizing them into products of states in $N=4$ 
gauge theory, as discussed in \sec{KLTSubSection}; the addition of 
helicities in the equation, $\lambda_i+\tilde\lambda_i$, corresponds to 
this factorization.  For example, the pure graviton splitting amplitudes
are obtained by substituting the values of the pure gluon splitting 
amplitudes (\ref{YMTreeSplit}) into \eqn{GravYMSplit}, yielding
$$
\eqalign{
 \SplitGravp(z,a^{+},b^{+})\ &=\ 0\,,\cr
 \SplitGravm(z,a^{+},b^{+})\ &=\ -{ 1 \over z(1-z) }
                 { \spb{a}.{b} \over \spa{a}.b }\, ,\cr
 \SplitGravp(z,a^{-},b^{+})\ &=\ - { z^3 \over 1-z }
                 { \spb{a}.{b} \over \spa{a}.b } \,. \cr}
\equn\label{GravTreeSplit}
$$
As a second example, the splitting amplitudes for a graviton into 
two gravitinos ($h^+ \rightarrow \tilde h^- \tilde h^+$) 
follow from eqs. (\ref{GravYMSplit}), (\ref{YMTreeSplit}) and
(\ref{QuarkTreeSplit}), 
$$
\SplitGravp(z,a_{\tilde h}^-, b_{\tilde h}^+) = 
- s_{ab} \times \Split^\tree_{+}(z,a^{-},b^{+}) \times
 \Split^\tree_{+}(z, b_{q}^{+}, a_{\bar{q}}^{-})
           =  - \sqrt{{z^5 \over 1-z}}{\spb{a}.b\over\spa{a}.b}\,,
\equn\label{GravitinoTreeSplit}
$$
and so forth.

\begin{figure}[ht]
\centerline{\epsfxsize 1.5 truein \epsfbox{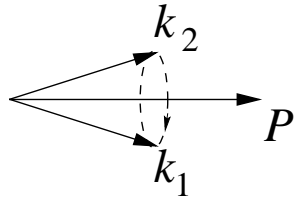}}
\vskip -.2 cm
\caption[]{
\label{RotateColFigure}
\small As two momenta become collinear, the gravity $S$-matrix element
develops a phase singularity which can be detected by rotating the 
two momenta about the axis formed by their sum.
}
\end{figure}

In terms of its implication for subleading terms, \eqn{GravLoopColl} has
a slightly different meaning from the corresponding equations for 
the collinear limits in gauge theory, or for the soft limits in either
gauge theory or gravity.  In these other limits, the leading
power-law behavior is determined; subleading, non-universal behavior is 
down by a power of either $\sqrt{k_s}$ or $\sqrt{s_{ab}}$.  In the case of
\eqn{GravLoopColl}, there are other terms of the same order as
$\spb{a}.{b}/\spa{a}.b$ as $s_{ab} \to 0$, namely any term that does
not vanish as $s_{ab} \to 0$.  However, these terms do not acquire any
phase as $\vec k_a$ and $\vec k_b$ are rotated around their sum $\vec
P$, as depicted in \fig{RotateColFigure}, and thus they can be meaningfully
separated from the terms described by \eqn{GravLoopColl}.  (In
space-time signature $(2,2)$, the spinor products $\spa{a}.{b}$ and
$\spb{a}.{b}$ are not complex conjugates of each other, so that $\spa{a}.{b}$
can be taken to zero independently of $\spb{a}.{b}$, in order to separate out
the $\spb{a}.{b}/\spa{a}.b$ terms~\cite{CSUnpublished}.)

%

For example, consider the two factors,
$$
\hbox{(a)} \quad {\spb1.2 \over \spa1.2} \,, \hskip 2 cm 
\hbox{(b)} \quad {\spb1.3 \over \spa1.3} \,.
\equn
$$
If we take $\vec k_1$ to be nearly collinear with $\vec k_2$ and
rotate $\vec k_1$ and $\vec k_2$ around the vector $\vec P= \vec k_1 +
\vec k_2$ the factor (b) undergoes only a slight numerical variation.
On the other hand, from \eqn{SpinorExplicit}, the factor (a)
undergoes a large phase variation, proportional to the angle of rotation.
Thus a Fourier analysis in this azimuthal rotation angle will extract
the universal terms in \eqn{GravLoopColl} from the (approximately)
constant non-universal terms, giving meaning to this equation.

\begin{figure}[ht]
\centerline{\epsfxsize 1.8 truein \epsfbox{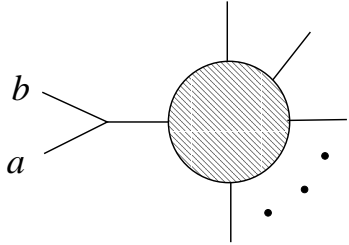}}
\vskip -.2 cm
\caption[]{
\label{CollFactFigure}
\small The class of tree diagrams in a gravity theory that can have a
phase singularity factorizes in the collinear limit $a\parallel b$.  The
appearance of the same three-vertex for any number of external legs
implies the universality of the tree-level splitting amplitudes. The
splitting amplitudes are given by evaluating the three-vertex in the
collinear limit in a helicity basis. }
\end{figure}

The universality of the tree-level splitting amplitudes for gravity
amplitudes with any number of external legs may be understood in terms of
Feynman diagrams in any gauge which does not introduce extra singularities
into the vertices or propagators, besides the usual $1/p^2$ propagator
factor (for example, de Donder gauge~\cite{DeWitt}).  Although use of
Feynman diagrams generally obscures the relationships between gravity and
gauge theory scattering amplitudes, here we only require the diagrams'
factorization properties.  Terms with a phase singularity for $a \parallel
b$ must contain a factor of the form $\spb{a}.{b}^2/s_{ab}$.  The only
tree-level Feynman diagrams that contain a pole in $s_{ab}$ 
(from a propagator) are of the type shown in \fig{CollFactFigure};
they all contain the same three-point vertex.%
\footnote{In the helicity formalism~\cite{SpinorHelicity}, a reference
momentum entering the polarization vector or tensor could produce a pole 
in $s_{ab}$ in other diagrams, but this is easily avoided by choosing the 
reference momenta to be neither $k_a$ nor $k_b$.}
The splitting amplitudes are given by a straightforward evaluation of 
the three-vertex (multiplied by the $s_{ab}$ pole) in a helicity basis 
in the collinear limit.  (For the analogous gauge theory computation,
see refs.~\cite{ManganoReview,Factorization}.)

%

\begin{figure}[ht]
\centerline{\epsfxsize 2.1 truein \epsfbox{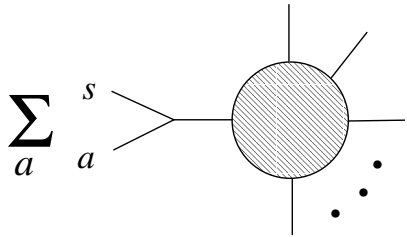}}
\vskip -.2 cm
\caption[]{
\label{SoftFactFigure}
\small The class of tree diagrams in a gravity theory that contribute 
in the soft limit, where leg $s$ is soft.  The soft functions are found 
by summing over all three-vertices containing a soft leg.}
\end{figure}

Similarly, the validity of the soft factor (\ref{GravTreeSoftFactor})
for an arbitrary number of external legs also
follows from the factorization properties of Feynman diagrams. In this
case the tree diagrams that contribute to the soft factors are of the
form shown in \fig{SoftFactFigure}, and the complete soft factor is given 
by summing over all three-point vertices with a soft leg (multiplied
by the respective propagator).

%

One may also prove the universality of the tree-level 
splitting amplitudes using the $n$-point version of the KLT relations
given in \app{UniversalAppendix}.  Alternatively, one may obtain
the tree-level soft and collinear splitting functions from string theory by
extending the gauge theory discussion given by Mangano and Parke
\cite{ManganoReview} to the case of gravity, using a closed string
instead of an open string.  The factorization of the closed string
integrands into products of open string integrands ensures that the
gravity splitting functions are given in terms of products of the
corresponding gauge theory splitting functions, as given in 
\eqn{GravYMSplit}.


\subsection{Absence of Loop Corrections}
\label{LoopSoftSplitSubSection}

\begin{figure}[ht]
\centerline{\epsfxsize 4.2 truein \epsfbox{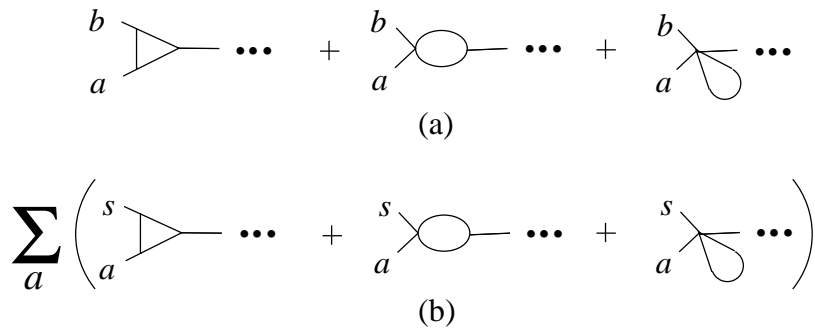}}
\vskip -.2 cm
\caption[]{
\label{LoopSoftSplitFigure}
\small One-loop factorizing corrections to (a) the collinear splitting
amplitudes and (b) the soft functions.}
\end{figure}

We now show that the soft and collinear splitting amplitudes for gravity
--- in contrast to those for gauge theory --- do not have any higher loop
corrections. In general, in covariant gauges the splitting and soft 
functions may be classified into two categories: factorizing and
non-factorizing contributions~\cite{Factorization}.  Diagrams for
the factorizing one-loop corrections to the splitting and soft 
functions are shown in \fig{LoopSoftSplitFigure}.  Non-factorizing 
contributions can arise whenever infrared divergences do not behave 
smoothly in the soft or collinear limits, as discussed in 
ref.~\cite{Factorization}.

First consider the factorizing contributions.  Since each Feynman
diagram in \fig{LoopSoftSplitFigure} has a power of $\kappa^2$ as
compared to the tree-level functions, dimensional analysis requires
that the diagram carry an extra power of $|s_{ab}|$ in the collinear
case or an extra power of $|s_{as}|$ in the soft limit.  This
suppresses potential one-loop corrections to either the collinear
($s_{ab} \rightarrow 0$) or soft ($s_{as} \rightarrow 0$) limits.

%

Now consider the non-factorizing contributions.  In one-loop gauge
theory amplitudes the infrared divergences, for e.g. a pure
gluon amplitude, are of the form
$$
- A_n^\tree \sum_{i\not = j}^n
\Bigl[{1\over \eps^2}  - {\ln(-s_{ij})\over\eps}\Bigr] \,.
\equn\label{YMIRSing}
$$
The mismatch between the infrared divergence of the $n$- and
$(n-1)$-point one-loop amplitudes on the left- and right-hand-sides of
the collinear limit~(\ref{OneLoopSplit}) implies that there must be 
a non-trivial contribution to the one-loop gauge splitting amplitude.  
This may be contrasted with the case of gravity:
a pure graviton amplitude has infrared divergences of the form~\cite{DNb}
$$
M_n^\tree 
\sum_{i\not = j}^n
\Bigl[ s_{ij}{\ln(-s_{ij})\over\eps}\Bigr] \,.
\equn\label{GravIRSing}
$$
In this case the infrared divergences exhibit smooth behavior in 
soft or collinear limits, because of the extra
power of $s_{ij}$ in each term; as any kinematic variable vanishes,
the infrared divergent term containing that variable goes smoothly to zero.
Thus there are no one-loop contributions to soft
or collinear splitting amplitudes arising from non-factorizing
contributions.  Again this difference in behavior between the gauge
and the gravity case is due to the dimensionful coupling in gravity
theories.

More generally, the appearance of a dimensionful coupling in gravity
implies that the contributions of the form that appear in gauge theory
splitting amplitudes and soft functions will be suppressed by
additional powers of vanishing $s_{ij}$ at all loop orders.  For the
factorizing contributions the argument is the same as for the one-loop
case.  For non-factorizing contributions, which involve infrared
divergences, e.g. $(-s_{ij})^{-\eps}/\eps^2$, a closer inspection is
required.

\begin{figure}[ht]
\centerline{\epsfxsize 2.9 truein \epsfbox{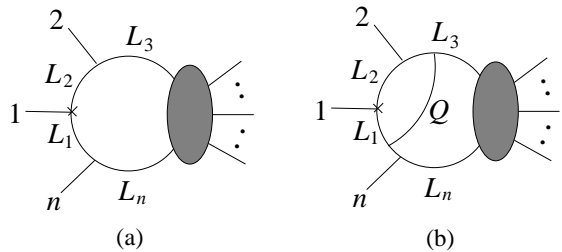}}
\vskip -.2 cm
\caption[]
{\label{NonFactFigure}
\small One- and two-loop examples of diagrams that can produce a 
pole in $k_1$ from the loop integration 
in the gauge theory case, but not in the gravity case.}
\end{figure}

In particular, in the gauge theory case it is possible for the loop
integration to generate a pole in $s_{ij}$, leading to a
non-factorizing contribution to soft factors or splitting amplitudes.  A
one-loop example of a diagram where this can happen is shown in
\fig{NonFactFigure}a.  In the soft limit where $k_1 \rightarrow 0$,
the region of loop integration that can produce a kinematic pole in
$k_1$ is where an extra propagator diverges.

As an example, before taking $k_1 \rightarrow 0$, in the region $L_1
\approx 0$ only the three propagators with momenta $L_n$, $L_1$ and
$L_2$ in \fig{NonFactFigure}a diverge.  As $k_1 \rightarrow 0$, the
propagator with momentum $L_3 = L_1 - k_1 - k_2$ also diverges since
$L_3^2 \approx 2 k_1 \cdot k_2 - 2 L_1 \cdot(k_1 + k_2)$.  Since we
are interested only in the leading behavior as $k_1\rightarrow 0$ we
may set $L_1 = 0$ in the remaining part of the diagram, effectively
leaving only a box diagram to be analyzed.

%

For the case of gravity, the graviton vertex attached to leg 1 (and
marked by a cross in the figure) contains one extra power of $k_1$ or
$L_1$, as compared to the gauge theory vertex, and the contribution is
therefore suppressed compared to gauge theory.  Since one can obtain
at most a single power of $1/k_1$ in the gauge theory case, 
the gravity case cannot have a pole in $k_1$ and is therefore
suppressed compared to the tree gravity soft $\Soft$ function which
does contain a single pole in $k_1$.  The cases where three
propagators diverge can also be analyzed by observing that these cases
effectively reduce to triangle integrals.

These arguments extend to the multi-loop case.  Consider, for example,
the two-loop diagram in \fig{NonFactFigure}b. In the gauge theory
case, in order to obtain a contribution analogous to the one-loop one
discussed above we must also take $Q \approx 0$. Again the extra
powers of $k_1$, $L_1$ or $Q$ in the vertices suppress any potential
gravity contribution.

Similarly, for other potential non-factorizing soft contributions, and
also in the case of collinear limits, one may show that the extra
powers of momenta in the vertices suppress all potential loop
contributions.  Thus the appearance of a dimensionful coupling in
gravity theories implies that the tree-level soft and collinear
functions are exact to all orders of perturbation theory, so that
\eqns{GravLoopColl}{GravLoopSoft} hold at any loop order.


\section{Ansatze for an Arbitrary Number of External Legs}
\label{AnsatzSection}

In this section we make use of the soft and collinear limits to
construct ansatze for both the all-plus and $N=8$ MHV
amplitudes for an arbitrary number of external legs.  In a previous
letter we presented the ansatz for the all-plus
(self-dual) case~\cite{GravAllPlus}; here we provide some of the
details of the derivation as well as an alternate representation of
the amplitude that has manifest symmetry under relabelings of
external legs.  We also present a new ansatz for the $N=8$ supergravity
amplitudes.  The constraints that the amplitudes have the correct poles 
in all channels are rather restrictive and very
likely determine the unique form of the amplitude, although we do not
have a proof that this is so. Analogous constructions in the gauge
theory case have been proven to lead to the correct 
results~\cite{SusyFour,AllPlus,MahlonAllPlus}.


\subsection{Functions with Simple Soft Properties}
\label{HalfSoftSubSection}

The first step in constructing ansatze for the amplitudes is to find 
a set of functions which have simple behavior in the soft limits.
A good starting point is the three $h$ functions defined in 
equations~(\ref{Half1}), (\ref{Half2}) and (\ref{Half3}), which appear 
in the coefficients of the box integrals in the explicit expressions 
for the four-, five- and six-point all-plus and $N=8$ MHV amplitudes.
We collect them again here, 
$$
\eqalign{
h(a,\{1\},b) &= {1\over \spa{a}.{1}^2 \spa{1}.{b}^2} \,, \cr 
h(a,\{1,2\},b) &=  
{ \spb1.2 \over 
  \spa1.2 \spa{a}.{1}\spa{1}.{b}\spa{a}.{2}\spa{2}.{b}} \,, \cr
h(a,\{1,2,3\},b) &=
{ \spb1.2\spb2.3\over
  \spa1.2\spa2.3 \spa{a}.{1}\spa{1}.{b}\spa{a}.{3}\spa{3}.{b} }
+ { \spb2.3\spb3.1\over
    \spa2.3\spa3.1 \spa{a}.{2}\spa{2}.{b}\spa{a}.{1}\spa{1}.{b} } \cr
&\hskip1cm
+ { \spb3.1\spb1.2\over
    \spa3.1\spa1.2 \spa{a}.{3}\spa{3}.{b}\spa{a}.{2}\spa{2}.{b}} \,. \cr
}
\equn\label{HalfExamples}
$$

It is easy to verify that these functions satisfy the following soft limits,
$$
h(a,M,b)\ \mathop{\longrightarrow}^{k_m\to0}
\ -\Soft_m(a,M,b) \times h(a,M-m,b),
\hskip2cm \hbox{for $m\in M$}.
\equn\label{HSoft}
$$
Here the `half-soft' factor,
$$
\Soft_m(a,M,b) \equiv { -1 \over \spa{a}.{m}\spa{m}.{b} } 
\sum_{j \in M} 
\spa{a}.{j}\spa{j}.{b} {\spb{j}.{m}\over\spa{j}.{m}}  \,, 
\equn\label{HalfSoftn}
$$
is closely related to the gravity soft function 
$\Soft_n \equiv \SoftGrav(n^+)$ defined in \eqn{GravTreeSoftFactor},
except that the sum in \eqn{HSoft} is over only a subset of the
legs in the amplitude.   Thus $h$ obeys soft limits very similar to the 
tree-level gravity amplitudes, except that there is no momentum conservation
constraint on $h$ or $\Soft_m(a,M,b)$, so the $h$ functions may be thought of
as off-shell extensions of the tree amplitudes.

Here we wish to find explicit forms for `half-soft' functions $h(a,M,b)$, 
which satisfy \eqn{HSoft} for an arbitrary number of external legs.
This is accomplished by using \eqn{HalfExamples} to motivate a guess
for the general form of $h$, and then using the soft properties to fix
its components.  From the form of $\Soft_m(a,M,b)$,
we see that $h(a,M,b)$ should be symmetric in $a \lr b$, and in the 
exchange of any members of $M$.  Also, \eqn{HalfExamples} suggests
that it can be written as sums of products of spinor phase factors
$\spb{j_1}.{j_2}/\spa{j_1}.{j_2}$, where $j_1,j_2 \in M$, multiplied
by appropriate powers of $\spa{a}.{j_l}\spa{j_l}.{b}$.
We write $h(a,M,b)$ as
$$
h(a,\{1,2, \ldots, m\},b)\ =\ \sum_{i_1,i_2,\ldots,i_m = 0}^{m-2} 
\phi(i_1,i_2,\ldots,i_m) 
\prod_{j=1}^m (\spa{a}.{j} \spa{j}.{b})^{i_j-1} \,,
\equn\label{Hgendef}
$$
where $\phi(i_1,i_2,\ldots,i_m)$ is defined to be a symmetric
function of its arguments, and nonzero only for $\sum_{j=1}^m i_j = m-2$.
Thus at least one of the arguments $i_j$ must equal zero, and using the
symmetry we can choose this to be the last argument.
Then, to incorporate the soft limits~(\ref{HSoft}), 
we define $\phi(i_1,i_2,\ldots,i_m)$ recursively by
$$
\eqalign{
 \phi(0,0) &= {\spb1.2\over\spa1.2}, \cr
 \phi(i_1,i_2,\ldots,i_{m-1},0) 
&= \sum_{j=1}^{m-1} \phi(i_1,i_2,\ldots,i_j-1,\ldots,i_{m-1})
      \times {\spb{j}.{m}\over\spa{j}.{m}}\,, \cr}
\equn\label{Phirecurse}
$$
where $\phi$ is also defined to be zero if any of its arguments
is negative.

We give a few examples of the factors $\phi$:
$$
\eqalign{
\phi(0,0) &= {\spb1.2\over\spa1.2}\,, \cr
\phi(1,0,0) &= {\spb1.2\spb1.3\over\spa1.2\spa1.3}\,, \cr
\phi(2,0,0,0) &= {\spb1.2\spb1.3\spb1.4\over\spa1.2\spa1.3\spa1.4}\,, \cr
\phi(1,1,0,0) &= {\spb1.2\spb2.3\spb1.4\over\spa1.2\spa2.3\spa1.4}
               + {\spb1.2\spb1.3\spb2.4\over\spa1.2\spa1.3\spa2.4}\,, \cr
\phi(2,1,0,0,0) &= 
  {\spb1.2\spb2.3\spb1.4\spb1.5\over\spa1.2\spa2.3\spa1.4\spa1.5} 
+ {\spb1.2\spb1.3\spb2.4\spb1.5\over\spa1.2\spa1.3\spa2.4\spa1.5}
+ {\spb1.2\spb1.3\spb1.4\spb2.5\over\spa1.2\spa1.3\spa1.4\spa2.5}\,. \cr
}\equn\label{PhiExamples}
$$

An interesting property of the $\phi$ functions is that they can
be generated from group theory Young tableaux, as an alternative to
\eqn{Phirecurse}. We can restrict our attention to the
$\phi(i_1,i_2,\ldots,i_m)$ with $i_1\geq i_2 \geq \cdots \geq i_m=0$, 
since all other $\phi$'s can be obtained by simple relabelings.  
The formula for $\phi$ can be schematically represented as
$$
\phi(i_1,i_2,\ldots,i_m) = \sum_{\rm NSYT} 
\underbrace{\left({[\hskip .3cm] \over \langle \hskip .3cm \rangle}\right)
\cdots\left({[\hskip .3cm] \over 
\langle \hskip .3cm \rangle}\right)}_{m-1 \hskip .2cm {\rm terms}} \,,
\equn\label{YTSchematic}
$$
where each term in the sum corresponds to a non-standard Young
tableaux (NSYT). The NSYT are defined as all possible labelings from
$1$ to $m-2$ of the tableaux (with $i_1$ boxes in the first row, $i_2$
boxes in the second row, etc...), with the restriction of ascending
order along rows. The standard requirement of descending order along
columns is relaxed. The rule for constructing each phase factor in
\eqn{YTSchematic} from the corresponding NSYT is best illustrated by
an example.  For $\phi(2,1,0,0,0)$, the three terms in \eqn{PhiExamples}
correspond to these NSYT:
$$
\begin{array}[c]{c}
\begin{picture}(700,30)
\put(150,0){$\stableau{1 & 2\\3}$}
\put(350,0){$\stableau{1 & 3\\2}$}
\put(550,0){$\stableau{2 & 3\\1}$}
\end{picture}
\end{array}
$$ 
\hskip .5cm
\vskip .5cm
\par\noindent
To obtain the phase factors, one first extends the YT vertically with
`empty' boxes until it has $m$ rows, in order to represent all $m$ 
arguments in $\phi$.  Then one removes both the last empty box (in row $j$,
say) and the full box containing the highest number (in row $i$),
writing a factor of $\spb{i}.{j}/\spa{i}.{j}$ for this step.  
Repeating the step until all boxes are gone yields the phase factor.
For example,
$$
\begin{array}[c]{c}
\begin{picture}(50,85)
\put(0,40){$\stableau{1 & 2\\3}$}
\put(0,22){\line(0,-1){54}}
\put(0,4){\line(1,0){7}}
\put(0,-14){\line(1,0){7}}
\put(0,-32){\line(1,0){7}}
\put(-10,44){\mbox{\small 1}}
\put(-10,26){\mbox{\small 2}}
\put(-10,8){\mbox{\small 3}}
\put(-10,-10){\mbox{\small 4}}
\put(-10,-28){\mbox{\small 5}}
\end{picture}
\end{array} 
= {\ph2.5} \times
\begin{array}[c]{c}
\begin{picture}(50,85)
\put(0,40){$\stableau{1 & 2}$}
\put(0,40){\line(0,-1){54}}
\put(0,22){\line(1,0){7}}
\put(0,4){\line(1,0){7}}
\put(0,-14){\line(1,0){7}}
\end{picture}
\end{array} 
= {\ph1.4}{\ph2.5} \times
\begin{array}[c]{c}
\begin{picture}(50,85)
\put(0,40){$\stableau{1}$}
\put(0,40){\line(0,-1){36}}
\put(0,22){\line(1,0){7}}
\put(0,4){\line(1,0){7}}
\end{picture}
\end{array} 
= {\ph1.3}{\ph1.4}{\ph2.5} \times
\begin{array}[c]{c}
\begin{picture}(50,85)
\put(0,58){\line(0,-1){36}}
\put(0,58){\line(1,0){7}}
\put(0,40){\line(1,0){7}}
\put(0,22){\line(1,0){7}}
\end{picture}
\end{array} 
\hskip -.5cm= {\ph1.2}{\ph1.3}{\ph1.4}{\ph2.5}
\equn
$$
\vskip .5cm
\par\noindent
gives the last term in $\phi(2,1,0,0,0)$ in \eqn{PhiExamples}.

The NSYT approach gives a simple formula for the number of terms in 
each $\phi$,
$$
\Bigl[ \mbox{\# of terms in $\phi(i_1,\ldots,i_m)$} \Bigr] 
 \ =\ {(m-2)! \over \prod_{j=1}^m i_j!} \,,
\equn
$$
which is the analog of the `hook' formula for standard Young
tableaux.  The $\phi$ functions can also be generated graphically 
from certain `elk diagrams', described in \app{HalfCoeffRecurseAppendix},
which are in one-to-one correspondence with the NSYT.

Another approach to constructing the $h$ functions is the recursive currents
method~\cite{RecursiveBG,RecursiveK,MahlonAllPlus} discussed in
\app{RecursiveAppendix}.  Using this method, we have found an explicit 
non-recursive form for the functions,
$$
\eqalign{
h(a,\{1,2,\ldots,n\},b) &\equiv {\spb1.2 \over \spa1.2}
 { \spab{a}.{\Ksl_{1,2}}.3  \spab{a}.{\Ksl_{1,3}}.4 
   \cdots \spab{a}.{\Ksl_{1,n-1}}.{n}
  \over \spa2.3\spa3.4 \cdots \spa{n-1,}.{n} 
  \, \spa{a}.1 \spa{a}.2\spa{a}.3 \cdots \spa{a}.{n} 
  \, \spa1.{b} \spa{n}.{b} }  \cr
&\hskip1cm + \Perm(2,3,\ldots,n), \cr}
\equn\label{NonRecursiveH}
$$
where $K_{i,j} = k_i + k_{i+1} + \cdots + k_j$.  In the
form~(\ref{NonRecursiveH}) the symmetry properties of $h$ under the
interchange of $a \leftrightarrow b$ and 
$1 \leftrightarrow j \in \{2,\ldots,n\}$ are not manifest.  
Nevertheless, in \app{HalfCoeffEquivAppendix} we show that the forms in
\eqns{Hgendef}{NonRecursiveH} are in fact equal.

As mentioned above, the $h$ functions can be thought of as off-shell
extensions of gravity tree amplitudes.  Using the non-recursive 
form~(\ref{NonRecursiveH}), it is not hard to show that they are related 
to the BGK expressions for the MHV tree amplitudes~(\ref{BGKMHV}) via,
$$
\eqalign{
{ h(n,\{n-1,n-2,\ldots,2\},1) \over \spa{n}.1^2 } 
  \Bigg|_{k_1+k_2+\cdots+k_n=0}
 &= (-1)^n { M_n^\tree(1^-,2^-,3^+,\ldots,n^+)
   \over i \, \, \spa{1}.{2}^8 }\ . \cr}
\equn\label{hMrelation}
$$
In this form, momentum conservation only has to be used in one factor
in $M_n^\tree$, in order to convert it into $h$.  

In light of \eqn{hMrelation}, it is perhaps not too surprising that
the $h$ functions satisfy a squaring relation to the $g$ 
functions (\ref{gDef}) appearing in the gauge theory amplitudes, 
analogous to the KLT relations for tree amplitudes.  For example,
$$
\eqalign{
h(a,\{1\},b) &= [g(a,\{1\},b)]^2\,, \cr
h(a,\{1,2\},b) &= s_{12} \, g(a,\{1,2\},b) \, g(a,\{2,1\},b)\,, \cr
h(a,\{1,2,3\},b) &= s_{12} \, s_{23} \, g(a,\{1,2,3\},b) \,
g(a,\{3,2,1\},b) + 
{\rm perms} \,, }
\equn\label{hIsgSquared}
$$
and so forth.  These relations are analogous to the KLT relations, in
\eqn{GravYM}, except that they hold for functions that appear at
one loop (and the $s_{ij}$ factors and permutations appearing
are not precisely the same).


\subsection{Ansatz for All-Plus Amplitudes}
\label{AllPlusAnsatzSubSection}

\begin{figure}[ht]
\centerline{\epsfxsize 1.1 truein \epsfbox{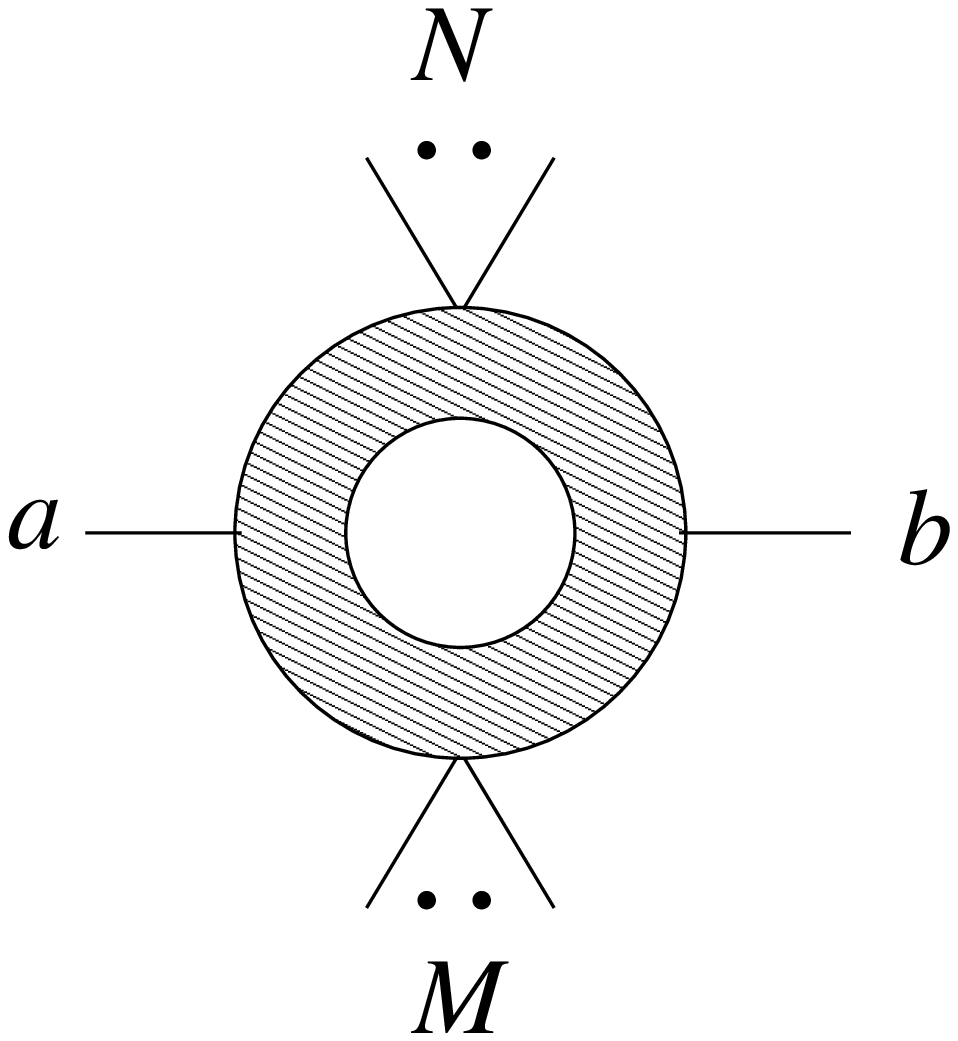}}
\vskip -.2 cm
\caption[]{
\label{LegsFigure}
\small The configurations of external legs that are summed over in 
\eqn{AllPlusSimple}.
}
\end{figure}

The forms of the four-, five- and six-point amplitudes in
eqs.~(\ref{FourGravAllPlusDFour}), (\ref{FiveGravAllPlusDFour}) and
(\ref{SixGravAllPlusDFour}), and the soft properties~(\ref{HSoft}) of
the $h$ functions have led us to the following ansatz (see also
ref.~\cite{GravAllPlus}) for the one-loop all-plus (self-dual)
amplitudes in $D=4$,
$$
M_n(1^+, 2^+, \ldots, n^+) = 
- {i \, (-1)^n \over (4 \pi)^2\cdot 960} \,  
\sum_{1 \leq a < b \leq n \atop M, N} h(a, M, b) h(b, N, a) 
 \tr^3[a\, M\, b\, N]\ +\ \Ord(\e),
\equn\label{AllPlusSimple}
$$
where $a$ and $b$ are massless legs, and $M$ and $N$ are two sets
forming a `distinct non-trivial partition' of the remaining $n-2$ legs;
i.e., $M$ and $N$ should both be non-empty, and the partition $(M,N)$
is not considered distinct from $(N,M)$.  This configuration of
external legs is depicted in \fig{LegsFigure}.   We do not have
an ansatz that works to all orders in $\e$.  The $D=4$ 
amplitudes~(\ref{AllPlusSimple}) are also generated by a self-dual
gravity action~\cite{SiegelSelfDualSugra,ChalmersSiegel,CSUnpublished}.

%

The fact that the amplitudes~(\ref{AllPlusSimple}) have the correct soft
limits is a consequence of the soft properties of the $h$ functions, 
\eqn{HSoft}.  As $k_n \to 0$, the term labeled by $(a,M,b,N)$ in $M_{n-1}$
gets contributions from two terms in $M_n$, those labeled by
$(a,M+n,b,N)$ and $(a,M,b,N+n)$.  Each of the factors $h(a,M+n,b)$ and
$h(b,N+n,a)$ in \eqn{AllPlusSimple} supplies `half' of the soft factor in
this limit, since
$$
\Soft_n = \Soft_n(a,M,b) + \Soft_n(b,N,a).
\equn\label{ReconSoft}
$$
The trace factors behave smoothly in the soft limit, serving only to
prevent the unwanted terms where $a$ or $b$ becomes soft.

The collinear properties are slightly more difficult to establish.  They 
rely on the two non-trivial collinear limits of the half-soft 
function $h$ (after taking into account its symmetries),
$$
\eqalign{
h(a,\{1,2,3,\ldots,n\},b)\ &\mathop{\longrightarrow}^{1 \parallel 2}\ 
 {1\over z(1-z)} {\spb1.2\over\spa1.2} 
 \times h(a,\{P,3,\ldots,n\},b) \,, \cr
h(1,\{2,3,\ldots,n\},b)\ &\mathop{\longrightarrow}^{1 \parallel 2}\ 
  {1 \over \spa{1}.{2}} \, 
{ \spa{1}.{b} \langle b^-| \Ksl_{3,n} | 2^-\rangle 
  \over \spa{2}.{b}^2 } \times h(1,\{3,\ldots,n\},b) \,, \cr}
\equn\label{HCol}
$$
where we have used the Schouten identity, \eqn{Schouten}, to derive
the second limit, dropping terms without phase singularities, in 
accordance with the discussion in \sec{CollinearGravSubSection}.

Consider the collinear limit $1 \parallel 2$ of
$M_n(1^+,2^+,\ldots,n^+)$.  For a term labeled by $(a,M,b,N)$
in the ansatz~(\ref{AllPlusSimple}), there are three independent 
possibilities:
\par\noindent
(1) 1 and 2 both belong to the same set, say $M$,
\par\noindent
(2) $a=1$ and $b\neq2$, so that 2 belongs to a massive set,
\par\noindent
(3) $a=1$ and $b=2$.

Case (3) is trivial; \eqn{NonRecursiveH} has no $1/\spa{a}.{b}$ factor,
and hence there is no contribution to the collinear (phase) singularity.
Case (1) is also simple.  The first~\eqn{HCol} shows that these terms
precisely account for all the terms in the expression (\ref{AllPlusSimple}) 
for the target amplitude $M_{n-1}(P^+, \ldots, n^+)$ 
in which $P \in M$.

The only remaining task is to show that the case (2) terms correctly 
give rise to the terms in the expression for the target amplitude in 
which $P$ does not belong to $M$ or $N$.  The second~\eqn{HCol} shows that 
an individual $h$ function has a `too singular' $1/\spa1.2$ 
behavior in case (2).  However, the combination of two different terms,
labeled by $(1,M+2,b,N)$ and $(1,M,b,N+2)$, cancels the singularity 
down to the desired level of $\spb1.2/\spa1.2$.  More concretely, 
by using momentum conservation, $K_M+K_N+k_1+k_2+k_b = 0$, 
where $K_M$ is the sum of the massless momenta in the set $M$, we have
$\langle 1^+| \Ksl_M | b^+\rangle = - \langle 1^+| \Ksl_N | b^+\rangle
- \spb{1}.2 \spa2.b$.
Also expanding the traces to first order in $\spb1.2$, we obtain
$$
\eqalign{
&h(1,M+2,b) h(b,N,1) \tr^3[1(M+2)bN] + h(1,M,b) h(b,N+2,1) \tr^3[1Mb(N+2)]
\cr
&\mathop{\longrightarrow}^{1 \parallel 2}\ 
 {z^3+3z^2(1-z)\over z(1-z)} {\spb1.2\over\spa1.2} \, 
 h(P,M,b) h(b,N,P) \tr^3[PMbN] \,. \cr}
\equn\label{LoopColIdent}
$$
Adding the analogous equation with the roles of 1 and 2 (and $z$ and $1-z$) 
exchanged gives
$$
\eqalign{
&h(1,M+2,b) h(b,N,1) \tr^3[1(M+2)bN] + h(1,M,b) h(b,N+2,1) \tr^3[1Mb(N+2)]
\ +\ (1\lr2) \cr
&\mathop{\longrightarrow}^{1 \parallel 2}\ 
 {1 \over z(1-z)} {\spb1.2\over\spa1.2} \, 
 h(P,M,b) h(b,N,P) \tr^3[PMbN] \,, \cr}
\equn\label{LoopColIdent2}
$$
which accounts properly for the terms in the target expression
$M_{n-1}$ in which $P=a$ (or similarly, $P=b$).  Thus
$M_n(1^+,2^+,\ldots,n^+)$ does obey the required collinear limits.


\subsection{Ansatz for $N=8$ MHV Amplitudes}
\label{N8AnsatzSubSection}

In \sec{N8CutSubSection} we used the dimension-shifting relation
(\ref{GravDimShiftRel}) to obtain the four-, five- and six-point
one-loop MHV amplitudes in $N=8$ supergravity, given in
\eqn{N8Explicit}, from the corresponding all-plus amplitudes.  These
results suggest the following ansatz for the $n$-point one-loop MHV
$N=8$ amplitudes:
$$
M_n^{N=8}(1^-, 2^-, 3^+, \ldots, n^+) = 
{(-1)^n\over 8} \, \spa1.2^8 
\sum_{1 \leq a < b \leq n \atop M, N} h(a, M, b) h(b, N, a) 
 \tr^2[a\, M\, b\, N]\, \I_4^{aMbN}\ +\ \Ord(\e)\,,
\equn\label{MHVAnsatz}
$$
where the notation is identical to that of
\eqn{AllPlusSimple}.  The scalar box integral functions,
$\I_4^{aMbN}$, are given through $\Ord(\eps^0)$ by
\eqn{EasyTwoMassAnswer}.  The sum in \eqn{MHVAnsatz} includes all
inequivalent two-mass scalar box integrals with diagonally-opposite
massive legs, as shown in \fig{EasyTwoMassFigure},
as well as all one-mass scalar box integrals arising
from the terms in \eqn{MHVAnsatz} where either $M$ or $N$ consists of
a single massless leg.  (In the four-point case the sum is over 6
boxes with all massless legs, of which only 3 are inequivalent because of
an extra degeneracy, leading to the extra factor of 2 in the first line of
\eqn{N8Explicit}.)  The $N=8$ SWI, \eqn{PermIdentityGrav}, requires
that $M_n^{N=8}(1^-, 2^-, 3^+, \ldots, n^+)/\spa1.2^8$ is totally
symmetric with respect to permutations of its arguments; this symmetry
is manifest in \eqn{MHVAnsatz}.

Just as in the case of the all-plus amplitudes, the soft behavior of
\eqn{MHVAnsatz} follows from the soft properties of the
$h$ functions.  The only real difference is that one of the powers of 
the trace $\tr[a\, M\, b\, N]$ is replaced by $\I_4^{aMbN}$.  But 
these integrals also transform smoothly in the limit $k_n \to 0$,
$$
\I_4^{aMb(N+n)} \hskip .5cm \mbox{and} \hskip .5cm \I_4^{a(M+n)bN}
\hskip .5cm \stackrel{k_n\to 0}{\longrightarrow} \hskip .5cm
\I_4^{aMbN}\,,
\equn
$$ 
and two powers of the traces suffice to kill the unwanted terms where
either $a$ or $b$ becomes soft ($\I_4^{aMbN}$ does not develop a
singularity as $k_a\to0$).

The ansatz (\ref{MHVAnsatz}) also must have universal collinear behavior.
In the limit $1\parallel2$, the analysis is again quite similar to that
presented in \sec{AllPlusAnsatzSubSection} for the all-plus ansatz.  The
same three cases are encountered; the only subtle case is case (2).
In this case, using the same labeling as in the all-plus discussion,
the coefficients of the scalar box integrals $\I_4^{1(M+2)bN}$, 
$\I_4^{1Mb(N+2)}$, $\I_4^{2(M+1)bN}$ and $\I_4^{2Mb(N+1)}$
each have a $1/\spa1.2$ singularity.  This singularity should cancel, since 
we only expect the phase singularity $\spb1.2/\spa1.2$.  The
cancellation can be demonstrated with the use of integral relations 
of the type,
$$  
z\, \I_4^{1(M+2)bN} + (1-z)\, \I_4^{2Mb(N+1)}\
{\buildrel {1 \parallel 2}\over\longrightarrow}\ \I_4^{PMbN}\ +
\ \Ord(\sqrt{s_{12}}) \,,
\equn\label{IntegralCollinear}
$$
where $k_1=zk_P$ and $k_2=(1-z)k_P$ in the collinear limit.  However,
to verify that phase singularity of \eqn{MHVAnsatz} matches that of
\eqn{GravTreeSplit} requires collinear analysis of the integrals to
one higher order in $\sqrt{s_{12}}$. The required integral relation
appears to be more subtle and involves a larger combination
of integrals. We have, however, verified numerically that the ansatz
(\ref{MHVAnsatz}) has the correct collinear limits for $n \leq 7$.
(We have also checked numerically through $n=8$ that the infrared
singularities of $M_n^{N=8}$ are correctly given by \eqn{GravIRSing}.)

Comparing the all-plus and $N=8$ MHV ansatze,
\eqns{AllPlusSimple}{MHVAnsatz}, it might appear that the former is
obtained from the latter (up to an overall factor) simply by substituting
the higher-dimensional values~(\ref{I4D12}) for the box integrals.
However, the dimension-shifting relation does not work that simply, for
$n=5$ or 6.  The quadratic polynomial for $\I_4^{aMbN}[\mu^8]$ in
\eqn{I4D12} is not the same as the extra trace factor $\tr[a\,M\,b\,N]$ in
the all-plus expression, and the pentagon and hexagon contributions to
\eqn{N8Explicit} give a non-vanishing contribution, which somehow
compensates for this discrepancy.  Presumably the required rearrangements
become yet more complicated for $n>6$.

It is instructive to compare the one-loop MHV $N=8$ supergravity
amplitudes, \eqn{MHVAnsatz}, with the corresponding amplitudes in
$N=4$ super-Yang-Mills theory, \eqn{N4YMAlphaDecomp}.  Note that both
amplitudes are expressed just in terms of scalar box integral functions. 
This result, though expected for $N=4$ super-Yang-Mills theory based
on power-counting grounds, is somewhat surprising for $N=8$ supergravity, 
since as remarked in \sec{N8PowerSubSection}, a naive power-count 
for $n\geq5$ does not exclude the appearance of triangle or bubble 
integrals.


\section{Discussion}
\label{ConclusionsSection}

In this paper we have exploited relations between gauge theory and
gravity to calculate the first three members of two infinite sequences of
one-loop gravity amplitudes: the all-plus helicity amplitudes of
non-supersymmetric gravity, and the maximally helicity-violating
amplitudes of $N=8$ supergravity.  From these results, and the analytic
properties of $n$-graviton amplitudes, we obtained ansatze for the
remaining members of both sequences.

In approaching any amplitude calculation in gravity we have found it
useful to first consider the corresponding gauge theory calculation.
Kawai, Lewellen and Tye (KLT) have given precise expressions for
closed-string tree amplitudes as (roughly speaking) the squares of 
open-string tree amplitudes.  In the field theory limit, this implies that
properties of gauge theory tree amplitudes should be reflected in
gravity tree amplitudes.  As an example of this, we derived the
properties of the gravity amplitudes as two momenta become collinear
from the corresponding properties of gauge theory amplitudes.  

The methods used to construct gravity amplitudes in this paper,
relying on the analytic properties of the amplitudes, i.e. their
(unitarity) cuts and poles, are essentially the same as those
developed previously for evaluating multi-leg one-loop amplitudes in
QCD~\cite{AllPlus,SusyFour,SusyOne,Review,Zjets}.  The analytic
properties for dimensions away from $D=4$ are useful since they remove
ambiguities normally associated with reconstruction of amplitudes from
their cuts.  When the KLT relations are combined with this analytic
approach, gravity loop amplitudes can be obtained without evaluating
even a single gravity Feynman diagram; the entire process relies only
on gauge theory tree amplitudes.

Unitarity implies that relationships between tree amplitudes should
somehow be reflected as relationships between loop amplitudes.  Here we
demonstrated that `dimension-shifting' relations between one-loop all-plus
and maximally helicity-violating $N=4$ supersymmetric gauge theory
amplitudes can be transformed into relations between one-loop all-plus and
maximally helicity-violating $N=8$ supersymmetric gravity amplitudes.  The
gravity relation is of practical value since it allows us to obtain the
exact forms for the four-, five- and six-point $N=8$ amplitudes without
performing an explicit calculation; instead we applied a simple `dimension
shift' to the integral functions appearing in the all-plus amplitudes.  We
have also seen that there are `squaring relations', \eqn{hIsgSquared},
reminiscent of the tree-level KLT relations, between the coefficients of
the box integrals for $N=8$ supergravity and $N=4$ super-Yang-Mills, which
have survived the integral reduction procedure.

The one-loop $N=8$ amplitudes first develop ultraviolet divergences in
$D=8$.  This is immediately apparent in the four-point amplitude
first evaluated by Green, Schwarz and Brink~\cite{GSB}.  In dimensional
regularization, there are no divergences in odd dimensions, because
odd-dimensional Lorentz-invariant quantities do not exist.  So the
next higher dimension one can inspect for one-loop ultraviolet divergences
in $N=8$ supergravity is $D=10$.  In the four-point amplitude, the $D=10$ 
divergence cancels~\cite{BDDPR}, but from the point of view of 
\eqn{FourGravAllPlus} the cancellation is again for a relatively trivial
reason; the $D=10$ divergences of the sum of the three integrals has to be
proportional to the only totally symmetric dimension-two four-point 
invariant, $s+t+u = 0$.  We have investigated the $D=10$ divergences
for the five- and six-point amplitudes as well, using \eqn{N8Explicit},
and find that they also cancel.  (Because $D$ is not close to 4, we 
need to use an expression good to all orders in $\e$.)
This cancellation may be connected with the fact that $D=10$ is the 
critical dimension for the superstring; however, it should be pointed out
that the two-loop divergence does not cancel, even at the four-point 
level~\cite{BDDPR}.

Currently the squaring relations that have been found between gravity 
and gauge theory are expressed entirely in terms of on-shell $S$-matrix 
elements.  It would be nice to find an off-shell field-theoretic 
formulation of gravity that also exhibits a squaring relation with 
gauge theory, and to investigate the implications of such a formalism
for the field equations.


\vskip .3 cm 
\noindent
{\bf Acknowledgments}

We thank D.C. Dunbar, A.K. Grant and D.A. Kosower for helpful discussions.

\appendix

\section{KLT Relations for an Arbitrary Number of External Legs}  
\label{UniversalAppendix}

The KLT relations~\cite{KLT} between gravity and gauge amplitudes are,
in the field theory limit and for an arbitrary number $n$ 
of external particles,
$$
\eqalign{
M_n^\tree(1,2,\ldots,n) 
&= i \, (-1)^{n+1} \, \Bigl[ A_n^\tree(1,2,\ldots,n) \sum_{\rm perms} 
f(i_1,\ldots,i_j) \bar{f} (l_1,\ldots,l_{j'}) \cr
& \hskip 4.5 cm
\times A_n^\tree(i_1,\ldots,i_j,1,n-1,l_1,\ldots,l_{j'},n) \Bigr]\cr
& \hskip 2 cm + \Perm(2,\ldots,n-2),
}
\equn\label{KLTGeneral}
$$
where `perms' are $(i_1,\ldots,i_j) \in \Perm(2,\ldots,n/2)$,
$(l_1,\ldots,l_{j'}) \in \Perm(n/2+1,\ldots,n-2)$, 
$j=n/2-1$, $j'=n/2-2$, giving a total of $(n/2-1)! \times (n/2-2)!$ 
terms in the square brackets.  We have assumed that $n$ is even here; 
the case of odd $n$ is completely analogous.  The functions $f$ and 
$\bar{f}$ are given by
$$
\eqalign{
f(i_1,\ldots,i_j)\ &=\ s(1,i_j) \prod_{m=1}^{j-1} 
\left( s(1,i_m)+ \sum_{k=m+1}^j g(i_m,i_k) \right), \cr
\bar{f}(l_1,\ldots,l_{j'})\ &=\ s(l_1,n-1) \prod_{m=2}^{j'} 
\left( s(l_m,n-1)+ \sum_{k=1}^{m-1} g(l_k,l_m) \right),
}
\equn\label{fDef}
$$
with 
$$
 g(i,j)\ =\ \cases{ s(i,j) \equiv s_{ij},  &$i>j$, \cr
                                       0,  &otherwise. \cr}
\equn\label{gijDef}
$$
These definitions are used to compute the first term in the `big'
permutation sum; the rest of the terms ($(n-3)$! all together) are
obtained by simply permuting the arguments of the $s(i,j)$'s and
$A_n^\tree$'s in the square brackets.  As a consistency check, we have
numerically verified that these relations correctly give the MHV
amplitudes of BGK, \eqn{BGKMHV}, up to $n=8$.


\section{Recursive Currents Approach}
\label{RecursiveAppendix}

The soft behavior of the half-soft functions $h(a,M,b)$ 
defined in \sec{HalfSoftSubSection} is similar to that of a 
tree amplitude, except that the momenta belonging to $h(a,M,b)$ do 
not have to satisfy any momentum-conservation relation.  Thus we
might suspect that the $h(a,M,b)$ could appear in off-shell quantities
related to tree amplitudes.  


\subsection{Off-Shell Currents $J$}
\label{JSubAppendix}

In gauge theory, Berends and Giele defined `currents', sums of
color-ordered Feynman diagrams where one leg is off-shell and the rest are
on-shell~\cite{RecursiveBG}.  These currents appear in recursive
algorithms for generating multi-leg tree amplitudes.  Mahlon used similar 
currents, but with two legs off-shell, to derive the one-loop gauge amplitudes
$A_{n;1}(1^\pm,2^+,3^+,\ldots,n^+)$~\cite{MahlonAllPlus}.  In this
appendix we discuss analogous currents in gravity, and show that the
half-soft functions also appear there.  The general factorization
properties of such currents --- which poles are present, with what
residues, etc. --- are very similar to the gauge case.  This is because
both tree-level gravity and gauge theory obey the supersymmetry Ward
identities~(\ref{SWIVanish}).  However, the lack of color-ordering in the
gravitational case leads to the appearance of more complicated structures.
As in the gauge-theory case, the off-shell currents will depend on the
choice of reference momenta for the on-shell gravitons.

We denote the generic gravitational current by $J^{\mu\alpha}(C)$, 
where the polarization tensor for the
off-shell graviton leg is described by the index pair $\mu\alpha$,
and $C \equiv \{1,2,\ldots,n\}$ stands for the set of on-shell
gravitons, with momenta $k_i$, $i=1,2,\ldots,n$.  By inspecting the
tree-level Feynman diagrams contributing to $J^{\mu\alpha}(C)$, it is easy
to derive the recursion relation,
$$
\eqalign{
  J^{\mu\alpha}(C) &= {1 \over K_C^2 } \Biggl\{
     \sum_{ {A \subset C \atop B \equiv C - A } } 
        V_{3,{\rm grav}}^{\mu\alpha,\nu\beta,\rho\gamma}(K_A,K_B)
        \ J^{\nu\beta}(A) J^{\rho\gamma}(B) \cr
&\hskip2cm
  +\ \hbox{terms from 4- and higher-point graviton vertices} \Biggr\} 
\,, \cr}
\equn\label{GenRecurse}
$$
where $K_A^\mu \equiv \sum_{i \in A} k_i^\mu$.
The sum here is over {\it distinct} partitions of the set $C$ into sets 
$A$ and $B$ (i.e. $B_0+A_0$ should not be counted in addition to $A_0+B_0$).

The three-graviton vertex $V_{3,{\rm grav}}$ is gauge-dependent in general.
Although not essential, it is possible to choose a gauge~\cite{GrantGauge}
in which $V_{3,{\rm grav}}$ can be written as the square of the 
color-ordered gauge 3-vertex in Gervais-Neveu gauge~\cite{GN}, 
plus terms with the cyclic ordering reversed,
$$
\eqalign{ 
V_{3,{\rm grav}}^{\mu\alpha,\nu\beta,\rho\gamma}(K_A,K_B)
 &= - {1\over2} \Biggl[ 
     \Bigl( \eta^{\rho\mu} K_B^\nu - \eta^{\mu\nu} (K_A+K_B)^\rho
          + \eta^{\nu\rho} K_A^\mu \Bigr)
     \Bigl( \eta^{\gamma\alpha} K_B^\beta 
          - \eta^{\alpha\beta} (K_A+K_B)^\gamma
          + \eta^{\beta\gamma} K_A^\alpha \Bigr)   \cr
&\hskip0.3cm
 +   \Bigl( - \eta^{\rho\mu} (K_A+K_B)^\nu + \eta^{\mu\nu} K_A^\rho
          + \eta^{\nu\rho} K_B^\mu \Bigr)
     \Bigl( - \eta^{\gamma\alpha} (K_A+K_B)^\beta 
            + \eta^{\alpha\beta} K_A^\gamma
            + \eta^{\beta\gamma} K_B^\alpha \Bigr) \Biggr] \,. \cr}
\equn\label{Grav3Vertex}
$$

The currents we consider in the following will all satisfy the 
conservation law,
$$
  K_C^\mu J^{\mu\alpha}(C) = K_C^\alpha J^{\mu\alpha}(C) = 0. 
\equn\label{JCons}
$$
In addition, they will vanish under the contractions,
$$
  J^{\nu\beta}(A) J^{\nu\gamma}(B) = J^{\nu\beta}(A) J^{\gamma\nu}(B)
    = J^{\beta\nu}(A) J^{\gamma\nu}(B) = J^{\mu\mu}(A) = 0.
\equn\label{JContract}
$$
For currents satisfying \eqn{JContract}, by using the fact that $m$-point 
vertices in gravity come from terms with two derivatives, i.e.,
$V_m \sim \partial^2 (h_{.\,.})^m$, it is easy to see that only the 3-vertex
can contribute to \eqn{GenRecurse}.  Furthermore, the 3-vertex can be 
simplified to 
$$
\eqalign{ 
V_{3,{\rm grav,eff}}^{\mu\alpha,\nu\beta,\rho\gamma}(K_A,K_B)
 &= - \Bigl( \eta^{\rho\mu} K_B^\nu - \eta^{\mu\nu} K_A^\rho \Bigr)
      \Bigl( \eta^{\gamma\alpha} K_B^\beta 
           - \eta^{\alpha\beta} K_A^\gamma \Bigr) \,. \cr}
\equn\label{NewGrav3Vertex}
$$

Define $J^{\mu\alpha}_{+q}(C)$ to be the all-plus current
$J^{\mu\alpha}(1^+,2^+,\ldots,n^+)$, where a common reference
momentum $q$ is chosen for all the gravitons; that is, their polarization
tensors in the spinor helicity formalism~\cite{SpinorHelicity} are chosen 
to be 
$$
  \pol^{\mu\alpha}_+(i) \equiv {1\over2}
  { \langle q^- | \gamma^\mu    |i^-\rangle
    \langle q^- | \gamma^\alpha |i^-\rangle  \over \spa{i}.{q}^2 }\ .
\equn\label{SameqTensor}
$$
This reference momentum choice implies the vanishing of 
$\pol^{\mu\alpha}(i)\pol^{\mu\beta}(j)$ for all pairs $i,j$,
which accounts for why eqs.~(\ref{JContract}) can be applied, and 
why 4- and higher-point vertices do not contribute in \eqn{GenRecurse}.

Now, $J^{\mu\alpha}_{+q}$ contains no multi-particle poles, because 
of the SWI~(\ref{SWIVanish}).  Explicit computation gives for the 
first few $J^{\mu\alpha}_{+q}$,
$$
\eqalign{
 J^{\mu\alpha}_{+q}(1^+) &\equiv 2\,\pol^{\mu\alpha}_+(1)
  = { \langle q^- | \gamma^\mu    |1^-\rangle
      \langle q^- | \gamma^\alpha |1^-\rangle  \over \spa1.{q}^2 } 
  = { \langle q^- | \gamma^\mu 1    |q^+\rangle
      \langle q^- | \gamma^\alpha 1 |q^+\rangle  \over \spa1.{q}^4 } \,, \cr
 J^{\mu\alpha}_{+q}(1^+,2^+) &= \langle q^- | \gamma^\mu (1+2) |q^+\rangle
      \langle q^- | \gamma^\alpha (1+2) |q^+\rangle \times
     { \spb1.2 \over \spa1.{q}^2 \spa2.{q}^2 \, \spa1.2 } \,, \cr
 J^{\mu\alpha}_{+q}(1^+,2^+,3^+) &= 
      \langle q^- | \gamma^\mu    (1+2+3) |q^+\rangle
      \langle q^- | \gamma^\alpha (1+2+3) |q^+\rangle \cr
&\hskip1cm \times
  \biggl( {\spb1.2\spb1.3\over \spa2.{q}^2\spa3.{q}^2 \spa1.2\spa1.3}
       +  {\spb1.2\spb2.3\over \spa1.{q}^2\spa3.{q}^2 \spa1.2\spa2.3}
       +  {\spb1.3\spb2.3\over \spa1.{q}^2\spa2.{q}^2 \spa1.3\spa2.3}
  \biggr) \,. \cr}
\equn\label{JPlusExplicit}
$$
The result for $J^{\mu\alpha}_{+q}(1^+,2^+,3^+)$ has been processed into
this form using the knowledge that the $1/(k_1+k_2+k_3)^2$ pole
must cancel.  

Comparing \eqn{JPlusExplicit} with the first few half-soft functions in
\eqn{HalfExamples}, it is clear that a specialized version of the 
half-soft coefficients appears in $J^{\mu\alpha}_{+q}(C)$, those where
$a$ and $b$ are both equal.  We are led to the ansatz
$$
J^{\mu\alpha}_{+q}(C) = 
      \langle q^- | \gamma^\mu    \Ksl_C |q^+\rangle
      \langle q^- | \gamma^\alpha \Ksl_C |q^+\rangle 
      \times h(q,C,q) \,.
\equn\label{JPlusAnsatz}
$$
Given this ansatz, \eqns{JCons}{JContract} hold, and the recursion 
relation~(\ref{GenRecurse}) reduces to a relation for $h$,
$$
 \sum_{ {A \subset C \atop B \equiv C - A } } 
   h(q,A,q) h(q,B,q) \langle q^-| \Ksl_A \Ksl_B | q^+\rangle^2
  = -K_C^2 \times h(q,C,q) \,.
\equn\label{hRecurse}
$$

In \app{HalfCoeffRecurseAppendix} we prove that the more general half-soft
functions that actually appear in the one-loop amplitudes, namely 
$h(q,M,r)$ with $q\neq r$, obey an obvious extension of \eqn{hRecurse},
$$
 \sum_{ {A \subset C \atop B \equiv C - A } } 
   h(q,A,r) h(q,B,r) \langle q^-| \Ksl_A \Ksl_B | q^+\rangle
                     \langle r^-| \Ksl_A \Ksl_B | r^+\rangle
  = -K_C^2 \times h(q,C,r) \,.
\equn\label{hRecurseqr}
$$
This recursion relation is what one might expect for an all-plus 
off-shell current,
$$
J^{\mu\alpha}_{+q,r}(C) \equiv 
      \langle q^- | \gamma^\mu    \Ksl_C |q^+\rangle
      \langle r^- | \gamma^\alpha \Ksl_C |r^+\rangle 
      \times h(q,C,r) \,,
\equn\label{JPlusAnsatzqr}
$$
corresponding to choosing graviton polarization tensors of the form
$$
  \pol^{\mu\alpha}_+(i) \equiv {1\over2}
  { \langle q^- | \gamma^\mu    |i^-\rangle
    \langle r^- | \gamma^\alpha |i^-\rangle  \over \spa{i}.{q}\spa{i}.{r} }\ .
\equn\label{Tensorqr}
$$
However, in contrast to the special case $q = r$, we have been unable to 
show for $q\neq r$ that \eqn{hRecurseqr} actually follows from the 
vertices in the Einstein action.  The difficulty is that for 
$q \neq r$ the 4- and higher-point graviton vertices also contribute, 
at least in a general gauge, because eqs.~(\ref{JContract}) are no longer 
obeyed by the ansatz.


\subsection{Double Off-Shell Scalar Currents $S$}
\label{SSubAppendix}

We consider one additional tree-level off-shell structure, the
doubly-off-shell scalar current $S_q(\ell;C)$, which contains
$n$ on-shell positive helicity gravitons for $C = \{1,2,\ldots,n\}$,
plus a massless scalar line with outgoing momenta $\ell$ and 
$-\ell-K_C$ at the two ends of the line.  By sewing the two off-shell ends of
the scalar line to each other (in conjunction with an additional
off-shell current), we can construct the one-loop all-plus gravity 
amplitudes $M_n(+,+,\ldots,+)$, in direct analogy to Mahlon's construction
of the one-loop all-plus gauge amplitudes 
$A_{n;1}(+,+,\ldots,+)$~\cite{MahlonAllPlus}.

In practice, the one-loop expressions generated by this procedure are 
rather complicated (because of the high superficial degrees of divergence
encountered), so in the end we have used them only as a numerical check.  
However, as a byproduct of calculating $S$ we have obtained a compact, 
non-recursive form for $h(q,M,r)$.
 
The recursion relation obeyed by $S_q$ is
$$
\eqalign{
  S_q(\ell;C) &= { -1 \over (\ell+K_C)^2 }
     \sum_{ {A \subset C \atop B \equiv C - A } } 
      S_q(\ell;A) \, (\ell+K_A)^\mu (\ell+K_A)^\alpha 
      J_{+q}^{\mu\alpha}(B) \cr
  &= { -1 \over (\ell+K_C)^2 }
     \sum_{ {A \subset C \atop B \equiv C - A } } 
      S_q(\ell;A) h(q,B,q) \langle q^-| (\lsl+\Ksl_A) \Ksl_B | q^+\rangle^2
\,, \cr}
\equn\label{SRecurse}
$$
where the contribution of higher-point scalar$+$graviton vertices can be
neglected using~\eqn{JContract}.  In \eqn{SRecurse} the partition
$B_0+A_0$ {\it is} distinct from $A_0+B_0$.  Also, the set $A$ (but not
$B$) is allowed to be the empty set $\{ \}$ (no graviton legs).  The
recursion is initiated by $S_q(\ell;\{ \}) = 1$.

As with the current $J$, it is useful to generalize $S_q(\ell;C)$ slightly
to $S_{q,r}(L;C)$, which is defined to obey
$$
\eqalign{
  S_{q,r}(L;C) &= { -1 \over (L+K_C)^2 }
     \sum_{ {A \subset C \atop B \equiv C - A } } 
      S_{q,r}(L;A) h(q,B,r) 
        \langle q^-| (\lsl+\Ksl_A) \Ksl_B | q^+\rangle
        \langle r^-| (\lsl+\Ksl_A) \Ksl_B | r^+\rangle \,, \cr}
\equn\label{SRecurseqr}
$$
and $S_{q,r}(L;\{ \}) = 1$.  Here $L$ is a $(4-2\e)$-dimensional 
momentum, with $L=\ell+\mu$ its decomposition into 4-dimensional and
$(-2\e)$-dimensional pieces.  We have found the following solution to
the recursion relation, 
$$
\eqalign{ 
S_{q,r}(L;C) &= L^2 \Biggl\{
   { 1 \over \spa1.{r} \spa{n}.{r} \, \prod_{j=1}^{n-1} \spa{j,}.{j+1} }
   \prod_{j=1}^n \left( { \langle q^-| (\lsl+\Ksl_{1,j}) | j^-\rangle 
                         \over \spa{q}.{j} } \right) \cr
&\hskip1cm 
\times \sum_{m=1}^n { 1 \over (L+K_{1,m-1})^2 \, (L+K_{1,m})^2 } 
   \biggl[ \langle r^-| (\lsl+\Ksl_{1,m}) \ksl_m | r^+\rangle 
    - \mu^2 { \langle r^-| \Ksl_{m+1,n} \ksl_m | r^+\rangle
              \over (L+K_{1,n})^2 } \biggr] \cr
&\hskip1cm 
+\ \Perm(1,2,\ldots,n) \Biggr\}. \cr}
\equn\label{SAnsatzMuqr}
$$
\Eqn{SAnsatzMuqr} is not actually the full solution for arbitrary $\mu$,
but rather the solution where the $\mu$-dependence is retained only in the
last step of the recursion, and is dropped in previous steps.  This
approximation in $\mu$ turns out to be good enough to allow one to sew up
$S_q \equiv S_{q,q}$ into the one-loop all-plus gravity amplitudes,
$M_n(+,+,\ldots,+)$ (see below).  The solution~(\ref{SAnsatzMuqr}) was
found by explicit calculation of the first few cases, making use of the
factorization properties.  (For example, for $\mu=0$, at most two
different multi-particle poles can co-exist in a given term, and they
should be of the form $1/(L+K_A)^2$ and $1/(L+K_A+k_m)^2$, where $A$
is some subset of $C$.)  \Eqn{SAnsatzMuqr} has been checked directly
up to $n=3$, and numerically for $\mu=0$ up through $n=8$; it also 
satisfies other consistency checks.

From the similarity between the recursion relations~(\ref{hRecurseqr}) 
and (\ref{SRecurseqr}), it is easy to show that $h$ may be constructed
from $S_{q,r}(L;C)$ by taking the four-dimensional limit ($\mu\to0$,
$L\to\ell$), and then the on-shell limit $\ell^2 \to 0$,
$$
h(q,\{\ell,C\},r) = { 1 \over \spa{\ell}.{q}^2 \spa{\ell}.{r}^2 } 
   \lim_{\ell^2 \to 0 \atop \mu \to 0} S_{q,r}(L;C).
\equn\label{SingleScalarAnsatz}
$$
This limit gives a compact, non-recursive, form for $h$, to be
contrasted with the recursive construction in \sec{HalfSoftSubSection}.
We get
$$
\eqalign{
h(q,\{1,2,\ldots,n\},r) &\equiv {\spb1.2 \over \spa1.2}
 { \spab{q}.{\Ksl_{1,2}}.3  \spab{q}.{\Ksl_{1,3}}.4 
   \cdots \spab{q}.{\Ksl_{1,n-1}}.{n}
  \over \spa2.3\spa3.4 \cdots \spa{n-1,}.{n} 
  \, \spa{q}.1 \spa{q}.2\spa{q}.3 \cdots \spa{q}.{n} 
  \, \spa1.{r} \spa{n}.{r} }  \cr
&\hskip1cm + \Perm(2,3,\ldots,n). \cr}
\equn\label{NonRecursiveHqr}
$$
Although this expression is not manifestly symmetric in $q \lr r$ or in 
$1 \lr j \in \{2,3,\ldots,n\}$, in \app{HalfCoeffEquivAppendix}
we show that \eqn{NonRecursiveHqr} is equivalent to the recursive 
form~(\ref{Hgendef}) in which these symmetries are manifest.


\subsection{Sewing of $S$ and $J$ to Obtain a Loop Amplitude}
\label{SewAppendix}

Because the current $S_q$ contains two off-shell ends of a scalar line,
one should be able to equate the momenta of the two ends, and integrate
over $L$, in order to obtain the scalar loop contribution to
$M_n(+,+,\ldots,+)$.  Using the SWI~(\ref{SWIVanish}), the scalar
contribution is the same, up to trivial overall factors, as the
contribution of a fermion or a graviton in the loop.
There are two subtleties:
\par
\begin{figure}[ht]
\centerline{\epsfxsize 2.7 truein \epsfbox{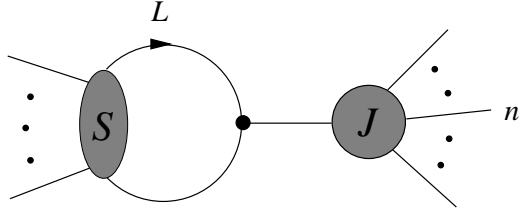}}
\vskip -.2 cm
\caption[]{
\label{SewSJFigure}
\small Sewing of a double-off-shell current $S$ and a single-off-shell
current $J$ to obtain the all-plus loop amplitude
$M_n(1^+,2^+,\ldots,n^+)$.  The particle circulating in the loop is a
scalar, while the remaining lines are gravitons.  The $n$th leg must
belong to $J$.}
\end{figure}

\noindent
(1) To avoid double-counting problems, one has to insert an additional
current $J$ into the sewing operation.  That is, one focuses on a
particular leg, say $n$, and defines the tree by which it is attached to
the loop to be $J^{\mu\alpha}(B)$ for some subset $B$, as depicted in  
\fig{SewSJFigure}.  The graviton legs 
not belonging to $B$ then belong to the complementary set $A \equiv C-B$, 
where $C = \{1,2,\ldots,n\}$.  They can all be assigned to the 
double-off-shell scalar current $S_q(L;A)$.  Thus we obtain, after
substituting in \eqn{JPlusAnsatz} for $J^{\mu\alpha}_{+q}$,
$$
M_n(1^+,2^+,\ldots,n^+) = \int {d^D L \over (2\pi)^D} 
  \sum_{B \subset C;\  n \in B  \atop A \equiv C - B}
  { S_q(L;A) \over L^2 } 
  \, \langle q^-| \lsl \Ksl_B | q^+\rangle^2 \, h(q,B,q).
\equn\label{SewAllPlus}
$$
\par\noindent
(2) As explained by Mahlon~\cite{MahlonAllPlus} for the analogous
gauge-theory case, one cannot set $D=4$ (i.e., $\mu=0$) immediately in 
\eqn{SewAllPlus}.   Even though the loop amplitudes
$M_n(1^+,2^+,\ldots,n^+)$ are ultraviolet finite, they are not finite
term-by-term in the form given by \eqn{SewAllPlus}.
Indeed, the $\mu$-independent terms in $S_q(L;C)$ (\eqn{SAnsatzMuqr}) 
give a vanishing contribution:  
They each have only two propagators, separated by a single 
external leg.  For these terms, the loop integral in \eqn{SewAllPlus} 
becomes a two-point (bubble) integral with a massless external momentum, 
which is zero in dimensional regularization.   

%

Thus we have to retain the $\mu$-dependence in the last step of the
recursion relation for $S_q$.  The $\mu^2$ terms in \eqn{SAnsatzMuqr} have
an additional propagator, so they correspond to a triangle integral in
$D=6-2\e$, multiplied by a factor of $-\e$.  Because such a triangle
integral has an ultraviolet $1/\e$ pole, we get a finite contribution as
$\e \to 0$.  The reason one does not have to retain $\mu$-dependence for
more than the last step of the $S_q$ recursion is related to the
finiteness of the triangle integral in $D=4$~\cite{MahlonAllPlus}.

It is important to note in the preceding arguments that every loop
momentum appears in a spinor string 
$\langle q^- | \lsl \ldots \rangle$.  Normally,
contraction terms between different loop momenta in the numerator
($\ell^\mu\ell^\nu \to \ell^2 \delta^{\mu\nu}$, etc.) give the leading
ultraviolet behavior.  Here, however, a Fierz identity implies that
all such contraction terms vanish:
$$
\langle q^- | \gamma^\mu \ldots \rangle
\langle q^- | \gamma_\mu \ldots \rangle
= 2 \spa{q}.{q} \ \langle \ldots \rangle = 0.
\equn\label{NoContract}
$$
The surviving terms in the triangle integrals are those obtained by shifting
the loop momenta by the usual amount depending on the Feynman parameters.

We have evaluated the relevant Feynman-parametrized triangle integrals
as $\e \to 0$ for both $M_4$ and $M_5$.  The analytic expressions obtained
are rather complicated; in particular the answers are not manifestly
independent of the reference momentum $q$.  However, numerical evaluation
shows that both answers are indeed independent of $q$, and that they 
agree with  the $D=4$ limits of $M_4$ and $M_5$ as obtained via cuts,
\eqns{FourGravAllPlusDFour}{FiveGravAllPlusDFour}.  

We have carried out a similar numerical check for $n=4$, in the case that 
$q \neq r$, using $J^{\mu\alpha}_{+q,r}$ and $S_{q,r}$.  This check is 
more complicated because: 
(a) there are a number of contraction terms now, proportional to $\spa{q}.{r}$;
(b) because of this, the $\mu^2$ approximation given for $S_{q,r}$ 
in \eqn{SAnsatzMuqr} is inadequate; and (c) also because of this,
box integrals as well as triangle integrals have to be evaluated.
Nevertheless, after making all three modifications the check works, 
which provides some evidence that the $q \neq r$ recursion relations 
we have used for $h(q,M,r)$ and $S_{q,r}$, \eqns{hRecurseqr}{SRecurseqr}, 
do actually follow from the Einstein action (in some gauge).


\section{Properties of the Half-Soft Functions}
\label{HalfCoeffAppendix}


\subsection{Equivalence of the Two Forms of $h(q,M,r)$}
\label{HalfCoeffEquivAppendix}

In this appendix, we shall show that the non-recursive
formula~(\ref{NonRecursiveHqr}) (or \eqn{NonRecursiveH}) for the half-soft
functions $h(q,M,r)$, where $M = \{1,2,\ldots,n\}$, is equivalent to the
recursive definition~(\ref{Hgendef}) in \sec{HalfSoftSubSection}.  
In the recursive solution, $h(q,M,r)$ is represented as a sum of terms 
of the form,
$$
{ \bX \over \aX } 
  \prod_{j=1}^{n} (\spa{q}.{j} \spa{j}.{r})^{i_j-1},
\equn\label{SquareBracketDecomp}
$$
where $\bX/\aX$ is one of the terms in the sum for the 
factor $\phi(i_1,i_2,\ldots,i_n)$.  That is, $\bX$ is a product 
(obeying certain rules) of $n-1$ square brackets $\spb{j}.{k}$, 
with $j,k \in M$; 
$\aX$ is the product of the corresponding angle brackets; and 
$i_j=$ (number of appearances of $j$ in $\bX$) $-\,1$. 
Through the recursive definition~(\ref{Phirecurse}) of the $\phi$'s,
each term~(\ref{SquareBracketDecomp}) can be built recursively, 
starting from
$$
\phi(0,0) {1 \over \spa{q}.{1} \spa{1}.{r}} 
          {1 \over \spa{q}.{2} \spa{2}.{r}}
\ =\ 
{\spb1.2 \over \spa1.2} {1 \over \spa{q}.{1} \spa{1}.{r}} 
                        {1 \over \spa{q}.{2} \spa{2}.{r}}\ ,
\equn\label{OldForm12}
$$
and multiplying by
$$
{\spb{m}.{l} \over \spa{m}.{l}} {\spa{q}.{m} \over \spa{q}.{l}}
{\spa{m}.{r} \over \spa{l}.{r}}
\equn\label{RecursiveMultFactor}
$$
at each step of the recursion, where $m$ is one of the `legs' 
added in a previous step, and $l$ is the new `leg' we add at this step.

Now, let us look at \eqn{NonRecursiveHqr}.   In each of the $(n-1)!$ terms in 
the permutation sum we can expand each factor, 
$$
\langle q^-| \Ksl_{1,l-1} | l^- \rangle 
\ =\ \sum_{j=1}^{l-1} \spa{q}.{j} \spb{j}.{l}
\ =\ \spa{q}.1 \spb1.{l} + \spa{q}.2 \spb2.{l} + \cdots
   + \spa{q,}.{l-1} \spb{l-1,}.{l}  \,,
\equn\label{ExpandFactor}
$$
and collect all terms having a given product of $(n-1)$ square brackets,
which we shall also call $\bX$.  We wish to show that 
\par\noindent
(a) the types of square bracket products $\bX$ that appear here are the 
same as in the recursive construction, and 
\par\noindent
(b) the products of angle bracket factors multiplying them are also the same.  
\par\noindent
To prove part (b) we will have to apply the Schouten identity on angle
brackets,
$$
\spa{i}.{j} \spa{k}.{l}\ =\ 
  \spa{i}.{k} \spa{j}.{l} + \spa{i}.{l} \spa{k}.{j} \,.
\equn\label{Schouten}
$$
We will {\it not} need the corresponding identity for square brackets;
i.e. the square brackets are already in the correct form.  
Our strategy for part (b) will be a recursive one, based on demonstrating
that the multiplicative factor~(\ref{RecursiveMultFactor}) arises in going
from $n-1$ to $n$.

First we establish part (a).  Note from \eqn{NonRecursiveHqr} that each $j
\in M$ (for $j\neq1$) appears in $\bX$ once `from the back' (i.e. from
$\spab{q}.{\s{K}_{\ldots}}.{j}$), and the remaining $i_j \geq0$ times
`from the front' (i.e. from $\spab{q}.{\ldots + \ksl_j + \ldots}.{l}$).
Because there are $n-1$ square brackets in $\bX$, each containing two
arguments, there are a total of $2(n-1) = \sum_{j=1}^n (i_j+1)$ arguments
in $\bX$, and we have $\sum_{j=1}^n i_j = n-2$, just as in the recursive
formula.  Since all the $i_j$ are non-negative, at least two of the $i_j$
must vanish.  Thus we can always find some $l$, with $l\neq1$, such that
$i_l=0$; i.e. $l$ appears exactly once in $\bX$, from the back.  It
appears in the factor $\spb{m}.{l}$, for some $m\neq l$.  Consider the
product $\bX$ with $\spb{m}.{l}$ removed from it, which no longer contains
$l$.  That is, define $\btX \equiv \bX / \spb{m}.{l}$, and for $j\neq l$
let $\tilde{i}_j =$ (number of appearances of $j$ in $\btX$) $-\,1$.  We
see that $\btX$ obeys the same `counting rules' as $\bX$, except with
$n-1$ arguments, because $l$ is no longer present: $\sum_{j\neq l}
\tilde{i}_j = n-3$, with all $\tilde{i}_j \geq0$, and each $j\neq l$
($j\neq1$) still appears at least once from the back.  Hence we can again
find some $l'$, with $l'\neq1$, such that $\tilde{i}_{l'}=0$; i.e. $l'$
appears exactly once in $\btX$, from the back.  Clearly, this procedure
can be repeated until we have only two arguments left, at which stage we
obtain the square bracket structure $\spb1.{j}$ corresponding to
$\phi(0,0)$.  We have thus established part (a) by working backwards
through the recursive construction of the $\phi$ factors,
\eqn{Phirecurse}.

To prove part (b), first note that the $\spa{q}.{j}$ factors in the
non-recursive formula already agree precisely with the recursive form, 
in {\it every} contributing permutation.  As noted above, in the numerator
of each term in \eqn{NonRecursiveHqr} each $j$ appears once from the back
and $i_j$ times from the front.  (For $j=1$, we count its appearance
in $\spb1.2$ as `from the back'.)   In the second case, it is 
always accompanied by $\spa{q}.{j}$, so we get ${\spa{q}.{j}}^{i_j}$. 
But we also have $\spa{q}.{j}$ in the denominator for each $j$, so overall
we get ${\spa{q}.{j}}^{i_j-1}$, as required by the recursive \eqn{Hgendef}.
Thus we only have to rearrange the spinor products of the form
$\spa{j}.{k}$ and $\spa{k}.{r}$, where $j,k\in M$.

In general, an individual term in the permutation sum for $h(q,M,r)$ in 
\eqn{NonRecursiveHqr} either contains the square bracket structure 
$\bX$ {\it once}, or does not contain it at all.  (The fact that $j<l$ in
\eqn{ExpandFactor} means that, for a given permutation, one can uniquely
determine which of the arguments in $\bX$ come from the front, and which
from the back.)   However, $\bX$ can appear in many {\it different} terms 
in the sum over $(n-1)!$ permutations.  The coefficient of 
$$
\bX \prod_{j=1}^{n} {\spa{q}.{j}}^{i_j-1}  
\equn\label{AlreadyGotRight}
$$
is
$$
{1 \over \spa1.{r}} \sum_{\sigma \in \Perm_{\bX}} 
{1\over \langle 1,\sigma_2\ldots \sigma_n, r \rangle}\,,
\equn\label{GoodPermSum}
$$
where the sum is over permutations $\sigma$ containing $\bX$,  
$\{\sigma_2\ldots\sigma_n\}$ is a permutation of $\{2\ldots n\}$,
and 
$$
\langle a,bc\ldots d,e\rangle 
\equiv \spa{a}.{b}\spa{b}.{c}\cdots\spa{d}.{e} \,.
\equn\label{ChainDef}
$$

As in the proof of (a), let $l\neq1$ be the leg appearing exactly once
in $\bX$, in the combination $\spb{m}.{l}$.  Since $l$ appears from the
back, only the permutations where $l$ appears {\it after} $m$ will
contribute.  In the permutation sum in \eqn{GoodPermSum}, let us 
hold fixed the positions of all legs except $l$, and just sum over
the insertions of $l$ after $m$. 
From the Schouten identity~(\ref{Schouten}) it is easy to 
derive the `eikonal' identity 
$$
\sum_{i=m}^{r-1}  
   { \spa{i,}.{i+1} \over \spa{i}.{l} \spa{l,}.{i+1} } 
\ =\  { \spa{m}.{r} \over \spa{m}.{l} \spa{l}.{r} }\ .
\equn\label{EikonalIdentity}
$$
This identity allows us to simplify the sum over $l$ insertions,
$$
\sum_{{l\ {\rm after}\ m} \atop 
       {\sigma_2\ldots \hat{l} \ldots \sigma_n\ {\rm fixed}}} 
{1\over \langle 1, \sigma_2 \ldots l \ldots \sigma_n, r \rangle} 
\ =\ {\spa{m}.{r} \over \spa{m}.{l} \spa{l}.{r}} 
{1\over \langle 1, \sigma_2 \ldots \hat{l} \ldots \sigma_n ,r \rangle} \,,
\equn\label{SimplifylSum}
$$
where a hat over $l$ signifies that it is no longer present.

We find that the terms in \eqn{NonRecursiveHqr} containing $\bX$ can 
be rewritten in terms of $\btX \equiv \bX / \spb{m}.{l}$ and $\tilde{i}_j$
as,
$$
{\spb{m}.{l} \over \spa{m}.{l}} {\spa{q}.{m} \over \spa{q}.{l}}
{\spa{m}.{r} \over \spa{l}.{r}}
\times \btX \prod_{j=1 \atop j\neq l}^{n} {\spa{q}.{j}}^{\tilde{i}_j-1}  
\times  {1 \over \spa1.{r}} \sum_{\sigma \in \Perm_{\btX}} 
{1\over \langle 1,\sigma_2\ldots \hat{l} \ldots \sigma_n, r \rangle}\,.
\equn\label{NewGoodPermSum}
$$
Note that the first factor is exactly the desired
factor~(\ref{RecursiveMultFactor}) appearing in the 
recursion relation when $l$ is added, and that the third factor is 
precisely the same as \eqn{GoodPermSum} with $l$ removed.
As in the proof of part (a), we can repeat this argument for
$\btX/\spb{m'}.{l'}$, etc., until we arrive at the case $n=2$, which
works by inspection.  Thus we have proven part (b), by showing that the 
coefficient of $\bX$ is correct, inductively in $n$.   


\subsection{Proof of $h(q,M,r)$ Recursion Relation}
\label{HalfCoeffRecurseAppendix}

In this appendix, we prove that the recursion relation~(\ref{hRecurseqr})
is satisfied by the $h(q,M,r)$.  The key observation is (as in
\app{HalfCoeffEquivAppendix}) that we do not need to use any Schouten 
identities on the square brackets (after we expand both sides so that 
they only contain square and angle brackets).  So, our strategy will be 
to look at the coefficients of a given square bracket structure on 
both sides of \eqn{hRecurseqr} and show that they are equal.

First we need to identify which square bracket structures can appear.  The
easiest way to do this is by representing such structures graphically.  
For each structure, we draw an `elk diagram', according to following rules:
\par\noindent
(a) Each element of $M$ corresponds to a node on the diagram. 
\par\noindent
(b) If the structure contains $\spb{i}.{j}$, draw a line connecting nodes
$i$ and $j$.
\par\noindent
Each elk diagram corresponds to one of the non-standard Young tableaux 
discussed in \sec{HalfSoftSubSection}.  The recursive
construction of the square bracket structures in $\phi$, \eqn{Phirecurse},
always adds a new node at each step.  Therefore every elk diagram
corresponding to a square bracket structure for $h(q,M,r)$ is a tree graph,
with at least one line attached to each node.

Now look at the right-hand side of \eqn{hRecurseqr}.  Any square bracket
structure appearing there is obtained by multiplying some structure from
$h(q,C,r)$ by one extra square bracket from $-K_C^2 = \sum_{i<j\in C}
\spa{i}.{j} \spb{i}.{j}$.  The corresponding elk diagram is obtained by
adding one line to the tree-like elk diagram for $h$.  The new diagram
must have one, and only one, closed loop.  Thus the new square bracket
structure will have exactly one closed cycle.  There are two possible
cases:%
\par\noindent%
(i) The closed cycle consists of just two square
brackets, $\spb{i}.{j}^2$.  In this case, it is clear that one of the
$\spb{i}.{j}$'s comes from $K_C^2$, while the other one comes from
$h$. The coefficient of a structure of this type consists of a single
term, which is:
$$
{\bX \over \aX} {\spb{i}.{j} \over \spa{i}.{j}} \times \spa{i}.{j}^2
\times \prod_{k=1}^n (\spa{q}.{k} \spa{k}.{r})^{i_k-1}, 
\equn\label{ShortCycle}
$$ 
where $\bX$ comes from $h$ (and contains $\spb{i}.{j}$), and $i_k=$ (number 
of appearances of $k$ in $X$) $-1$.%
\par\noindent%
(ii) The closed cycle contains several square brackets, e.g. $
[[1 \ldots k]] \equiv \spb1.2 \spb2.3 \ldots \spb{k-1,}.{k} \spb{k}.1 $. 
In this case, it is not possible to identify
uniquely which of the square brackets in the cycle came from $K_C^2$.  The 
coefficient of such a structure will contain $k$ terms. It is equal to:
$$
{[[1 \ldots k]] \over \langle\langle1 \ldots k\rangle\rangle} \times
\biggl( \sum_{\spb{i}.{j}\in {\rm cycle}} 
{\spa{i}.{j}^2 \over \spa{q}.{i} \spa{i}.{r} \spa{q}.{j} \spa{j}.{r}} \biggr)
\times {[Y] \over \langle Y\rangle} 
\prod_m (\spa{q}.{m} \spa{m}.{r})^{i_m-1}
\times \prod_{m \in {\rm cycle}} (\spa{q}.{m}\spa{m}.{r})^2 \,,
\equn\label{LongCycle}
$$
where $[Y]$ are the square brackets in $[X]$ that {\it do not} enter
the closed cycle, and $i_m=$ (number of appearances of $m$ in $Y$)
$-1$.

Next look at the left-hand side of \eqn{hRecurseqr}.  First, consider square
bracket structures of type (i). Note that only the terms with $i\in A$, 
$j\in B$ contribute, and both $\spb{i}.{j}$'s must come from
$\langle q^-| \Ksl_A \Ksl_B | q^+\rangle 
 \langle r^-| \Ksl_A \Ksl_B | r^+\rangle$, while
the $h$'s give rise to $[\hat{X}] \equiv [X]/\spb{i}.{j}$. There is only one 
such term. This is again easily seen using the elk diagram: cutting the 
`double link' $(ij)$ separates the original diagram into two unique tree
structures. Picking out the terms in $h(q,A,r)$ and $h(q,B,r)$ that contain
these structures, we get:
$$
{[\hat{X}] \over \langle \hat{X} \rangle } 
\times {\spb{i}.{j}^2 \over \spa{i}.{j}^2}
\times \spa{i}.{j}^2 \times \spa{q}.{i} \spa{i}.{r} \spa{q}.{j} \spa{j}.{r}
\times \prod_{l} (\spa{q}.{l} \spa{l}.{r})^{i_l-1}, 
\equn
$$ 
where $i_l=$ (number of appearances of $l$ in $\hat{X}$) $-1$. This is 
precisely eq.~(\ref{ShortCycle}).

Finally, consider square bracket structures of type (ii). Here, the square 
brackets coming from $\langle q^-| \Ksl_A \Ksl_B | q^+\rangle$
and $\langle r^-| \Ksl_A \Ksl_B | r^+\rangle$ {\it must} belong to the cycle. 
Suppose these square brackets are $\spb{i}.{j}$ and $\spb{k}.{l}$. If we cut
the links $(ij)$ and $(kl)$ in the original elk diagram, we separate it into
two tree-like diagrams; we can then pick up the unique term in 
$h(q,A,r) h(q,B,r)$ which gives rise to these structures. To get the full
coefficient of the original square bracket structure, we need to sum over
various combinations of $(ij)$ and $(kl) \not= (ij)$:
$$
{[[1 \ldots k]] \over \langle\langle1 \ldots k \rangle\rangle} \times
\biggl( \sum_{(ij) \not= (kl)} 
 {\spa{i}.{j} \spa{k}.{l} 
  \over \spa{r}.{i} \spa{r}.{j} \spa{q}.{k} \spa{q}.{l}} \biggr)  
\times {[Y] \over \langle Y\rangle} 
\prod_m (\spa{q}.{m} \spa{m}.{r})^{i_m-1}
\times \prod_{m \in {\rm cycle}} (\spa{q}.{m} \spa{m}.{r})^2 \,.
\equn\label{LClhs}
$$
Using the eikonal identity,
$$
\sum_{(kl) \not= (ij)} {\spa{k}.{l} \over \spa{q}.{k} \spa{q}.{l}} =
{\spa{i}.{j} \over \spa{q}.{i} \spa{q}.{j}} \,,
\equn
$$
we see that eqs.~(\ref{LongCycle}) and (\ref{LClhs}) are indeed the same. 


\section{Integrals}
\label{IntegralAppendix}

In this appendix we define our notation, and collect explicit 
expressions (through $\Ord(\eps^0)$) for the loop momentum integrals
used in the paper.  The one-loop scalar $m$-point integral in $D$
dimensions is defined by
$$
\eqalign{
\I_m^{K_1K_2\cdots K_m}
&\equiv \int { d^D L \over (2\pi)^D }
  { 1 \over L^2 (L-K_1)^2 (L-K_1-K_2)^2 \ldots 
    (L-\sum_{i=1}^{m-1} K_i)^2 } \cr
&\equiv \int {d^{4}\ell \over (2\pi)^{4} }
    {d^{-2\e} \mu \over (2\pi)^{-2\e} }
  { 1 \over (\ell^2 - \mu^2)((\ell-K_1)^2 - \mu^2) \ldots (
    (\ell-\sum_{i=1}^{m-1} K_i )^2 - \mu^2) } \,, \cr}
\equn\label{IntegralDef}
$$
where $K_1,K_2,\ldots,K_m$ are the (four-dimensional) external momenta for
the integral, and $L$ is the $D$-dimensional loop momentum, decomposed in
the second line into 4- and $(-2\e)$-dimensional components, $L=\ell+\mu$.
In general, the $K_i$ may be either individual massless external momenta
$k_j$ for the amplitude under consideration, or else sums of such external
momenta.  To simplify the notation, in the former case we will replace
$K_i$ in the argument of $\I_m$ simply by the appropriate integer index
$j$; in the latter case we will often replace it by the set of integers
entering the momentum sum, enclosed in parentheses.  For example, one of the
box integrals encountered in six-point amplitudes contains two
diagonally-opposite massive external legs, with masses $s_{23}$ and
$s_{56}$, and is given by
$$
\I_4^{1(23)4(56)} \equiv \int { d^D L \over (2\pi)^D }
  { 1 \over L^2 (L-k_1)^2 (L-k_1-k_2-k_3)^2 (L+k_5+k_6)^2 } \,.
\equn\label{TwoMassBoxIntegralDef}
$$

Integrals where an additional factor of $(\mu^2)^r \equiv \mu^{2r}$ 
has been inserted into the loop integrand are denoted by
$$
\I_m^{K_1K_2\cdots K_m}[\mu^{2r}] 
\equiv \int {d^{4}\ell \over (2\pi)^{4} }
    {d^{-2\e} \mu \over (2\pi)^{-2\e} }
  { \mu^{2r} \over (\ell^2 - \mu^2)((\ell-K_1)^2 - \mu^2) \cdots (
    (\ell-\sum_{i=1}^{m-1} K_i )^2 - \mu^2) } \,,
\equn\label{MuIntegralDefApp}
$$
These integrals can be expressed, via \eqn{DimShiftIntegral}, in terms
of the integrals $\I_m^{D=4+2r-2\e}$, that is, $\I_m$ in \eqn{IntegralDef}
with $D$ replaced by $D+2r$.

The $N=4$ super-Yang-Mills and $N=8$ supergravity amplitudes contain 
scalar box integrals with one and two external masses.  
The two-mass box integral with two diagonally-opposite massive legs 
(see \fig{EasyTwoMassFigure}), evaluated in $D=4-2\eps$, is~\cite{Integrals}
$$
\eqalign{
  \I_4^{a K_1 b K_2} &= 
i \, { \cg \over S_{1a} S_{1b} - K_1^2 K_2^2 } 
\biggl\{
  {2\over\e^2} \Bigl[ (-S_{1a})^{-\e} + (-S_{1b})^{-\e}
              - (-K_1^2)^{-\e} - (-K_2^2)^{-\e} \Bigr] \cr
  &\ -\ 2\ \Li_2\left(1-{K_1^2\over S_{1a}}\right)
   \ -\ 2\ \Li_2\left(1-{K_1^2\over S_{1b}}\right)
   \ -\ 2\ \Li_2\left(1-{K_2^2\over S_{1a}}\right)
   \ -\ 2\ \Li_2\left(1-{K_2^2\over S_{1b}}\right) \cr
  &\ +\ 2\ \Li_2\left(1-{K_1^2 K_2^2 \over S_{1a} S_{1b}}\right)
   \ - \ \ln^2\left({S_{1a} \over S_{1b}}\right) \biggr\}
\ +\ \Ord(\e)\,, \cr }
\equn\label{EasyTwoMassAnswer}
$$
where $S_{1a} = (K_1 + k_a)^2$, $S_{1b} = (K_1 + k_b)^2$, and
$$
\cg = { 1 \over (4 \pi)^{2-\e} }
{ \Gamma(1+\e) \Gamma^2(1-\e) \over \Gamma(1-2\e) }\,.
\equn\label{cGammaPrefactor}
$$
One-mass box integrals also appear in the amplitudes, corresponding to the
case where $K_1$ reduces to a single external momentum.  They are also
given by \eqn{EasyTwoMassAnswer} --- one may set $K_1^2 = 0$ after
dropping the $(-K_1^2)^{-\e}$ term.  (One may similarly set $K_1^2 =
K_2^2 = 0$ to obtain the box integral with no external masses
encountered in the four-point amplitudes.)

For the all-plus gauge amplitudes~(\ref{ExactYM}) and gravity
amplitudes~(\ref{FourGravAllPlus}), (\ref{FiveGravAllPlus}) and 
(\ref{SixGravAllPlus}), only the ultraviolet-singular parts ($1/\e$ poles)
of the higher-dimensional integrals $\I_m^{D=4+2r-2\e}$ contribute in the
four-dimensional limit $\e\to0$, due to the overall prefactor of 
$\e$ in \eqn{DimShiftIntegral}.  Such terms are given by elementary 
integrals of polynomials in the Feynman parameters.

In the gauge theory case, the only divergent integrals that appear 
are $D=8-2\eps$ box integrals and $D=10-2\eps$ pentagon integrals. 
The box integral is 
$$
\eqalign{ \I_4 [\mu^4] & = -\eps (1-\eps) (4 \pi)^2 \I_4^{D=8-2\e} \cr
&= -i \eps (1-\eps) {\Gamma(\e)\over (4\pi)^{2-\e}} \int d^4a_i\ 
\delta\Bigl(1-\sum_i a_i\Bigr)\ 
  (-s_{12}a_1a_3-s_{23}a_2a_4+\cdots)^{-\e} \cr & =
-{i\over(4\pi)^2}\, {1\over 6} + \Ord(\eps) \,. \cr}
\equn\label{I4D8}
$$
Since $\I_4^{D=8-2\e}$ is dimensionless, the 
pole in $\eps$ does not depend on the particular kinematic
configuration, so we have suppressed the labels describing the 
kinematics.  (External masses would contribute to the
`$+\cdots$' terms in \eqn{I4D8}.)
Similarly, the pentagon integral that appears is
$$
\eqalign{
\I_5[\mu^6] & = -\eps(1-\eps)(2-\eps) (4\pi)^3 \, \I_5^{D=10-2\e}\cr
& = i \eps(1-\eps)(2-\eps) {\Gamma(\e) \over (4 \pi)^{2-\e}}
\int d^5a_i\ \delta\Bigl(1-\sum_i a_i\Bigr)\ 
(-s_{45}a_4a_1+\hbox{cyclic}+\cdots)^{-\e} \cr
& = {i\over(4\pi)^2}\, {1\over 12}
 + \Ord(\eps)  \,, \cr}
\equn\label{I5D10}
$$
where the leading term again does not depend on the kinematic
configuration.  The hexagon integral in \eqn{ExactYM} is
$$
\I_{6}[\mu^6] =  -\eps(1-\eps)(2-\eps) (4\pi)^3 \I_6^{D=10-2\e} 
= \Ord(\eps)\,,
\equn
$$
since $\I_6^{D=10-2\e}$ is finite.

In the all-plus gravity amplitudes, the required higher-dimensional
integrals are dimensionful, and therefore depend on the kinematics.
We give here the box and pentagon integrals with the maximum required
number of external masses; those with fewer masses can be obtained by
setting masses to zero.  The two-mass box integral that
appears in the amplitudes is
$$
\eqalign{
\I_4^{a K_1 b K_2}[\mu^8] & = 
- \e (1-\e) (2-\e) (3-\e) (4 \pi)^4\, \I_4^{a K_1 b K_2,\, D=12-2\e} \cr
& =  - i \e (1-\e) (2-\e) (3-\e) {\Gamma(-2+\e) \over (4\pi)^{2-\e}} \cr
& \hskip 1.5 cm \times 
 \int d^4a_i\ \delta\Bigl(1-\sum_i a_i\Bigr)\ 
( - S_{1a} \, a_1a_3 - S_{1b} \, a_2a_4 
  - K_1^2 \, a_2 a_3 - K_2^2 \, a_4 a_1)^{2-\e} \cr
& =  -  {i\over (4\pi)^{2} }{1\over840} \Bigl[
 2 S_{1a}^2 + 2 S_{1b}^2 + 2 (K_1^2)^2 + 2 (K_2^2)^2 
 + S_{1a} S_{1b} + K_1^2 K_2^2 \cr 
& \hskip 3 cm 
 + 2 (S_{1a} + S_{1b})(K_1^2 + K_2^2) \Bigr] +\Ord(\eps)\,, \cr}
\equn\label{I4D12}
$$
where the kinematic configuration is depicted in \fig{EasyTwoMassFigure}.

The one-mass pentagon integral that appears in the six-point all-plus
gravity amplitude is
$$
\eqalign{
\I_5^{1234(56)}[\mu^{10}] &= 
 - \e (1-\e) \cdots (4-\eps) (4 \pi)^5 \,
\I_5^{1234(56),\,D=14-2\e} \cr
&=   i \e (1-\e) \cdots (4-\eps)\, 
{\Gamma(-2+\e) \over (4 \pi)^{2-\e}} 
\int d^5a_i\ \delta\Bigl(1-\sum_i a_i\Bigr) \cr
& \hskip 1.4 cm \times 
(-s_{12} a_1 a_3 - s_{23} a_2 a_4 - s_{34}a_3 a_5 - t_{456} a_4 a_1
 - t_{561} a_5 a_2 - s_{56} a_5 a_1)^{2-\e} \cr
& =  {i\over (4 \pi)^2} {1\over 1680}\Bigl[
   2 s_{12}^2 + 2 s_{23}^2 + 2 s_{34}^2 + 2 s_{56}^2 
 + 2 t_{456}^2 + 2 t_{561}^2 
 + 2 s_{12} s_{34} + 2 s_{12} s_{56} + 2 s_{34} s_{56} \cr
& \hskip 3cm
 + 2 s_{12} t_{456} + 2 s_{34} t_{561} 
 + 2 (s_{23}+s_{56}) (t_{456}+t_{561}) \cr
& \hskip 3cm
 + s_{12} s_{23} + s_{23} s_{34} + s_{23} s_{56} 
 + s_{12} t_{561} + s_{34} t_{456} + t_{456} t_{561}
 \Bigr]  + \Ord(\e). \cr}
\equn\label{OneMassPentPole}
$$
To obtain the massless pentagon integral appearing in the 
five-point amplitude from \eqn{OneMassPentPole},
simply replace $k_5+k_6$ by $k_5$, i.e., $s_{56}\to 0$, 
$t_{456}\to s_{45}$, and $t_{561}\to s_{51}$.


\end{document}